\documentclass[aps,pre,showpacs,noshowkeys,amsmath,amssymb,amsfonts,superscriptaddress,reprint]{revtex4-1}

\usepackage{graphicx}
\usepackage{bm}
\usepackage{physics}
\usepackage{mathtools}
\setcitestyle{super}
\usepackage{subcaption}
\usepackage{caption}
\DeclareCaptionLabelSeparator{bar}{~\rule[-0.4ex]{0.2ex}{1em}~}
\DeclareCaptionLabelFormat{subfor}{\textbf{#2}}
\newcommand*\bfcaption[2]{\caption[#1]{\textbf{#1.}#2}}
\usepackage{xcolor}
\definecolor{UBcolor}{HTML}{007CC1}
\usepackage[colorlinks=true,pdfnewwindow=true,linkcolor=UBcolor,citecolor=UBcolor,urlcolor=UBcolor,breaklinks=true,linktocpage]{hyperref}
\usepackage[all]{hypcap}
\usepackage[nameinlink,capitalise]{cleveref}
\begin{document}

\title{Active wetting of epithelial tissues}

\author{Carlos P\'{e}rez-Gonz\'{a}lez}
\altaffiliation{These authors contributed equally to this work.}
\affiliation{Institute for Bioengineering of Catalonia, The Barcelona Institute for Science and Technology (BIST), 08028 Barcelona, Spain}
\affiliation{Facultat de Medicina, University of Barcelona, 08028 Barcelona, Spain}

\author{Ricard Alert}
\altaffiliation{These authors contributed equally to this work.}
\affiliation{Departament de F\'{i}sica de la Mat\`{e}ria Condensada, Universitat de Barcelona, Av. Diagonal 647, 08028 Barcelona, Spain}
\affiliation{Universitat de Barcelona Institute of Complex Systems (UBICS), Universitat de Barcelona, 08028 Barcelona, Spain}

\author{Carles Blanch-Mercader}
\affiliation{Laboratoire Physico Chimie Curie, Institut Curie, PSL Research University - Sorbonne Universit\'{e}s, UPMC-CNRS UMR 168, 26 rue d'Ulm, 75005 Paris, France}
\affiliation{Department of Biochemistry, Faculty of Sciences II, University of Geneva, 30 Quai Ernest-Ansermet, 1205 Gen\`{e}ve, Switzerland}

\author{Manuel G\'{o}mez-Gonz\'{a}lez}
\affiliation{Institute for Bioengineering of Catalonia, The Barcelona Institute for Science and Technology (BIST), 08028 Barcelona, Spain}

\author{Tomasz Kolodziej}
\affiliation{Faculty of Physics, Astronomy and Applied Computer Science, Jagiellonian University in Krak\'{o}w, 30-348 Krak\'{o}w, Poland}

\author{Elsa Bazelli\`{e}res}
\affiliation{Institute for Bioengineering of Catalonia, The Barcelona Institute for Science and Technology (BIST), 08028 Barcelona, Spain}

\author{Jaume Casademunt}
\email{jaume.casademunt@ub.edu}
\affiliation{Departament de F\'{i}sica de la Mat\`{e}ria Condensada, Universitat de Barcelona, Av. Diagonal 647, 08028 Barcelona, Spain}
\affiliation{Universitat de Barcelona Institute of Complex Systems (UBICS), Universitat de Barcelona, 08028 Barcelona, Spain}

\author{Xavier Trepat}
\email{xtrepat@ibecbarcelona.eu}
\affiliation{Institute for Bioengineering of Catalonia, The Barcelona Institute for Science and Technology (BIST), 08028 Barcelona, Spain}
\affiliation{Facultat de Medicina, University of Barcelona, 08028 Barcelona, Spain}
\affiliation{Instituci\'{o} Catalana de Recerca i Estudis Avan\c{c}ats (ICREA), Barcelona, Spain}
\affiliation{Centro de Investigaci\'{o}n Biom\'{e}dica en Red en Bioingenier\'{i}a, Biomateriales y Nanomedicina, Spain}

\date{\today}

\begin{abstract}
Development, regeneration and cancer involve drastic transitions in tissue morphology. In analogy with the behavior of inert fluids, some of these transitions have been interpreted as wetting transitions. The validity and scope of this analogy are unclear, however, because the active cellular forces that drive tissue wetting have been neither measured nor theoretically accounted for. Here we show that the transition between 2D epithelial monolayers and 3D spheroidal aggregates can be understood as an active wetting transition whose physics differs fundamentally from that of passive wetting phenomena. By combining an active polar fluid model with measurements of physical forces as a function of tissue size, contractility, cell-cell and cell-substrate adhesion, and substrate stiffness, we show that the wetting transition results from the competition between traction forces and contractile intercellular stresses. This competition defines a new intrinsic lengthscale that gives rise to a critical size for the wetting transition in tissues, a striking feature that has no counterpart in classical wetting. Finally, we show that active shape fluctuations are dynamically amplified during tissue dewetting. Overall, we conclude that tissue spreading constitutes a prominent example of active wetting --- a novel physical scenario that may explain morphological transitions during tissue morphogenesis and tumor progression.
\end{abstract}

\maketitle

Living tissues are active materials with the ability to undergo drastic transitions in shape and dimensionality\cite{Gonzalez-Rodriguez2012}. When properly controlled, such morphological transitions enable development and regeneration. When regulation fails, however, aberrant morphological transitions underlie developmental defects and tumour formation\cite{Friedl2009,Julicher2017}. Transitions in tissue shape are regulated by a myriad of molecular processes that act upon a limited number of physical properties to ultimately determine tissue dynamics. To understand the nature of these physical properties and their impact on tissue shape, extensive research has focused on how a three-dimensional cell aggregate spreads on a substrate\cite{Ryan2001,Douezan2011,Douezan2012c,Beaune2014,Beaune2017}. Besides mimicking biological processes such as epiboly in zebrafish\cite{Behrndt2012,Campinho2013,Morita2017,Wallmeyer2018}, the spreading of a cell aggregate is amenable to theoretical and experimental access, and has become a widespread model process.

Given the fluid behaviour of cell aggregates at long times, their spreading on a substrate has been studied as a wetting problem\cite{Gonzalez-Rodriguez2012}. In analogy with the case of a fluid drop, the extent to which the aggregate spreads on the substrate has been proposed to rely on a competition between cell-cell ($W_{\text{cc}}$) and cell-substrate ($W_{\text{cs}}$) adhesion energies\cite{Ryan2001,Douezan2011} encoded in the so-called spreading parameter $S=W_{\text{cs}}-W_{\text{cc}}$. This parameter changes sign at the wetting transition that separates tissue spreading ($S>0$) from retraction into a droplet-like aggregate ($S<0$)\cite{Douezan2011,Douezan2012c,Douezan2012a,Smeets2016}. This analogy with the classical theory of wetting has successfully explained aspects of tissue wetting such as changes in contact angle as a function of cell-cell and cell-extracellular matrix (ECM) adhesion\cite{Ravasio2015a}. However, this conceptual framework overlooks the active nature of living tissues and, hence, it does not explicitly account for the ability of cells to polarize, generate traction forces, and couple such forces with adhesion dynamics. To a great extent, this limitation stems from the lack of direct measurements of cell-cell and cell-matrix forces during tissue wetting and dewetting.

To overcome these experimental and theoretical limitations, we performed a systematic quantitative study of the mechanics of tissue wetting as a function of cell-cell and cell-matrix adhesion, ECM ligand density, ECM stiffness, tissue size, and contractility. Our results cannot be explained solely in terms of the physics of passive fluids. Instead, we show that the tissue wetting transition and dewetting dynamics are well captured by a new framework for active wetting based on an active polar fluid model of tissue spreading.

\vskip0.5cm
\noindent\textbf{E-cadherin expression induces dewetting of a cell monolayer}

\begin{figure*}[p]
\begin{center}
\includegraphics[width=0.75\textwidth]{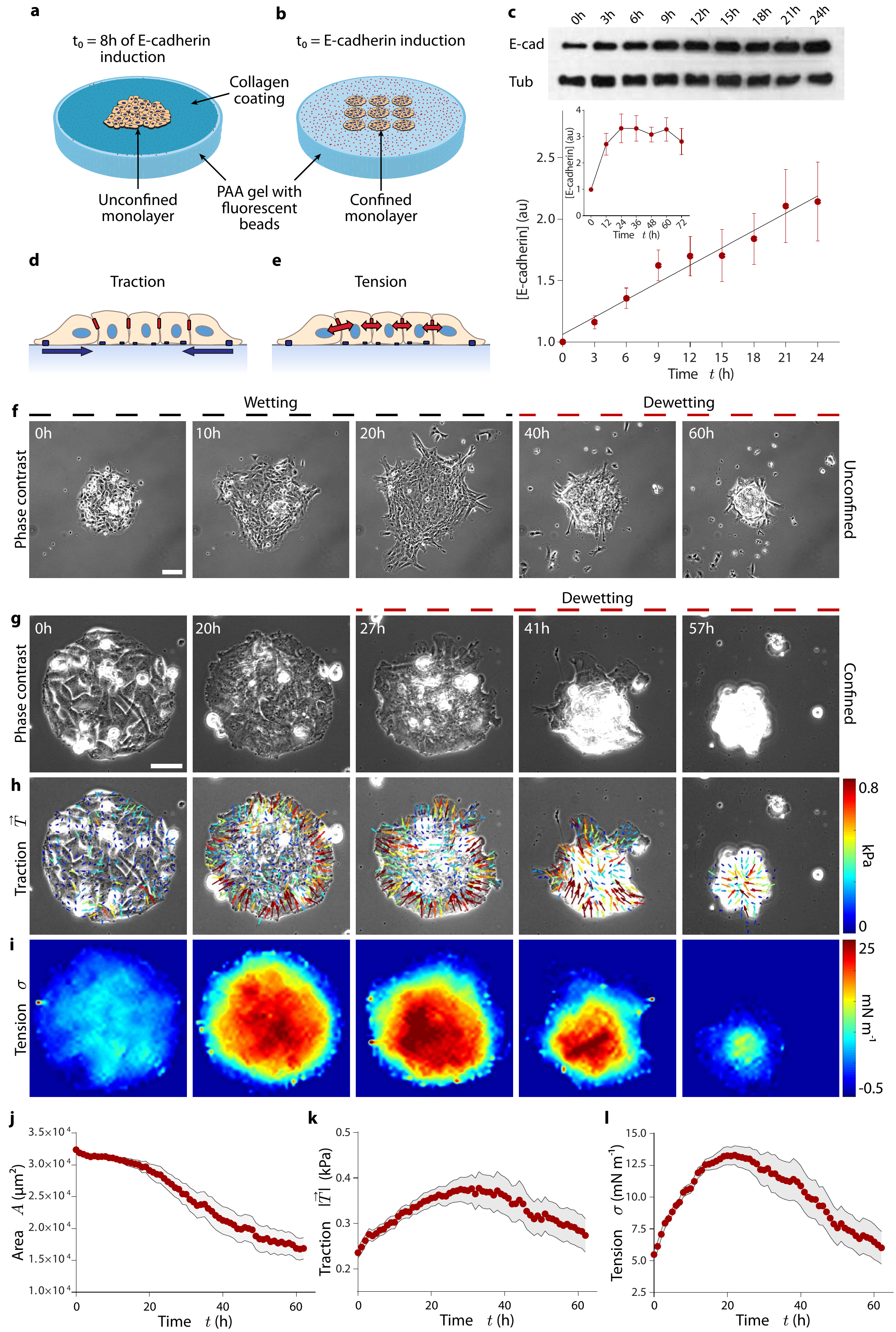}
\end{center}
  {\phantomsubcaption\label{Fig1a}}
  {\phantomsubcaption\label{Fig1b}}
  {\phantomsubcaption\label{Fig1c}}
  {\phantomsubcaption\label{Fig1d}}
  {\phantomsubcaption\label{Fig1e}}
  {\phantomsubcaption\label{Fig1f}}
  {\phantomsubcaption\label{Fig1g}}
  {\phantomsubcaption\label{Fig1h}}
  {\phantomsubcaption\label{Fig1i}}
  {\phantomsubcaption\label{Fig1j}}
  {\phantomsubcaption\label{Fig1k}}
  {\phantomsubcaption\label{Fig1l}}
\bfcaption{E-cadherin expression causes an increase in traction forces and monolayer tension, and induces dewetting}{ \subref*{Fig1a},\subref*{Fig1b}, Scheme of the experimental setups. For spreading experiments, cells form a monolayer within the circular opening of a PDMS membrane. After 8 hours in the dexamethasone-containing medium, the PDMS membrane is removed and the monolayer spreads on the collagen-coated substrate (\subref*{Fig1a}). For confined monolayers, cells are seeded in circular islands of collagen on the substrate and allowed to cover them for 3 hours. Dexamethasone is then added to induce E-cadherin expression and time-lapse imaging starts (\subref*{Fig1b}). \subref*{Fig1c}, Quantification of E-cadherin upon addition of dexamethasone (inset, up to 3 days). \subref*{Fig1d},\subref*{Fig1e}, Illustration of traction forces (\subref*{Fig1d}) and monolayer tension (\subref*{Fig1e}). \subref*{Fig1f}, Spreading monolayer exhibiting a wetting transition at time $t=25$ h. Scale bar $= 100$ $\mu$m. \subref*{Fig1g}-\subref*{Fig1i}, Phase contrast images (\subref*{Fig1g}), and maps of traction forces (\subref*{Fig1h}) and average normal monolayer tension (\subref*{Fig1i}) for a representative confined cell island of radius $100$ $\mu$m. Monolayer dewetting starts at $\sim 25$ h. Scale bar $= 40$ $\mu$m. \subref*{Fig1j}-\subref*{Fig1l}, Evolution of monolayer area (\subref*{Fig1j}), mean traction magnitude (\subref*{Fig1k}) and average normal monolayer tension (\subref*{Fig1l}). Data are presented as mean $\pm$ s.e.m. $n=18$ cell islands.} \label{Fig1}
\end{figure*}

We designed an experimental assay to study wetting transitions of epithelial clusters induced by controlled changes in tissue mechanics. The idea behind the experimental approach is to progressively increase cell-cell adhesion in an epithelial monolayer while measuring its effect on cellular forces and tissue spreading. To this end, we use human breast adenocarcinoma cells (MDA-MB-231) transfected with a dexamethasone-inducible vector containing human E-cadherin mRNA\cite{Sarrio2009}. In the absence of dexamethasone, these metastatic cells do not express significant levels of cell-cell adhesion proteins. Upon adding dexamethasone, the concentration of E-cadherin increases almost linearly in time for $\sim 24$ h and plateaus thereafter (\cref{Fig1c,FigS1}). To study tissue wetting, MDA-MB-231 cells are seeded on a $12$ kPa polyacrylamide gel coated with collagen I (\cref{Fig1a}). Initially, the cells form a monolayer within a circular opening of a PDMS membrane deposited on the gel. 8 hours after E-cadherin induction, the confining membrane is removed and the monolayer spreads. However, after $\sim 20$ hours, the monolayer often starts retracting, eventually becoming a spheroidal cell aggregate (\cref{Fig1f,FigS2}; Supplementary Movies 1, 2). Thus, the monolayer undergoes a transition from wetting to dewetting, which we refer to hereafter generically as wetting transition.

To reproducibly study this transition, we seed cells on adherent (collagen I-coated) circular islands of controlled size ($100$ $\mu$m in radius) surrounded by an uncoated surface that cells cannot invade (\cref{Fig1b}). We use Traction Force Microscopy to measure traction forces on the substrate (\cref{Fig1d})\cite{Trepat2009}, and Monolayer Stress Microscopy to measure tension within and between cells (\cref{Fig1e})\cite{Serra-Picamal2012,Tambe2011}. A few hours after E-cadherin induction, monolayers become cohesive (\cref{Fig1g}). Cells at the edge polarize by extending lamellipodia towards the exterior of the island, generating radially-oriented inwards-pointing tractions (\cref{Fig1h,FigS3})\cite{Mertz2013,Mertz2012}. Monolayer tension increases from the edge of the monolayer and reaches a maximum at the center (\cref{Fig1i}). Note that monolayer tension is a bulk property and should not be confused with the interfacial surface tension that plays a central role in classical wetting phenomena. During the first $\sim 25$ hours of the experiment, tractions (\cref{Fig1k}) and tension (\cref{Fig1l}) rise in parallel with the increase in E-cadherin. As for unconfined spreading monolayers, the monolayer eventually retracts, decreasing its area (\cref{Fig1j}) and dewetting the substrate to form a spheroidal aggregate, thus completing a transition from 2D to a 3D tissue geometry (\cref{FigS4}; Supplementary Movies 3, 4).

\vskip0.5cm
\noindent\textbf{Formation of E-cadherin junctions activates myosin}

\begin{figure*}[p]
\begin{center}
\includegraphics[width=0.85\textwidth]{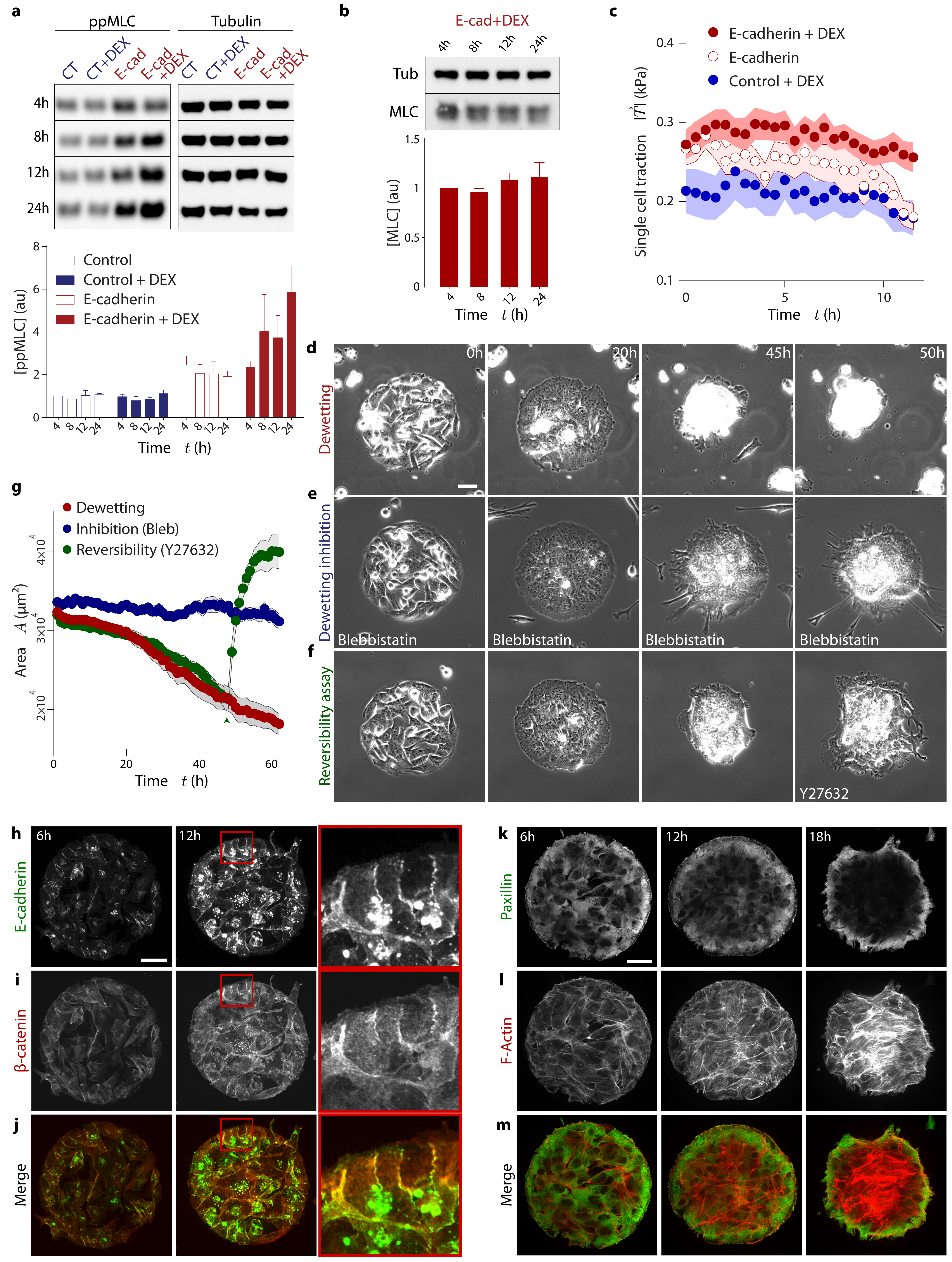}
\end{center}
  {\phantomsubcaption\label{Fig2a}}
  {\phantomsubcaption\label{Fig2b}}
  {\phantomsubcaption\label{Fig2c}}
  {\phantomsubcaption\label{Fig2d}}
  {\phantomsubcaption\label{Fig2e}}
  {\phantomsubcaption\label{Fig2f}}
  {\phantomsubcaption\label{Fig2g}}
  {\phantomsubcaption\label{Fig2h}}
  {\phantomsubcaption\label{Fig2i}}
  {\phantomsubcaption\label{Fig2j}}
  {\phantomsubcaption\label{Fig2k}}
  {\phantomsubcaption\label{Fig2l}}
  {\phantomsubcaption\label{Fig2m}}
\bfcaption{Formation of E-cadherin junctions induces myosin phosphorylation, and hence the increase in tension that is responsible for monolayer dewetting}{ \subref*{Fig2a}, Evolution of active myosin light chain (ppMLC) concentration. Control = Mock transduced cells; E-cad = cells transduced with E-cadherin under the dex-inducible promoter; DEX= treatment with dexamethasone. \subref*{Fig2b}, Evolution of total myosin light chain (MLC) concentration. \subref*{Fig2c}, Evolution of the average traction magnitude in single cells. \subref*{Fig2d}-\subref*{Fig2f}, Phase contrast images of a dewetting experiment (\subref*{Fig2d}), a cell island treated with blebbistatin (\subref*{Fig2e}) and a cell island treated with Y27632 once dewetting has started ($t=46$ h) (\subref*{Fig2f}). (\subref*{Fig2g}) Evolution of monolayer area for dewetting, dewetting inhibition and reversibility assays (green arrow indicates addition of Y27632). \subref*{Fig2h}-\subref*{Fig2j}, Immunostaining of E-cadherin (\subref*{Fig2h}), Beta-catenin (\subref*{Fig2i}) and merge images (\subref*{Fig2j}) at $6$ h and $12$ h after induction of E-cadherin expression (red square indicates the inset). \subref*{Fig2k}-\subref*{Fig2m}, Immunostaining of Paxillin (\subref*{Fig2j}), Actin (\subref*{Fig2k}) and merge images (\subref*{Fig2m}) at $6$, $12$ and $18$ hours after induction of E-cadherin expression. Scale bars $= 40$ $\mu$m. Data are presented as mean $\pm$ s.e.m. Single cell tractions: $n=24$ cells. Dewetting inhibition: $n=9$ cell islands. Reversibility assay: $n=16$ cell islands.} \label{Fig2}
\end{figure*}

To study the mechanisms underlying the increase of tension we measure myosin levels and activity. During the first 24 hours of E-cadherin expression, myosin levels remain constant but diphosphorylated myosin light chain (ppMLC) exhibits a $\sim 3$-fold increase (\cref{Fig2a,Fig2b}). Untransfected cells (CT) or transfected cells lacking dexamethasone in their medium (labelled E-cad) show constant ppMLC levels (\cref{Fig2a}), indicating that the observed response is not attributable to a secondary effect of dexamethasone addition or to transfection artifacts. Unlike in cohesive monolayers, expression of E-cadherin does not lead to an increase of tension in single cells (\cref{Fig2c}). Consistently, monolayers show higher levels of ppMLC than single cells several hours after induction (\cref{FigS5}). Moreover, abrogating cell-cell adhesions with EGTA (2 mM) prevents the buildup of traction and tension, as well as the wetting transition (\cref{FigS6}, Supplementary Movie 5). Thus, we conclude that E-cadherin regulates myosin-generated contractility through a mechanism dependent on cell-cell junction formation. Notably, E-cadherin not only affects intercellular forces but also tractions, ultimately determining the global mechanics of the monolayer\cite{Harris2014,Bazellieres2015,Lecuit2015,Muhamed2016}.

\vskip0.5cm
\noindent\textbf{Tissue tension induces the wetting transition}

Next, we study the reorganization of adhesive and cytoskeletal structures in the monolayer. Upon induction of its expression, E-cadherin progressively accumulates at cell-cell contacts (\cref{Fig2h}) and colocalizes with $\beta$-catenin (\cref{Fig2i,Fig2j}), confirming the formation of adherens junctions. In parallel, the focal adhesion protein paxillin redistributes to the periphery of the monolayer (\cref{Fig2k}), and supracellular stress fibers rich in active myosin massively form (\cref{FigS7,Fig2l,Fig2m}). These results suggest that dewetting is not directly caused by an increase in cell-cell adhesion, but rather by an increase in tension, which eventually causes the failure of cell-substrate adhesions. To test this hypothesis, we incubate the cells with blebbistatin (25 $\mu$M) to hinder the increase in contractility without impairing the over-expression of E-cadherin (\cref{Fig2e}). This treatment reduces cellular forces (\cref{FigS8}) and delays dewetting (\cref{Fig2g}; Supplementary Movie 6). Conversely, the addition of the ROCK inhibitor Y27632 (25 $\mu$M) during dewetting causes the monolayer to rewet the substrate (\cref{Fig2f,Fig2g}; Supplementary Movie 6), thus demonstrating the active origin and reversibility of the transition. Together, these results show that the wetting transition results from a competition between active cellular forces, rather than simply between cell-cell and cell-substrate adhesion energies.

\vskip0.5cm
\noindent\textbf{An active polar fluid model of tissue wetting}

To understand how the wetting transition emerges from active cellular forces, we build upon a continuum mechanical model of epithelial spreading\cite{Blanch-Mercader2017}. Given the long time scales of the wetting/dewetting processes, we neglect the elastic response of the tissue\cite{Mertz2012,Kopf2013,Banerjee2015,Notbohm2016}, assuming that it has a purely viscous behavior\cite{Blanch-Mercader2017,Lee2011a,Lee2011,Vig2017,Saw2017,Blanch-Mercader2017c,Yabunaka2017d}. Thus, taking a coarse-grained approach, the model describes the cell monolayer as a two-dimensional (2D) active polar fluid\cite{Kruse2005,Julicher2011,Marchetti2013,Prost2015}, namely in terms of a polarity field $\mathbf{p}(\mathbf{r},t)$ and a velocity field $\mathbf{v}(\mathbf{r},t)$ (Supplementary Note). Our 2D model does not aim at describing the out-of-plane flows and shape of the tissue, nor the dynamics of the contact angle. However, it allows us to predict the onset and initial dynamics of the wetting transition, which is the focus of our study.

\begin{figure*}[tb]
\begin{center}
\includegraphics[width=0.75\textwidth]{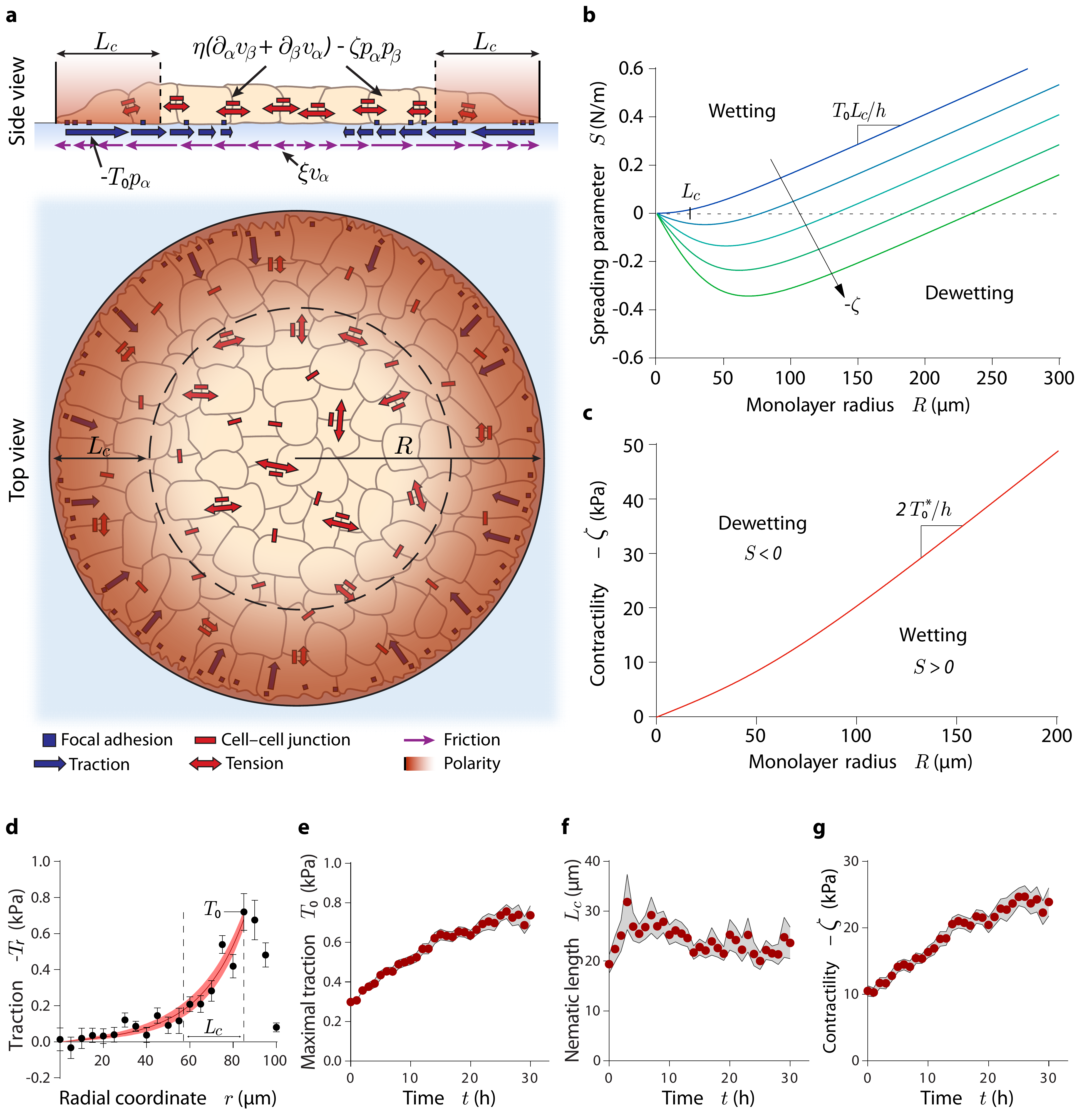}
\end{center}
  {\phantomsubcaption\label{Fig3a}}
  {\phantomsubcaption\label{Fig3b}}
  {\phantomsubcaption\label{Fig3c}}
  {\phantomsubcaption\label{Fig3d}}
  {\phantomsubcaption\label{Fig3e}}
  {\phantomsubcaption\label{Fig3f}}
  {\phantomsubcaption\label{Fig3g}}
\bfcaption{Active polar fluid model of tissue wetting}{ \subref*{Fig3a}, Scheme of the model. \subref*{Fig3b}, Spreading parameter of the monolayer as a function of its radius at increasing contractility (blue to green). The point at which $S = 0$ indicates the critical radius for tissue wetting. \subref*{Fig3c}, Predicted critical contractility for the wetting transition as a function of monolayer radius. \subref*{Fig3d}, Representative example of a fit of the radial traction profile, from which we infer the evolution of the model parameters. \subref*{Fig3e}-\subref*{Fig3g}, Evolution of the maximal traction (\subref*{Fig3e}), nematic length (\subref*{Fig3f}) and contractility (\subref*{Fig3g}) in islands of radius $100$ $\mu$m. Data are presented as mean $\pm$ s.e.m. $n=18$ cell islands.} \label{Fig3}
\end{figure*}

The cell monolayer is unpolarized in the bulk and polarized at the edge (see \cref{Fig1e,Fig1f}). Hence, we take a free energy for the polarity field that favours the unpolarized state $\mathbf{p}=0$ with a restoring coefficient $a>0$, and that introduces a cost for polarity gradients, with $K$ the Frank constant of nematic elasticity in the one-constant approximation\cite{DeGennes-Prost}:
\begin{equation}
F=\int\left[\frac{a}{2} p_\alpha p_\alpha + \frac{K}{2} (\partial_\alpha p_\beta)(\partial_\alpha p_\beta)\right]\dd^3\vec{r}.
\end{equation}
We assume that the polarity field is set by flow-independent mechanisms, so that it follows a purely relaxational dynamics, and that it equilibrates fast compared to the spreading dynamics. Hence, $\delta F/\delta p_\alpha =0$, which yields
\begin{equation}
L_c^2 \nabla^2 p_\alpha = p_\alpha,
\end{equation}
where $L_c=\sqrt{K/a}$ is the characteristic length with which the polarity decays from $p_r (R)=1$ at the edge of the monolayer of radius $R$ to $p_r (0)=0$ at the center (red shade in \cref{Fig3a}).

Then, force balance imposes
\begin{equation}
\partial_\beta \sigma_{\alpha\beta} = T_\alpha,
\end{equation}
where $\sigma_{\alpha\beta}$ and $T_\alpha$ are the components of the monolayer tension and traction stress fields, respectively. We relate these forces to the polarity and velocity fields via the following constitutive equations for a compressible active polar fluid\cite{Oriola2017}:
\begin{equation}
\sigma^{\text{s}}_{\alpha\beta} = \frac{\sigma_{\alpha\beta}}{h} = \eta (\partial_\alpha v_\beta + \partial_\beta v_\alpha) - \zeta p_\alpha p_\beta,
\end{equation}
\begin{equation}
f_\alpha = -\frac{T_\alpha}{h} = -\xi v_\alpha + \zeta_i p_\alpha.
\end{equation}
Here, $h$ is the monolayer height, $\eta$ is the monolayer viscosity, $\zeta$ is the active stress coefficient, $\xi$ is the cell-substrate viscous friction coefficient, and $\zeta_i$ is the contact active force coefficient. These parameters are assumed to be time-dependent to account for the evolving mechanical properties of the monolayer. Note that $\zeta<0$ for contractile behaviour, and hence we call $-\zeta$ `contractility'. In addition, we define the maximal traction stress exerted by polarized cells, $T_0=\zeta_i h$.

Assuming radial symmetry, neglecting cell-substrate viscous friction, and imposing stress-free boundary conditions, we analytically solve the model (Supplementary Note). Thus, we obtain the spreading velocity $V=v_r (R)=\dd R/\dd t$ and, hence, the spreading parameter\cite{Beaune2014} $S=\eta V$. In the experimentally relevant limit $L_c\ll R$, it reads
\begin{equation} \label{eq spreading-parameter-main}
S\approx \frac{T_0 L_c}{h}R + \left(\zeta - \frac{3 T_0 L_c}{h}\right)\frac{L_c}{2}.
\end{equation}
Strikingly, the spreading parameter depends on the monolayer radius $R$, which entails the existence of a critical radius
\begin{equation}
R^* \approx \frac{1}{2}\left(3L_c - \frac{\zeta h}{T_0}\right) \sim -\frac{1}{2}\frac{\zeta}{\zeta_i}
\end{equation}
above which the tissue spreads ($S>0$) driven by traction forces $T_0>0$ and below which it retracts ($S<0$) driven by tissue contractility $\zeta<0$ (\cref{Fig3b}). The competition between bulk and contact active forces defines a novel intrinsic lengthscale $L_p\equiv -\zeta/\zeta_i$ of active polar fluids that naturally gives rise to the critical radius for the wetting transition, a striking property that has no counterpart in the classical wetting scenario.

Unlike for ordinary fluids, the wetting properties of tissues are not determined by local forces at the contact line but by the balance of forces across the entire monolayer, which results in the size-dependent wetting. Specifically, the internal, unpolarized region of the monolayer is subject to almost no external forces, and hence it is under a uniform tension set by traction forces at the polarized boundary layer. Because of the viscous rheology of the tissue, this uniform tension generates an outwards-directed flow with a linearly increasing velocity profile (\cref{FigS9,FigS10}; Supplementary Movie 7). Thus, larger monolayers exhibit a larger velocity right behind the boundary layer, which requires a higher contractility to induce monolayer dewetting (\cref{Fig3c}; Supplementary Note). Finally, we suggest that the predicted non-monotonous flow profiles might induce the formation of 3D cell rims observed at the edge of epithelial monolayers\cite{Deforet2014,Kaliman2014}.

\vskip0.5cm
\noindent\textbf{Tissue wetting depends on tissue size and substrate properties}

\begin{figure*}[p]
\begin{center}
\includegraphics[width=0.85\textwidth]{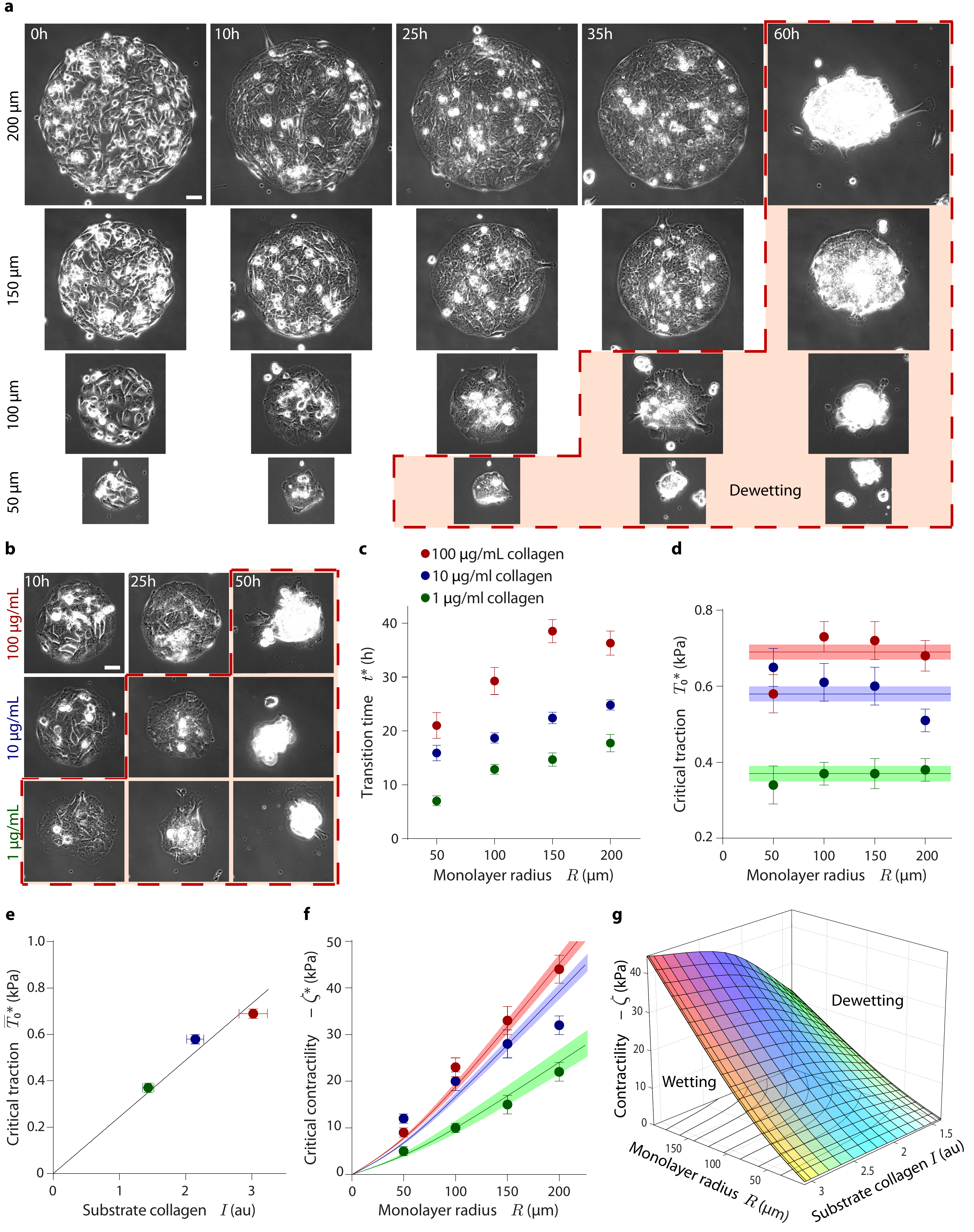}
\end{center}
  {\phantomsubcaption\label{Fig4a}}
  {\phantomsubcaption\label{Fig4b}}
  {\phantomsubcaption\label{Fig4c}}
  {\phantomsubcaption\label{Fig4d}}
  {\phantomsubcaption\label{Fig4e}}
  {\phantomsubcaption\label{Fig4f}}
  {\phantomsubcaption\label{Fig4g}}
\bfcaption{The wetting transition depends on substrate ligand density and monolayer radius}{ \subref*{Fig4a}, Time evolution of epithelial monolayers of different initial radius. Larger monolayers dewet later. \subref*{Fig4b}, Time evolution of monolayers on substrates with different ligand density. Monolayers on substrates with higher ligand density dewet later. Islands are $100$ $\mu$m in radius. The red dashed line and shade in (\subref*{Fig4a}) and (\subref*{Fig4b}) indicate dewetting. Scale bars $= 40$ $\mu$m. \subref*{Fig4c}, Wetting transition time as a function of monolayer radius and substrate ligand density. \subref*{Fig4d}, Critical traction as a function of monolayer radius and substrate ligand density. Horizontal lines show the average critical tractions at different collagen concentrations, with shadows indicating error margins. \subref*{Fig4e}, Average critical traction as a function of the relative amount of collagen on the substrate. \subref*{Fig4f}, Critical contractility as a function of monolayer radius and substrate ligand density. Lines show the critical contractility corresponding to the average critical traction for each collagen concentration, with shadows indicating error margins. \subref*{Fig4g}, Phase diagram of tissue wetting as a function of monolayer radius, contractility and substrate ligand density. The plotted surface corresponds to the observed wetting-dewetting transition. Data are presented as mean $\pm$ s.e.m. For islands on $100$ $\mu$g/mL collagen: $n=17$ ($200$ $\mu$m radius), $n=15$ ($150$ $\mu$m radius), $n=18$ ($100$ $\mu$m radius), and $n=11$ ($50$ $\mu$m radius). For islands on $10$ $\mu$g/mL collagen: $n=17$ ($200$ $\mu$m radius), $n=15$ ($150$ $\mu$m radius), $n=17$ ($100$ $\mu$m radius), and $n=10$ ($50$ $\mu$m radius). For islands on $1$ $\mu$g/mL collagen: $n=11$ ($200$ $\mu$m radius), $n=10$ ($150$ $\mu$m radius), $n=8$ ($100$ $\mu$m radius), and $n=8$ ($50$ $\mu$m radius).} \label{Fig4}
\end{figure*}

The model predicts that the wetting transition depends on monolayer size and tissue forces (\cref{Fig3b,Fig3c}). To assess the role of these variables in the experiments, we generate circular islands of different radii ($50$, $100$, $150$, and $200$ $\mu$m) on substrates of different ECM ligand densities ($100$, $10$, and $1$ $\mu$g/mL of collagen in the coating solution) (\cref{Fig4a,Fig4b}). We also study cell monolayers on substrates of different rigidities ($3$, $12$, and $30$ kPa) (\cref{FigS11}). With the only exception of $30$ kPa gels, on which dewetting does not occur in the time scale of the experiment, monolayers in all conditions feature a tension buildup phase and a dewetting phase (\crefrange{FigS11b}{FigS11d}, \crefrange{FigS12a}{FigS12k}; Supplementary Movies 8, 9). However, the duration of each phase presents large quantitative differences (\cref{Fig4a,Fig4b,FigS11a}). To assess these differences, we implement a robust user-blind method to measure the time $t^*$ at which dewetting starts (\cref{FigS13}; see Methods). This analysis establishes that smaller monolayers dewet earlier than larger ones, and so do monolayers on softer and/or less densely coated substrates (\cref{Fig4c,FigS11h}). Therefore, tissue size as well as substrate adhesion and stiffness are key parameters in determining the wetting transition (Supplementary Movie 10). Of note, transition times span from $7$ to $40$ h after E-cadherin induction and monolayers on stiff substrates do not dewet in the experimental time window ($85$ h), which implies that changes in E-cadherin levels alone (\cref{Fig1c}) cannot account for tissue dewetting.

In our experiments, tissue forces increase with time until the wetting transition takes place. As a consequence, larger monolayers not only dewet later than smaller ones but also at higher tension (\crefrange{FigS12a}{FigS12c}). This finding is consistent with our prediction (\cref{Fig3c}) that larger monolayers require higher contractility to dewet or, equivalently, that for a given contractility only sufficiently large monolayers will wet. Thus, our experimental results do not directly establish but support the existence of a critical radius for tissue wetting. Future work should further assess the size dependence of the wetting transition using direct control of cell contractility.

To infer the values of model parameters, we fit the predicted traction profiles to the experimental data (\cref{Fig3a}, see Methods). Hence, we obtain the time evolution of the model parameters $T_0 (t)$ and $L_c (t)$. In addition, by imposing that the velocity of the tissue boundary vanishes during the wetting phase, we obtain the time evolution of the contractility $-\zeta(t)$ (\cref{eq contractility-methods}, see Supplementary Note). The maximal traction $T_0$ and the contractility $-\zeta$ experience a $~3$-fold increase, whereas the nematic length $L_c$ remains constant (\crefrange{Fig3e}{Fig3g}). This analysis is performed for all experimental conditions (\cref{FigS14}, \crefrange{FigS11e}{FigS11g}), from which we obtain the critical values of the parameters at the wetting transition, namely at time $t^*$. The nematic length has a similar value of $L_c\approx 25$ $\mu$m for all the experimental conditions (\cref{FigS11e}, \crefrange{FigS14d}{FigS14f}), suggesting that, as considered in our model, it is an intrinsic property of the cell monolayer. Critical tractions $T_0^*=T_0 (t^*)$ are largely independent of monolayer radius (\cref{Fig4d}), but they increase with substrate rigidity (\cref{FigS11i}). Critical tractions also increase linearly with measured substrate ligand density (\cref{Fig4e,FigS12l}), suggesting that collagen is fully saturated with integrins at the wetting transition. The critical traction $T_0^*$ should thus be interpreted as the maximum force that cells can withstand before focal adhesions fail\cite{Schwarz2013}, and hence tissue spreading is not possible above it. Like critical tractions, the critical contractility $-\zeta^*=-\zeta(t^*)$ increases with substrate rigidity (\cref{FigS11j}) and with ligand density (\cref{Fig4f}). However, unlike critical tractions, the critical contractility also increases with monolayer radius (\cref{Fig4f}). We summarize our results in a phase diagram for the tissue wetting transition as a function of contractility, substrate ligand density, and monolayer radius (\cref{Fig4g}).

\begin{figure*}[p]
\begin{center}
\includegraphics[width=0.75\textwidth]{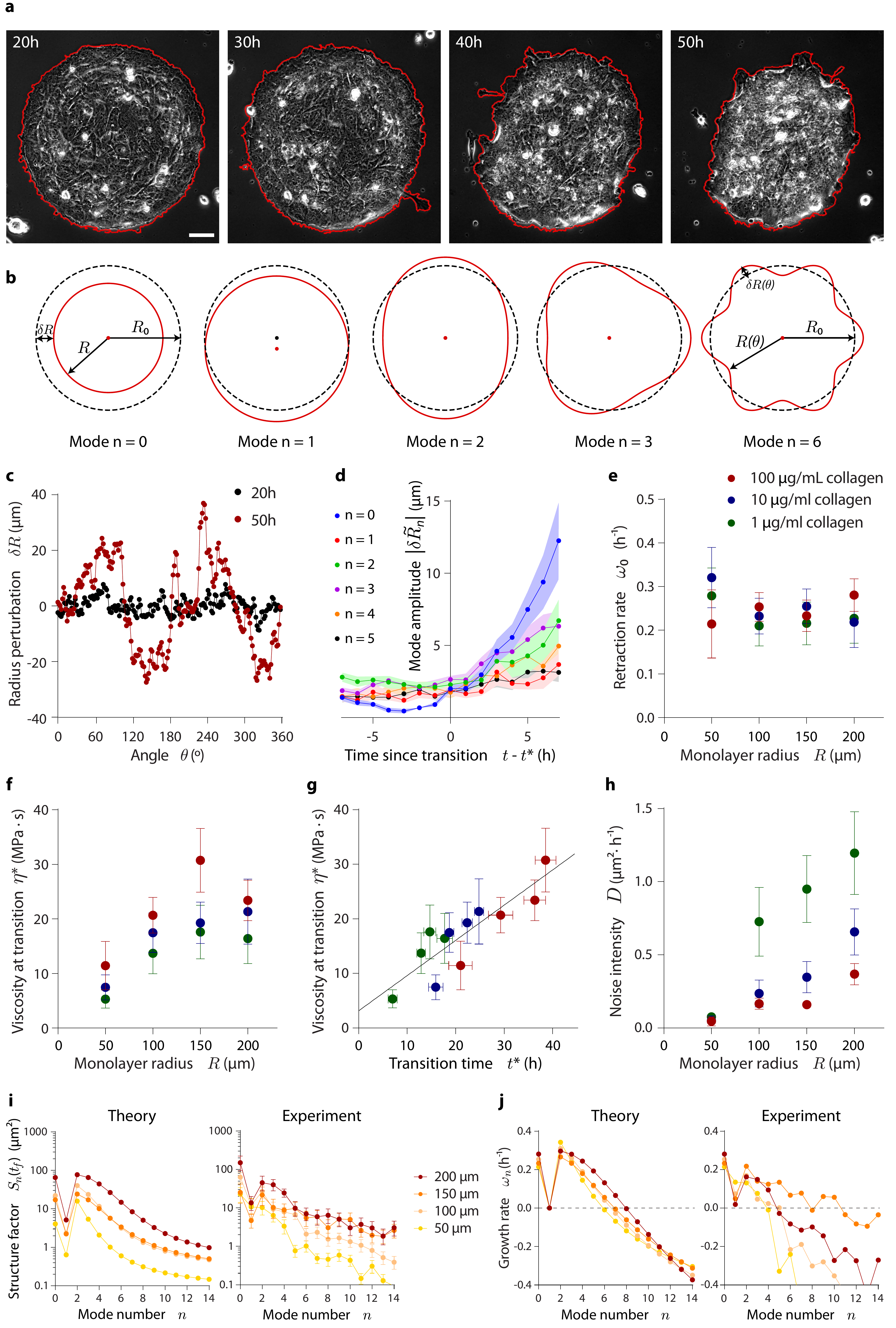}
\end{center}
  {\phantomsubcaption\label{Fig5a}}
  {\phantomsubcaption\label{Fig5b}}
  {\phantomsubcaption\label{Fig5c}}
  {\phantomsubcaption\label{Fig5d}}
  {\phantomsubcaption\label{Fig5e}}
  {\phantomsubcaption\label{Fig5f}}
  {\phantomsubcaption\label{Fig5g}}
  {\phantomsubcaption\label{Fig5h}}
  {\phantomsubcaption\label{Fig5i}}
  {\phantomsubcaption\label{Fig5j}}
\bfcaption{Evolution of monolayer morphology during dewetting}{ \subref*{Fig5a}, Phase contrast images of a $200$ $\mu$m radius island that loses its circular symmetry during dewetting, as shown by its contour (red line). Scale bar $= 30$ $\mu$m. \subref*{Fig5b}, Illustration of the lowest shape perturbation modes of a circle. \subref*{Fig5c}, Initial and final radius perturbation profiles of the island shown in (\subref*{Fig5a}). Note that the final time point is well after the onset of dewetting, into the nonlinear regime of the instability not captured by our analysis. \subref*{Fig5d}, Evolution of the average amplitude of the lowest shape perturbation modes for $200$ $\mu$m radius islands around the wetting-dewetting transition. \subref*{Fig5e}-\subref*{Fig5h}, Retraction rate, namely the growth rate of mode $n = 0$ (\subref*{Fig5e}), monolayer viscosity at the wetting transition (\subref*{Fig5f}), and noise intensity of mode amplitudes (\subref*{Fig5h}) as a function of monolayer radius and substrate ligand density. Monolayer viscosity correlates with transition time (\subref*{Fig5g}). \subref*{Fig5i}-\subref*{Fig5j}, Structure factor of the monolayer boundary (\subref*{Fig5i}), and growth rate of shape perturbation modes (\subref*{Fig5j}) for islands of all different radii on substrates coated with $100$ $\mu$g/mL of collagen. Theoretical predictions are shown along with average experimental data. Data are presented as mean $\pm$ s.e.m. Analyzed islands are the same as in \cref{Fig4}, but some islands were discarded due to imperfections in the patterning introducing initial biases towards some perturbations modes (see Methods). For islands on $100$ $\mu$g/mL collagen: $n=12$ ($200$ $\mu$m radius), $n=9$ ($150$ $\mu$m radius), $n=16$ ($100$ $\mu$m radius), and $n=11$ ($50$ $\mu$m radius). For islands on $10$ $\mu$g/mL collagen: $n=17$ ($200$ $\mu$m radius), $n=12$ ($150$ $\mu$m radius), $n=13$ ($100$ $\mu$m radius), and $n=6$ ($50$ $\mu$m radius). For islands on $1$ $\mu$g/mL collagen: $n=9$ ($200$ $\mu$m radius), $n=10$ ($150$ $\mu$m radius), $n=8$ ($100$ $\mu$m radius), and $n=7$ ($50$ $\mu$m radius).} \label{Fig5}
\end{figure*}

\vskip0.5cm
\noindent\textbf{Active forces govern tissue morphology during dewetting}

Our analysis thus far shows that an active polar fluid model captures the onset of dewetting as a function of the material properties and geometry of the tissue. Next, we focus on the early dynamics of tissue dewetting. Immediately after the onset of dewetting, the monolayer loses its circular symmetry, acquiring an elliptic-like shape before collapsing into a spheroidal cell aggregate (\cref{Fig5a}; Supplementary Movie 11). This striking symmetry breaking is in stark contrast with the known isotropic dewetting of passive fluids\cite{Edwards2016,DeGennes1985,Bonn2009}. Although pinning of the contact line\cite{DeGennes1985,Bonn2009,Chepizhko2016} may contribute to breaking the circular symmetry, the fact that monolayer retraction systematically tends to start at diametrically opposed points of the tissue boundary (Supplementary Movie 11) suggests the presence of a morphological instability of active origin. Indeed, from our active polar fluid model, we analytically predict a long-wavelength instability of monolayer shape during dewetting (Supplementary Note).

To test the predictions, we characterize the evolution of tissue morphology by tracking the contour of the monolayer (\cref{Fig5a}). The local radius perturbation $\delta R(\theta,t)=R(\theta,t)-R_0$ quantifies the loss of circular symmetry (\cref{Fig5c}), and its Fourier transform dissects the contribution of each perturbation mode to the overall shape of the monolayer (\cref{Fig5b}). Consistent with the predicted instability, the amplitudes $|\delta\tilde{R}_n|$ of the long-wavelength modes increase with time upon the onset of dewetting (\cref{Fig5d}). Their predicted growth rates $\omega_n$ depend on a single yet-unmeasured parameter, the monolayer viscosity at the wetting transition, $\eta^*$. Its value can be inferred from the retraction rate of the monolayer, $\omega_0$, which we experimentally measure by fitting the exponential growth of the zeroth perturbation mode: $\delta\tilde{R}_0 (t)=\delta\tilde{R}_0 (t^*)\,e^{\omega_0 (t-t^*)}$ (\cref{Fig5e,FigS15}). Comparing with the theoretical prediction (Supplementary Note)
\begin{equation}
\omega_0\approx \frac{T_0L_c}{2\eta h}
\end{equation}
we obtain viscosities that increase with monolayer radius and substrate ligand density, spanning from $3$ to $30$ MPa·s (\cref{Fig5f}). The tendency exhibited by the viscosity is similar to that of transition times (\cref{Fig4c}). In fact, these two quantities linearly correlate (\cref{Fig5g}), which suggests that monolayer viscosity increases with time in our experiment, likely due to a combination of cell-cell junction formation\cite{Blanch-Mercader2017,Garcia2015}, increasing contractility\cite{Stirbat2013}, and increasing cell density\cite{Garcia2015}.

Once the theoretical growth rates $\omega_n$ are known, we can predict the amplitudes of the different shape modes. Assuming that monolayer shape fluctuations are fast compared to the dewetting dynamics, we compute the structure factor of the monolayer boundary
\begin{equation} \label{eq structure-factor}
S_n(t)=\langle |\delta\tilde{R}_n (t)|^2\rangle = \frac{D}{\omega_n}\left[ e^{2\omega_n (t-t^*)} - 1\right],
\end{equation}
where $D$ is the noise intensity of mode amplitudes (Supplementary Note). By fitting this prediction to the experimental data (\cref{Fig5i,FigS16a}), we infer the value of $D$, which increases with tissue size but decreases with substrate ligand density (\cref{Fig5h}). This behavior is consistent with active shape fluctuations driven by the total traction force in the tissue, which scales linearly with monolayer radius (see \cref{eq spreading-parameter-main}), and damped by cell-substrate friction, which increases with substrate ligand density. Finally, we obtain experimental growth rates from the structure factors (\cref{Fig5j,FigS16b}; Supplementary Note). Despite their expected variability, our experimental results agree with the predictions, confirming that the growth of shape-changing perturbations, especially of mode $n=2$, is responsible for the elliptic-like shape of the monolayer (\cref{Fig5i,Fig5j}). Overall, these results show that active forces and shape fluctuations determine the morphological evolution during monolayer dewetting.

\vskip0.5cm
\noindent\textbf{Discussion and outlook}

Our results illustrate how E-cadherin adhesion regulates tissue mechanical properties and forces and, in turn, how these forces determine tissue shape, dynamics, fluctuations and dimensionality as a function of tissue size, contractility, substrate stiffness, and cell-cell and cell-substrate adhesion. In vivo, transitions in tissue morphology are characterized by changes in cell contractility\cite{Rodriguez-Hernandez2016,Ouderkirk2014}, cell adhesion\cite{Paredes2012,Paschos2009} and ECM composition\cite{Clark2015,Lu2012}. It is appealing to think that these changes translate into different wetting states. For example, this could explain increased tumor invasiveness when contractility decreases or critical traction increases due to an enhanced cell-substrate adhesion, ECM deposition or ECM stiffening. This scenario is supported by previous experiments associating E-cadherin-dependent epithelial retraction and suppression of tumor invasion in vivo\cite{Cortina2007}. Furthermore, tumor growth per se implies an increase in tissue radius, possibly leading to a dewetting-wetting transition even if contractility and critical traction remain unaltered. In this line, the nucleation of a spreading monolayer from a growing cell aggregate has been previously reported\cite{Kaliman2014}.

Our analysis unveils fundamental features of tissue wetting that differ qualitatively from the classical wetting paradigm. We account for these differences by developing a theoretical framework for active wetting, which explicitly relates the wetting properties of tissues to active cellular forces. This framework, based on active gel theory, captures the mechanics of the wetting transition as well as the dynamics of monolayer morphology during the early stages of tissue dewetting. Furthermore, it allows the quantification of active stresses, viscosity, and active fluctuations in the tissue. In light of these results, we propose that tissue spreading can be understood as the wetting process of an active polar fluid, constituting the defining example of the general phenomenon of active wetting.

\bibliography{All}

\begin{thebibliography}{10}
\expandafter\ifx\csname url\endcsname\relax
  \def\url#1{\texttt{#1}}\fi
\expandafter\ifx\csname urlprefix\endcsname\relax\def\urlprefix{URL }\fi
\providecommand{\bibinfo}[2]{#2}
\providecommand{\eprint}[2][]{\url{#2}}

\bibitem{Gonzalez-Rodriguez2012}
\bibinfo{author}{Gonzalez-Rodriguez, D.}, \bibinfo{author}{Guevorkian, K.},
  \bibinfo{author}{Douezan, S.} \& \bibinfo{author}{Brochard-Wyart, F.}
\newblock \bibinfo{title}{{Soft Matter Models of Developing Tissues and
  Tumors}}.
\newblock \emph{\bibinfo{journal}{Science}} \textbf{\bibinfo{volume}{338}},
  \bibinfo{pages}{910--917} (\bibinfo{year}{2012}).

\bibitem{Friedl2009}
\bibinfo{author}{Friedl, P.} \& \bibinfo{author}{Gilmour, D.}
\newblock \bibinfo{title}{{Collective cell migration in morphogenesis,
  regeneration and cancer}}.
\newblock \emph{\bibinfo{journal}{Nat. Rev. Mol. Cell Biol.}}
  \textbf{\bibinfo{volume}{10}}, \bibinfo{pages}{445--457}
  (\bibinfo{year}{2009}).

\bibitem{Julicher2017}
\bibinfo{author}{J{\"{u}}licher, F.} \& \bibinfo{author}{Eaton, S.}
\newblock \bibinfo{title}{{Emergence of tissue shape changes from collective
  cell behaviours}}.
\newblock \emph{\bibinfo{journal}{Semin. Cell Dev. Biol.}}
  \textbf{\bibinfo{volume}{67}}, \bibinfo{pages}{103--112}
  (\bibinfo{year}{2017}).

\bibitem{Ryan2001}
\bibinfo{author}{Ryan, P.~L.}, \bibinfo{author}{Foty, R.~A.},
  \bibinfo{author}{Kohn, J.} \& \bibinfo{author}{Steinberg, M.~S.}
\newblock \bibinfo{title}{{Tissue spreading on implantable substrates is a
  competitive outcome of cell-cell vs. cell-substratum adhesivity}}.
\newblock \emph{\bibinfo{journal}{Proc. Natl. Acad. Sci. U. S. A.}}
  \textbf{\bibinfo{volume}{98}}, \bibinfo{pages}{4323--4327}
  (\bibinfo{year}{2001}).

\bibitem{Douezan2011}
\bibinfo{author}{Douezan, S.} \emph{et~al.}
\newblock \bibinfo{title}{{Spreading dynamics and wetting transition of
  cellular aggregates}}.
\newblock \emph{\bibinfo{journal}{Proc. Natl. Acad. Sci. U. S. A.}}
  \textbf{\bibinfo{volume}{108}}, \bibinfo{pages}{7315--7320}
  (\bibinfo{year}{2011}).

\bibitem{Douezan2012c}
\bibinfo{author}{Douezan, S.}, \bibinfo{author}{Dumond, J.} \&
  \bibinfo{author}{Brochard-Wyart, F.}
\newblock \bibinfo{title}{{Wetting transitions of cellular aggregates induced
  by substrate rigidity}}.
\newblock \emph{\bibinfo{journal}{Soft Matter}} \textbf{\bibinfo{volume}{8}},
  \bibinfo{pages}{4578--4583} (\bibinfo{year}{2012}).

\bibitem{Beaune2014}
\bibinfo{author}{Beaune, G.} \emph{et~al.}
\newblock \bibinfo{title}{{How cells flow in the spreading of cellular
  aggregates}}.
\newblock \emph{\bibinfo{journal}{Proc. Natl. Acad. Sci. U. S. A.}}
  \textbf{\bibinfo{volume}{111}}, \bibinfo{pages}{8055--8060}
  (\bibinfo{year}{2014}).

\bibitem{Beaune2017}
\bibinfo{author}{Beaune, G.} \emph{et~al.}
\newblock \bibinfo{title}{{Reentrant wetting transition in the spreading of
  cellular aggregates}}.
\newblock \emph{\bibinfo{journal}{Soft Matter}} \textbf{\bibinfo{volume}{13}},
  \bibinfo{pages}{8474--8482} (\bibinfo{year}{2017}).

\bibitem{Behrndt2012}
\bibinfo{author}{Behrndt, M.} \emph{et~al.}
\newblock \bibinfo{title}{{Forces Driving Epithelial Spreading in Zebrafish
  Gastrulation}}.
\newblock \emph{\bibinfo{journal}{Science}} \textbf{\bibinfo{volume}{338}},
  \bibinfo{pages}{257--260} (\bibinfo{year}{2012}).

\bibitem{Campinho2013}
\bibinfo{author}{Campinho, P.} \emph{et~al.}
\newblock \bibinfo{title}{{Tension-oriented cell divisions limit anisotropic
  tissue tension in epithelial spreading during zebrafish epiboly}}.
\newblock \emph{\bibinfo{journal}{Nat. Cell Biol.}}
  \textbf{\bibinfo{volume}{15}}, \bibinfo{pages}{1405--1414}
  (\bibinfo{year}{2013}).

\bibitem{Morita2017}
\bibinfo{author}{Morita, H.} \emph{et~al.}
\newblock \bibinfo{title}{{The Physical Basis of Coordinated Tissue Spreading
  in Zebrafish Gastrulation}}.
\newblock \emph{\bibinfo{journal}{Dev. Cell}} \textbf{\bibinfo{volume}{40}},
  \bibinfo{pages}{354--366.e4} (\bibinfo{year}{2017}).

\bibitem{Wallmeyer2018}
\bibinfo{author}{Wallmeyer, B.}, \bibinfo{author}{Trinschek, S.},
  \bibinfo{author}{Yigit, S.}, \bibinfo{author}{Thiele, U.} \&
  \bibinfo{author}{Betz, T.}
\newblock \bibinfo{title}{{Collective Cell Migration in Embryogenesis Follows
  the Laws of Wetting}}.
\newblock \emph{\bibinfo{journal}{Biophys. J.}} \textbf{\bibinfo{volume}{114}},
  \bibinfo{pages}{213--222} (\bibinfo{year}{2018}).

\bibitem{Douezan2012a}
\bibinfo{author}{Douezan, S.} \& \bibinfo{author}{Brochard-Wyart, F.}
\newblock \bibinfo{title}{{Dewetting of cellular monolayers}}.
\newblock \emph{\bibinfo{journal}{Eur. Phys. J. E}}
  \textbf{\bibinfo{volume}{35}}, \bibinfo{pages}{34} (\bibinfo{year}{2012}).

\bibitem{Smeets2016}
\bibinfo{author}{Smeets, B.} \emph{et~al.}
\newblock \bibinfo{title}{{Emergent structures and dynamics of cell colonies by
  contact inhibition of locomotion}}.
\newblock \emph{\bibinfo{journal}{Proc. Natl. Acad. Sci. U. S. A.}}
  \textbf{\bibinfo{volume}{113}}, \bibinfo{pages}{14621--14626}
  (\bibinfo{year}{2016}).

\bibitem{Ravasio2015a}
\bibinfo{author}{Ravasio, A.} \emph{et~al.}
\newblock \bibinfo{title}{{Regulation of epithelial cell organization by tuning
  cell-substrate adhesion}}.
\newblock \emph{\bibinfo{journal}{Integr. Biol.}} \textbf{\bibinfo{volume}{7}},
  \bibinfo{pages}{1228--1241} (\bibinfo{year}{2015}).

\bibitem{Sarrio2009}
\bibinfo{author}{Sarri{\'{o}}, D.} \emph{et~al.}
\newblock \bibinfo{title}{{Functional characterization of E- and P-cadherin in
  invasive breast cancer cells}}.
\newblock \emph{\bibinfo{journal}{BMC Cancer}} \textbf{\bibinfo{volume}{9}},
  \bibinfo{pages}{74} (\bibinfo{year}{2009}).

\bibitem{Trepat2009}
\bibinfo{author}{Trepat, X.} \emph{et~al.}
\newblock \bibinfo{title}{{Physical forces during collective cell migration}}.
\newblock \emph{\bibinfo{journal}{Nat. Phys.}} \textbf{\bibinfo{volume}{5}},
  \bibinfo{pages}{426--430} (\bibinfo{year}{2009}).

\bibitem{Serra-Picamal2012}
\bibinfo{author}{Serra-Picamal, X.} \emph{et~al.}
\newblock \bibinfo{title}{{Mechanical waves during tissue expansion}}.
\newblock \emph{\bibinfo{journal}{Nat. Phys.}} \textbf{\bibinfo{volume}{8}},
  \bibinfo{pages}{628--634} (\bibinfo{year}{2012}).

\bibitem{Tambe2011}
\bibinfo{author}{Tambe, D.~T.} \emph{et~al.}
\newblock \bibinfo{title}{{Collective cell guidance by cooperative
  intercellular forces}}.
\newblock \emph{\bibinfo{journal}{Nat. Mater.}} \textbf{\bibinfo{volume}{10}},
  \bibinfo{pages}{469--475} (\bibinfo{year}{2011}).

\bibitem{Mertz2013}
\bibinfo{author}{Mertz, A.~F.} \emph{et~al.}
\newblock \bibinfo{title}{{Cadherin-based intercellular adhesions organize
  epithelial cell-matrix traction forces}}.
\newblock \emph{\bibinfo{journal}{Proc. Natl. Acad. Sci. U. S. A.}}
  \textbf{\bibinfo{volume}{110}}, \bibinfo{pages}{842--7}
  (\bibinfo{year}{2013}).

\bibitem{Mertz2012}
\bibinfo{author}{Mertz, A.~F.} \emph{et~al.}
\newblock \bibinfo{title}{{Scaling of Traction Forces with the Size of Cohesive
  Cell Colonies}}.
\newblock \emph{\bibinfo{journal}{Phys. Rev. Lett.}}
  \textbf{\bibinfo{volume}{108}}, \bibinfo{pages}{198101}
  (\bibinfo{year}{2012}).

\bibitem{Harris2014}
\bibinfo{author}{Harris, A.~R.}, \bibinfo{author}{Daeden, A.} \&
  \bibinfo{author}{Charras, G.~T.}
\newblock \bibinfo{title}{{Formation of adherens junctions leads to the
  emergence of a tissue-level tension in epithelial monolayers}}.
\newblock \emph{\bibinfo{journal}{J. Cell Sci.}}
  \textbf{\bibinfo{volume}{127}}, \bibinfo{pages}{2507--2517}
  (\bibinfo{year}{2014}).

\bibitem{Bazellieres2015}
\bibinfo{author}{Bazelli{\`{e}}res, E.} \emph{et~al.}
\newblock \bibinfo{title}{{Control of cell-cell forces and collective cell
  dynamics by the intercellular adhesome}}.
\newblock \emph{\bibinfo{journal}{Nat. Cell Biol.}}
  \textbf{\bibinfo{volume}{17}}, \bibinfo{pages}{409--420}
  (\bibinfo{year}{2015}).

\bibitem{Lecuit2015}
\bibinfo{author}{Lecuit, T.} \& \bibinfo{author}{Yap, A.~S.}
\newblock \bibinfo{title}{{E-cadherin junctions as active mechanical
  integrators in tissue dynamics}}.
\newblock \emph{\bibinfo{journal}{Nat. Cell Biol.}}
  \textbf{\bibinfo{volume}{17}}, \bibinfo{pages}{533--539}
  (\bibinfo{year}{2015}).

\bibitem{Muhamed2016}
\bibinfo{author}{Muhamed, I.} \emph{et~al.}
\newblock \bibinfo{title}{{E-cadherin-mediated force transduction signals
  regulate global cell mechanics}}.
\newblock \emph{\bibinfo{journal}{J. Cell Sci.}}
  \textbf{\bibinfo{volume}{129}}, \bibinfo{pages}{1843--54}
  (\bibinfo{year}{2016}).

\bibitem{Blanch-Mercader2017}
\bibinfo{author}{Blanch-Mercader, C.} \emph{et~al.}
\newblock \bibinfo{title}{{Effective viscosity and dynamics of spreading
  epithelia: a solvable model}}.
\newblock \emph{\bibinfo{journal}{Soft Matter}} \textbf{\bibinfo{volume}{13}},
  \bibinfo{pages}{1235--1243} (\bibinfo{year}{2017}).

\bibitem{Kopf2013}
\bibinfo{author}{K{\"{o}}pf, M.~H.} \& \bibinfo{author}{Pismen, L.~M.}
\newblock \bibinfo{title}{{A continuum model of epithelial spreading}}.
\newblock \emph{\bibinfo{journal}{Soft Matter}} \textbf{\bibinfo{volume}{9}},
  \bibinfo{pages}{3727} (\bibinfo{year}{2013}).

\bibitem{Banerjee2015}
\bibinfo{author}{Banerjee, S.}, \bibinfo{author}{Utuje, K. J.~C.} \&
  \bibinfo{author}{Marchetti, M.~C.}
\newblock \bibinfo{title}{{Propagating Stress Waves During Epithelial
  Expansion}}.
\newblock \emph{\bibinfo{journal}{Phys. Rev. Lett.}}
  \textbf{\bibinfo{volume}{114}}, \bibinfo{pages}{228101}
  (\bibinfo{year}{2015}).

\bibitem{Notbohm2016}
\bibinfo{author}{Notbohm, J.} \emph{et~al.}
\newblock \bibinfo{title}{{Cellular Contraction and Polarization Drive
  Collective Cellular Motion}}.
\newblock \emph{\bibinfo{journal}{Biophys. J.}} \textbf{\bibinfo{volume}{110}},
  \bibinfo{pages}{2729--2738} (\bibinfo{year}{2016}).

\bibitem{Lee2011a}
\bibinfo{author}{Lee, P.} \& \bibinfo{author}{Wolgemuth, C.~W.}
\newblock \bibinfo{title}{{Crawling Cells Can Close Wounds without Purse
  Strings or Signaling}}.
\newblock \emph{\bibinfo{journal}{PLoS Comput. Biol.}}
  \textbf{\bibinfo{volume}{7}}, \bibinfo{pages}{e1002007}
  (\bibinfo{year}{2011}).

\bibitem{Lee2011}
\bibinfo{author}{Lee, P.} \& \bibinfo{author}{Wolgemuth, C.}
\newblock \bibinfo{title}{{Advent of complex flows in epithelial tissues}}.
\newblock \emph{\bibinfo{journal}{Phys. Rev. E}} \textbf{\bibinfo{volume}{83}},
  \bibinfo{pages}{061920} (\bibinfo{year}{2011}).

\bibitem{Vig2017}
\bibinfo{author}{Vig, D.~K.}, \bibinfo{author}{Hamby, A.~E.} \&
  \bibinfo{author}{Wolgemuth, C.~W.}
\newblock \bibinfo{title}{{Cellular Contraction Can Drive Rapid Epithelial
  Flows}}.
\newblock \emph{\bibinfo{journal}{Biophys. J.}} \textbf{\bibinfo{volume}{113}},
  \bibinfo{pages}{1613--1622} (\bibinfo{year}{2017}).

\bibitem{Saw2017}
\bibinfo{author}{Saw, T.~B.} \emph{et~al.}
\newblock \bibinfo{title}{{Topological defects in epithelia govern cell death
  and extrusion}}.
\newblock \emph{\bibinfo{journal}{Nature}} \textbf{\bibinfo{volume}{544}},
  \bibinfo{pages}{212--216} (\bibinfo{year}{2017}).

\bibitem{Blanch-Mercader2017c}
\bibinfo{author}{Blanch-Mercader, C.} \& \bibinfo{author}{Casademunt, J.}
\newblock \bibinfo{title}{{Hydrodynamic instabilities, waves and turbulence in
  spreading epithelia}}.
\newblock \emph{\bibinfo{journal}{Soft Matter}} \textbf{\bibinfo{volume}{13}},
  \bibinfo{pages}{6913--6928} (\bibinfo{year}{2017}).

\bibitem{Yabunaka2017d}
\bibinfo{author}{Yabunaka, S.} \& \bibinfo{author}{Marcq, P.}
\newblock \bibinfo{title}{{Emergence of epithelial cell density waves}}.
\newblock \emph{\bibinfo{journal}{Soft Matter}} \textbf{\bibinfo{volume}{13}},
  \bibinfo{pages}{7046--7052} (\bibinfo{year}{2017}).

\bibitem{Kruse2005}
\bibinfo{author}{Kruse, K.}, \bibinfo{author}{Joanny, J.~F.},
  \bibinfo{author}{J{\"{u}}licher, F.}, \bibinfo{author}{Prost, J.} \&
  \bibinfo{author}{Sekimoto, K.}
\newblock \bibinfo{title}{{Generic theory of active polar gels: a paradigm for
  cytoskeletal dynamics}}.
\newblock \emph{\bibinfo{journal}{Eur. Phys. J. E}}
  \textbf{\bibinfo{volume}{16}}, \bibinfo{pages}{5--16} (\bibinfo{year}{2005}).

\bibitem{Julicher2011}
\bibinfo{author}{J{\"{u}}licher, F.}
\newblock \bibinfo{title}{{Active fluids and gels}}.
\newblock In \bibinfo{editor}{{Ben Amar}, M.}, \bibinfo{editor}{Goriely, A.},
  \bibinfo{editor}{M{\"{u}}ller, M.~M.} \& \bibinfo{editor}{Cugliandolo, L.}
  (eds.) \emph{\bibinfo{booktitle}{New Trends in the Physics and Mechanics of
  Biological Systems}}, chap.~\bibinfo{chapter}{4} (\bibinfo{publisher}{Oxford
  University Press}, \bibinfo{year}{2011}).

\bibitem{Marchetti2013}
\bibinfo{author}{Marchetti, M.~C.} \emph{et~al.}
\newblock \bibinfo{title}{{Hydrodynamics of soft active matter}}.
\newblock \emph{\bibinfo{journal}{Rev. Mod. Phys.}}
  \textbf{\bibinfo{volume}{85}}, \bibinfo{pages}{1143--1189}
  (\bibinfo{year}{2013}).

\bibitem{Prost2015}
\bibinfo{author}{Prost, J.}, \bibinfo{author}{J{\"{u}}licher, F.} \&
  \bibinfo{author}{Joanny, J.-F.}
\newblock \bibinfo{title}{{Active gel physics}}.
\newblock \emph{\bibinfo{journal}{Nat. Phys.}} \textbf{\bibinfo{volume}{11}},
  \bibinfo{pages}{111--117} (\bibinfo{year}{2015}).

\bibitem{DeGennes-Prost}
\bibinfo{author}{de~Gennes, P.-G.} \& \bibinfo{author}{Prost, J.}
\newblock \emph{\bibinfo{title}{{The Physics of Liquid Crystals}}}
  (\bibinfo{publisher}{Oxford University Press}, \bibinfo{year}{1993}),
  \bibinfo{edition}{2nd} edn.

\bibitem{Oriola2017}
\bibinfo{author}{Oriola, D.}, \bibinfo{author}{Alert, R.} \&
  \bibinfo{author}{Casademunt, J.}
\newblock \bibinfo{title}{{Fluidization and Active Thinning by Molecular
  Kinetics in Active Gels}}.
\newblock \emph{\bibinfo{journal}{Phys. Rev. Lett.}}
  \textbf{\bibinfo{volume}{118}}, \bibinfo{pages}{088002}
  (\bibinfo{year}{2017}).

\bibitem{Deforet2014}
\bibinfo{author}{Deforet, M.}, \bibinfo{author}{Hakim, V.},
  \bibinfo{author}{Yevick, H.}, \bibinfo{author}{Duclos, G.} \&
  \bibinfo{author}{Silberzan, P.}
\newblock \bibinfo{title}{{Emergence of collective modes and tri-dimensional
  structures from epithelial confinement}}.
\newblock \emph{\bibinfo{journal}{Nat. Commun.}} \textbf{\bibinfo{volume}{5}},
  \bibinfo{pages}{3747} (\bibinfo{year}{2014}).

\bibitem{Kaliman2014}
\bibinfo{author}{Kaliman, S.}, \bibinfo{author}{Jayachandran, C.},
  \bibinfo{author}{Rehfeldt, F.} \& \bibinfo{author}{Smith, A.-S.}
\newblock \bibinfo{title}{{Novel Growth Regime of MDCK II Model Tissues on Soft
  Substrates}}.
\newblock \emph{\bibinfo{journal}{Biophys. J.}} \textbf{\bibinfo{volume}{106}},
  \bibinfo{pages}{L25--8} (\bibinfo{year}{2014}).

\bibitem{Schwarz2013}
\bibinfo{author}{Schwarz, U.~S.} \& \bibinfo{author}{Safran, S.~A.}
\newblock \bibinfo{title}{{Physics of adherent cells}}.
\newblock \emph{\bibinfo{journal}{Rev. Mod. Phys.}}
  \textbf{\bibinfo{volume}{85}}, \bibinfo{pages}{1327--1381}
  (\bibinfo{year}{2013}).

\bibitem{Edwards2016}
\bibinfo{author}{Edwards, A. M.~J.}, \bibinfo{author}{Ledesma-Aguilar, R.},
  \bibinfo{author}{Newton, M.~I.}, \bibinfo{author}{Brown, C.~V.} \&
  \bibinfo{author}{McHale, G.}
\newblock \bibinfo{title}{{Not spreading in reverse: The dewetting of a liquid
  film into a single drop}}.
\newblock \emph{\bibinfo{journal}{Sci. Adv.}} \textbf{\bibinfo{volume}{2}}
  (\bibinfo{year}{2016}).

\bibitem{DeGennes1985}
\bibinfo{author}{de~Gennes, P.}
\newblock \bibinfo{title}{{Wetting: statics and dynamics}}.
\newblock \emph{\bibinfo{journal}{Rev. Mod. Phys.}}
  \textbf{\bibinfo{volume}{57}}, \bibinfo{pages}{827--863}
  (\bibinfo{year}{1985}).

\bibitem{Bonn2009}
\bibinfo{author}{Bonn, D.}, \bibinfo{author}{Eggers, J.},
  \bibinfo{author}{Indekeu, J.}, \bibinfo{author}{Meunier, J.} \&
  \bibinfo{author}{Rolley, E.}
\newblock \bibinfo{title}{{Wetting and spreading}}.
\newblock \emph{\bibinfo{journal}{Rev. Mod. Phys.}}
  \textbf{\bibinfo{volume}{81}}, \bibinfo{pages}{739--805}
  (\bibinfo{year}{2009}).

\bibitem{Chepizhko2016}
\bibinfo{author}{Chepizhko, O.} \emph{et~al.}
\newblock \bibinfo{title}{{Bursts of activity in collective cell migration}}.
\newblock \emph{\bibinfo{journal}{Proc. Natl. Acad. Sci. U. S. A.}}
  \textbf{\bibinfo{volume}{113}}, \bibinfo{pages}{11408--11413}
  (\bibinfo{year}{2016}).

\bibitem{Garcia2015}
\bibinfo{author}{Garcia, S.} \emph{et~al.}
\newblock \bibinfo{title}{{Physics of active jamming during collective cellular
  motion in a monolayer}}.
\newblock \emph{\bibinfo{journal}{Proc. Natl. Acad. Sci. U. S. A.}}
  \textbf{\bibinfo{volume}{112}}, \bibinfo{pages}{15314--15319}
  (\bibinfo{year}{2015}).

\bibitem{Stirbat2013}
\bibinfo{author}{Stirbat, T.~V.} \emph{et~al.}
\newblock \bibinfo{title}{{Fine Tuning of Tissues' Viscosity and Surface
  Tension through Contractility Suggests a New Role for alpha-Catenin}}.
\newblock \emph{\bibinfo{journal}{PLoS One}} \textbf{\bibinfo{volume}{8}},
  \bibinfo{pages}{e52554} (\bibinfo{year}{2013}).

\bibitem{Rodriguez-Hernandez2016}
\bibinfo{author}{Rodriguez-Hernandez, I.}, \bibinfo{author}{Cantelli, G.},
  \bibinfo{author}{Bruce, F.} \& \bibinfo{author}{Sanz-Moreno, V.}
\newblock \bibinfo{title}{{Rho, ROCK and actomyosin contractility in metastasis
  as drug targets}}.
\newblock \emph{\bibinfo{journal}{F1000Research}} \textbf{\bibinfo{volume}{5}},
  \bibinfo{pages}{783} (\bibinfo{year}{2016}).

\bibitem{Ouderkirk2014}
\bibinfo{author}{Ouderkirk, J.~L.} \& \bibinfo{author}{Krendel, M.}
\newblock \bibinfo{title}{{Non-muscle myosins in tumor progression, cancer cell
  invasion, and metastasis}}.
\newblock \emph{\bibinfo{journal}{Cytoskeleton (Hoboken).}}
  \textbf{\bibinfo{volume}{71}}, \bibinfo{pages}{447--63}
  (\bibinfo{year}{2014}).

\bibitem{Paredes2012}
\bibinfo{author}{Paredes, J.} \emph{et~al.}
\newblock \bibinfo{title}{{Epithelial E- and P-cadherins: Role and clinical
  significance in cancer}}.
\newblock \emph{\bibinfo{journal}{Biochim. Biophys. Acta}}
  \textbf{\bibinfo{volume}{1826}}, \bibinfo{pages}{297--311}
  (\bibinfo{year}{2012}).

\bibitem{Paschos2009}
\bibinfo{author}{Paschos, K.~A.}, \bibinfo{author}{Canovas, D.} \&
  \bibinfo{author}{Bird, N.~C.}
\newblock \bibinfo{title}{{The role of cell adhesion molecules in the
  progression of colorectal cancer and the development of liver metastasis}}.
\newblock \emph{\bibinfo{journal}{Cell. Signal.}}
  \textbf{\bibinfo{volume}{21}}, \bibinfo{pages}{665--674}
  (\bibinfo{year}{2009}).

\bibitem{Clark2015}
\bibinfo{author}{Clark, A.~G.} \& \bibinfo{author}{Vignjevic, D.~M.}
\newblock \bibinfo{title}{{Modes of cancer cell invasion and the role of the
  microenvironment}}.
\newblock \emph{\bibinfo{journal}{Curr. Opin. Cell Biol.}}
  \textbf{\bibinfo{volume}{36}}, \bibinfo{pages}{13--22}
  (\bibinfo{year}{2015}).

\bibitem{Lu2012}
\bibinfo{author}{Lu, P.}, \bibinfo{author}{Weaver, V.~M.} \&
  \bibinfo{author}{Werb, Z.}
\newblock \bibinfo{title}{{The extracellular matrix: A dynamic niche in cancer
  progression}}.
\newblock \emph{\bibinfo{journal}{J. Cell Biol.}}
  \textbf{\bibinfo{volume}{196}}, \bibinfo{pages}{395--406}
  (\bibinfo{year}{2012}).

\bibitem{Cortina2007}
\bibinfo{author}{Cortina, C.} \emph{et~al.}
\newblock \bibinfo{title}{{EphB--ephrin-B interactions suppress colorectal
  cancer progression by compartmentalizing tumor cells}}.
\newblock \emph{\bibinfo{journal}{Nat. Genet.}} \textbf{\bibinfo{volume}{39}},
  \bibinfo{pages}{1376--1383} (\bibinfo{year}{2007}).

\bibitem{Casares2015}
\bibinfo{author}{Casares, L.} \emph{et~al.}
\newblock \bibinfo{title}{{Hydraulic fracture during epithelial stretching}}.
\newblock \emph{\bibinfo{journal}{Nat. Mater.}} \textbf{\bibinfo{volume}{14}},
  \bibinfo{pages}{343--351} (\bibinfo{year}{2015}).

\bibitem{Tambe2013}
\bibinfo{author}{Tambe, D.~T.} \emph{et~al.}
\newblock \bibinfo{title}{{Monolayer Stress Microscopy: Limitations, Artifacts,
  and Accuracy of Recovered Intercellular Stresses}}.
\newblock \emph{\bibinfo{journal}{PLoS One}} \textbf{\bibinfo{volume}{8}},
  \bibinfo{pages}{e55172} (\bibinfo{year}{2013}).

\bibitem{Roux2016}
\bibinfo{author}{Roux, C.} \emph{et~al.}
\newblock \bibinfo{title}{{Prediction of traction forces of motile cells}}.
\newblock \emph{\bibinfo{journal}{Interface Focus}}
  \textbf{\bibinfo{volume}{6}}, \bibinfo{pages}{20160042}
  (\bibinfo{year}{2016}).

\bibitem{Mayor2010}
\bibinfo{author}{Mayor, R.} \& \bibinfo{author}{Carmona-Fontaine, C.}
\newblock \bibinfo{title}{{Keeping in touch with contact inhibition of
  locomotion}}.
\newblock \emph{\bibinfo{journal}{Trends Cell Biol.}}
  \textbf{\bibinfo{volume}{20}}, \bibinfo{pages}{319--328}
  (\bibinfo{year}{2010}).

\bibitem{Stramer2017}
\bibinfo{author}{Stramer, B.} \& \bibinfo{author}{Mayor, R.}
\newblock \bibinfo{title}{{Mechanisms and in vivo functions of contact
  inhibition of locomotion}}.
\newblock \emph{\bibinfo{journal}{Nat. Rev. Mol. Cell Biol.}}
  \textbf{\bibinfo{volume}{18}}, \bibinfo{pages}{43--55}
  (\bibinfo{year}{2017}).

\bibitem{Desai2009}
\bibinfo{author}{Desai, R.~A.}, \bibinfo{author}{Gao, L.},
  \bibinfo{author}{Raghavan, S.}, \bibinfo{author}{Liu, W.~F.} \&
  \bibinfo{author}{Chen, C.~S.}
\newblock \bibinfo{title}{{Cell polarity triggered by cell-cell adhesion via
  E-cadherin}}.
\newblock \emph{\bibinfo{journal}{J. Cell Sci.}}
  \textbf{\bibinfo{volume}{122}}, \bibinfo{pages}{905--911}
  (\bibinfo{year}{2009}).

\bibitem{Khalil2010}
\bibinfo{author}{Khalil, A.~A.} \& \bibinfo{author}{Friedl, P.}
\newblock \bibinfo{title}{{Determinants of leader cells in collective cell
  migration}}.
\newblock \emph{\bibinfo{journal}{Integr. Biol.}} \textbf{\bibinfo{volume}{2}},
  \bibinfo{pages}{568--574} (\bibinfo{year}{2010}).

\bibitem{Weber2012}
\bibinfo{author}{Weber, G.~F.}, \bibinfo{author}{Bjerke, M.~A.} \&
  \bibinfo{author}{DeSimone, D.~W.}
\newblock \bibinfo{title}{{A Mechanoresponsive Cadherin-Keratin Complex Directs
  Polarized Protrusive Behavior and Collective Cell Migration}}.
\newblock \emph{\bibinfo{journal}{Dev. Cell}} \textbf{\bibinfo{volume}{22}},
  \bibinfo{pages}{104--115} (\bibinfo{year}{2012}).

\bibitem{Theveneau2013}
\bibinfo{author}{Theveneau, E.} \& \bibinfo{author}{Mayor, R.}
\newblock \bibinfo{title}{{Collective cell migration of epithelial and
  mesenchymal cells}}.
\newblock \emph{\bibinfo{journal}{Cell. Mol. Life Sci.}}
  \textbf{\bibinfo{volume}{70}}, \bibinfo{pages}{3481--3492}
  (\bibinfo{year}{2013}).

\bibitem{Vedula2013a}
\bibinfo{author}{Vedula, S. R.~K.}, \bibinfo{author}{Ravasio, A.},
  \bibinfo{author}{Lim, C.~T.} \& \bibinfo{author}{Ladoux, B.}
\newblock \bibinfo{title}{{Collective Cell Migration: A Mechanistic
  Perspective}}.
\newblock \emph{\bibinfo{journal}{Physiology}} \textbf{\bibinfo{volume}{28}},
  \bibinfo{pages}{370--379} (\bibinfo{year}{2013}).

\bibitem{Ladoux2016}
\bibinfo{author}{Ladoux, B.}, \bibinfo{author}{M{\`{e}}ge, R.-M.} \&
  \bibinfo{author}{Trepat, X.}
\newblock \bibinfo{title}{{Front-Rear Polarization by Mechanical Cues: From
  Single Cells to Tissues}}.
\newblock \emph{\bibinfo{journal}{Trends Cell Biol.}}
  \textbf{\bibinfo{volume}{26}}, \bibinfo{pages}{420--433}
  (\bibinfo{year}{2016}).

\bibitem{Mayor2016}
\bibinfo{author}{Mayor, R.} \& \bibinfo{author}{Etienne-Manneville, S.}
\newblock \bibinfo{title}{{The front and rear of collective cell migration}}.
\newblock \emph{\bibinfo{journal}{Nat. Rev. Mol. Cell Biol.}}
  \textbf{\bibinfo{volume}{17}}, \bibinfo{pages}{97--109}
  (\bibinfo{year}{2016}).

\bibitem{Hakim2017}
\bibinfo{author}{Hakim, V.} \& \bibinfo{author}{Silberzan, P.}
\newblock \bibinfo{title}{{Collective cell migration: a physics perspective}}.
\newblock \emph{\bibinfo{journal}{Reports Prog. Phys.}}
  \textbf{\bibinfo{volume}{80}}, \bibinfo{pages}{076601}
  (\bibinfo{year}{2017}).

\bibitem{Zimmermann2016}
\bibinfo{author}{Zimmermann, J.}, \bibinfo{author}{Camley, B.~A.},
  \bibinfo{author}{Rappel, W.-J.} \& \bibinfo{author}{Levine, H.}
\newblock \bibinfo{title}{{Contact inhibition of locomotion determines
  cell-cell and cell-substrate forces in tissues}}.
\newblock \emph{\bibinfo{journal}{Proc. Natl. Acad. Sci. U. S. A.}}
  \textbf{\bibinfo{volume}{113}}, \bibinfo{pages}{2660--2665}
  (\bibinfo{year}{2016}).

\bibitem{Coburn2016}
\bibinfo{author}{Coburn, L.} \emph{et~al.}
\newblock \bibinfo{title}{{Contact inhibition of locomotion and mechanical
  cross-talk between cell-cell and cell-substrate adhesion determine the
  pattern of junctional tension in epithelial cell aggregates}}.
\newblock \emph{\bibinfo{journal}{Mol. Biol. Cell}}
  \textbf{\bibinfo{volume}{27}}, \bibinfo{pages}{3436--3448}
  (\bibinfo{year}{2016}).

\bibitem{Vincent2015}
\bibinfo{author}{Vincent, R.} \emph{et~al.}
\newblock \bibinfo{title}{{Active Tensile Modulus of an Epithelial Monolayer}}.
\newblock \emph{\bibinfo{journal}{Phys. Rev. Lett.}}
  \textbf{\bibinfo{volume}{115}}, \bibinfo{pages}{248103}
  (\bibinfo{year}{2015}).

\bibitem{Forgacs1998}
\bibinfo{author}{Forgacs, G.}, \bibinfo{author}{Foty, R.~A.},
  \bibinfo{author}{Shafrir, Y.} \& \bibinfo{author}{Steinberg, M.~S.}
\newblock \bibinfo{title}{{Viscoelastic Properties of Living Embryonic Tissues:
  a Quantitative Study}}.
\newblock \emph{\bibinfo{journal}{Biophys. J.}} \textbf{\bibinfo{volume}{74}},
  \bibinfo{pages}{2227--2234} (\bibinfo{year}{1998}).

\bibitem{Marmottant2009}
\bibinfo{author}{Marmottant, P.} \emph{et~al.}
\newblock \bibinfo{title}{{The role of fluctuations and stress on the effective
  viscosity of cell aggregates}}.
\newblock \emph{\bibinfo{journal}{Proc. Natl. Acad. Sci. U. S. A.}}
  \textbf{\bibinfo{volume}{106}}, \bibinfo{pages}{17271--17275}
  (\bibinfo{year}{2009}).

\bibitem{Guevorkian2010}
\bibinfo{author}{Guevorkian, K.}, \bibinfo{author}{Colbert, M.-J.},
  \bibinfo{author}{Durth, M.}, \bibinfo{author}{Dufour, S.} \&
  \bibinfo{author}{Brochard-Wyart, F.}
\newblock \bibinfo{title}{{Aspiration of Biological Viscoelastic Drops}}.
\newblock \emph{\bibinfo{journal}{Phys. Rev. Lett.}}
  \textbf{\bibinfo{volume}{104}}, \bibinfo{pages}{218101}
  (\bibinfo{year}{2010}).

\bibitem{Basan2013}
\bibinfo{author}{Basan, M.}, \bibinfo{author}{Elgeti, J.},
  \bibinfo{author}{Hannezo, E.}, \bibinfo{author}{Rappel, W.-J.} \&
  \bibinfo{author}{Levine, H.}
\newblock \bibinfo{title}{{Alignment of cellular motility forces with tissue
  flow as a mechanism for efficient wound healing}}.
\newblock \emph{\bibinfo{journal}{Proc. Natl. Acad. Sci. U. S. A.}}
  \textbf{\bibinfo{volume}{110}}, \bibinfo{pages}{2452--2459}
  (\bibinfo{year}{2013}).

\bibitem{Recho2016}
\bibinfo{author}{Recho, P.}, \bibinfo{author}{Ranft, J.} \&
  \bibinfo{author}{Marcq, P.}
\newblock \bibinfo{title}{{One-dimensional collective migration of a
  proliferating cell monolayer}}.
\newblock \emph{\bibinfo{journal}{Soft Matter}} \textbf{\bibinfo{volume}{12}},
  \bibinfo{pages}{2381--2391} (\bibinfo{year}{2016}).

\bibitem{Yabunaka2017a}
\bibinfo{author}{Yabunaka, S.} \& \bibinfo{author}{Marcq, P.}
\newblock \bibinfo{title}{{Cell growth, division, and death in cohesive
  tissues: A thermodynamic approach}}.
\newblock \emph{\bibinfo{journal}{Phys. Rev. E}} \textbf{\bibinfo{volume}{96}},
  \bibinfo{pages}{022406} (\bibinfo{year}{2017}).

\bibitem{Wyatt2016}
\bibinfo{author}{Wyatt, T.}, \bibinfo{author}{Baum, B.} \&
  \bibinfo{author}{Charras, G.}
\newblock \bibinfo{title}{{A question of time: tissue adaptation to mechanical
  forces}}.
\newblock \emph{\bibinfo{journal}{Curr. Opin. Cell Biol.}}
  \textbf{\bibinfo{volume}{38}}, \bibinfo{pages}{68--73}
  (\bibinfo{year}{2016}).

\bibitem{Khalilgharibi2016}
\bibinfo{author}{Khalilgharibi, N.}, \bibinfo{author}{Fouchard, J.},
  \bibinfo{author}{Recho, P.}, \bibinfo{author}{Charras, G.} \&
  \bibinfo{author}{Kabla, A.}
\newblock \bibinfo{title}{{The dynamic mechanical properties of cellularised
  aggregates}}.
\newblock \emph{\bibinfo{journal}{Curr. Opin. Cell Biol.}}
  \textbf{\bibinfo{volume}{42}}, \bibinfo{pages}{113--120}
  (\bibinfo{year}{2016}).

\bibitem{Julicher2009}
\bibinfo{author}{J{\"{u}}licher, F.} \& \bibinfo{author}{Prost, J.}
\newblock \bibinfo{title}{{Generic theory of colloidal transport}}.
\newblock \emph{\bibinfo{journal}{Eur. Phys. J. E}}
  \textbf{\bibinfo{volume}{29}}, \bibinfo{pages}{27--36}
  (\bibinfo{year}{2009}).

\bibitem{Cochet-Escartin2014}
\bibinfo{author}{Cochet-Escartin, O.}, \bibinfo{author}{Ranft, J.},
  \bibinfo{author}{Silberzan, P.} \& \bibinfo{author}{Marcq, P.}
\newblock \bibinfo{title}{{Border Forces and Friction Control Epithelial
  Closure Dynamics}}.
\newblock \emph{\bibinfo{journal}{Biophys. J.}} \textbf{\bibinfo{volume}{106}},
  \bibinfo{pages}{65--73} (\bibinfo{year}{2014}).

\bibitem{Eisenhoffer2012}
\bibinfo{author}{Eisenhoffer, G.~T.} \emph{et~al.}
\newblock \bibinfo{title}{{Crowding induces live cell extrusion to maintain
  homeostatic cell numbers in epithelia}}.
\newblock \emph{\bibinfo{journal}{Nature}} \textbf{\bibinfo{volume}{484}},
  \bibinfo{pages}{546--549} (\bibinfo{year}{2012}).

\bibitem{Marinari2012}
\bibinfo{author}{Marinari, E.} \emph{et~al.}
\newblock \bibinfo{title}{{Live-cell delamination counterbalances epithelial
  growth to limit tissue overcrowding}}.
\newblock \emph{\bibinfo{journal}{Nature}} \textbf{\bibinfo{volume}{484}},
  \bibinfo{pages}{542--545} (\bibinfo{year}{2012}).

\bibitem{Eisenhoffer2013}
\bibinfo{author}{Eisenhoffer, G.~T.} \& \bibinfo{author}{Rosenblatt, J.}
\newblock \bibinfo{title}{{Bringing balance by force: live cell extrusion
  controls epithelial cell numbers}}.
\newblock \emph{\bibinfo{journal}{Trends Cell Biol.}}
  \textbf{\bibinfo{volume}{23}}, \bibinfo{pages}{185--192}
  (\bibinfo{year}{2013}).

\bibitem{Joanny2012}
\bibinfo{author}{Joanny, J.-F.} \& \bibinfo{author}{Ramaswamy, S.}
\newblock \bibinfo{title}{{A drop of active matter}}.
\newblock \emph{\bibinfo{journal}{J. Fluid Mech.}}
  \textbf{\bibinfo{volume}{705}}, \bibinfo{pages}{46--57}
  (\bibinfo{year}{2012}).

\bibitem{Joanny2013}
\bibinfo{author}{Joanny, J.-F.}, \bibinfo{author}{Kruse, K.},
  \bibinfo{author}{Prost, J.} \& \bibinfo{author}{Ramaswamy, S.}
\newblock \bibinfo{title}{{The actin cortex as an active wetting layer}}.
\newblock \emph{\bibinfo{journal}{Eur. Phys. J. E}}
  \textbf{\bibinfo{volume}{36}}, \bibinfo{pages}{9866} (\bibinfo{year}{2013}).

\end{thebibliography}

\vskip0.5cm
\noindent\textbf{Acknowledgements}

We thank David Sarri\'{o} and Gemma Moreno-Bueno for providing the E-cadherin inducible cells; Natalia Castro for technical assistance; Alberto Elosegui, V\'{i}ctor Gonz\'{a}lez, Ernest Latorre, L\'{e}o Valon and Romaric Vincent for stimulating discussions. R.A. thanks Gen\'{i}s Torrents for assistance with mathematical details. C.P-G. and R.A. were funded by Fundaci\'{o} ``La Caixa''. R.A. thanks Jacques Prost and acknowledges EMBO (Short Term Fellowship ASTF 365-2015), The Company of Biologists (Development Travelling Fellowship DEVTF-151206), and Fundaci\'{o} Universit\`{a}ria Agust\'{i} Pedro i Pons for supporting visits to Institut Curie. This work was supported by the Spanish Ministry of Economy and Competitiveness/FEDER (BFU2015-65074-P to XT, FIS2016-78507-C2-2-P to JC), the Generalitat de Catalunya (2014-SGR-927 and CERCA program to XT, 2014-SGR-878 to JC), the European Research Council (CoG-616480 to XT), European Commission (H2020-FETPROACT-01-2016-731957 to XT), and Obra Social ``La Caixa''. IBEC is recipient of a Severo Ochoa Award of Excellence from the MINECO.

\vskip0.5cm
\noindent\textbf{Author contributions}

C.P-G., R.A., J.C. and X.T. conceived the study and designed experiments. C.P-G. performed the experiments with the help of T.K. and E.B. C.P-G. and M.G-G. developed computational analysis tools. C.P-G. processed and analyzed the experimental data. R.A. developed the active wetting theory with the help of C.B-M. and fitted the model predictions to the experimental data. J.C. and X.T. supervised the study. C.P-G., R.A., J.C. and X.T. wrote the manuscript. All authors contributed to the interpretation of the results and commented the manuscript.

\onecolumngrid 

\clearpage
\appendix

\twocolumngrid

\begin{center}
\textbf{\large Methods}
\end{center}

\textbf{MDA-MB-231 cell culture.} MDA-MB-231 cells were grown on Dulbecco Modified Eagle Medium‎ (DMEM) media supplemented with $10\%$ fetal bovine serum (FBS), $100$ U mL$^{-1}$ penicillin and $100$ $\mu$g mL$^{-1}$ streptomycin.

\textbf{E-cadherin induction.} Right before starting an experiment, normal cell media was replaced by media containing $10$ nM of dexamethasone to induce the expression of E-cadherin.

\textbf{Polyacrylamide gel substrate.} Polyacrylamide (PAA) gels of 3, 12 and 30 kPa (Young modulus) were produced as described previously\cite{Bazellieres2015}. Briefly, a solution containing $5.5\%$ acrylamide, $0.09\%$ bis-acrylamide (3 kPa); $7.5\%$ acrylamide, $0.16\%$ bis-acrylamide (12 kPa); or $12\%$ acrylamide, $0.15\%$ bis-acrylamide (30 kPa); plus $0.5\%$ ammonium persulphate, $0.05\%$ tetramethylethylenediamine and $0.64\%$ of 200-nm-diameter red fluorescent carboxylate-modified beads was prepared and allowed to polymerize. PAA gel surface was then incubated with a solution of 2 mg/mL Sulpho-SANPAH under UV light for 5 minutes (wavelength of 365 nm at a distance of 5 cm). After that, 3 washes of 3 minutes each were performed to remove the excess of Sulfo-SANPAH. At this point, the gel was ready to add the ECM protein.

\textbf{PDMS stencils.} Polydimethylsiloxane (PDMS) membranes were fabricated as explained previously\cite{Casares2015}. Briefly, SU8-50 masters containing arrays of circles of different sizes ($200$ $\mu$­m, $150$ $\mu$­­m, $100$ $\mu$­­m, and $50$ $\mu$­­m radius) were raised using conventional photolithography. Importantly, all the different sizes were included in the same array to allow having different conditions in the same gel, therefore decreasing experimental variability. Uncured PDMS was spin coated on top of the masters to a thickness lower than the SU8 features ($35$ $\mu$­­m) and cured at $80^\circ$'C for 2 hours. A thick border of PDMS was added for handling purposes. Finally, PDMS stencils were peeled off and stored in $96\%$ ethanol at $4^\circ$C until use.

\textbf{Cell patterning on PAA gels.} The PDMS stencils were incubated with a solution of pluronic acid F127 $2\%$ for one hour. After that, they were washed twice in Phosphate-Buffered Saline (PBS) and let dry for 20 minutes. For confined monolayers, the stencils were carefully placed on top of the PAA gels. Then a solution of rat tail type I collagen at the desired concentration was added on top of the PDMS openings and left at $4^\circ$C overnight. The day after, the collagen solution was washed and the PDMS stencils were removed. The PAA gels were washed twice with PBS. For cell seeding, the PBS was removed and a $75$ $\mu$­­L drop containing $\sim 500000$ cells was placed on top of the PAA gels. After 30 minutes, the unattached cells were washed away and more medium was added. Cells were allowed to spread for 3 hours before starting the experiment. In the case of unconfined monolayers, the PDMS stencil was placed on top of gels already coated with collagen. Cells fell in the openings and attached to the gel for 8 h before releasing the confinement.

\textbf{Time-lapse microscopy.} Multidimensional acquisitions were performed on an automatic inverted microscope (Nikon Eclipse Ti) using a 20X objective (NA 0.75, air) for TFM experiments. MetaMorph (Universal Imaging) was used to image every hour during the duration of the experiment. Around 50 cell islands were imaged in parallel using a motorized stage. In the case of the 3D reconstruction (\cref{FigS4}, Supplementary Movie 4) and nuclei position analysis (\cref{FigS10}, Supplementary Movie 7), multidimensional acquisitions were performed on a Nikon microscope with a spinning disk confocal unit (CSU-W1, Yokogawa) using a 40X objective (NA 0.75, air) and a 20X (NA 0.75, air) respectively. IQ3 (Andor) software was used to image every 15 minutes with a Z-step of $1$ $\mu$­­m. All microscopes were equipped with thermal, CO2, and humidity control.

\textbf{Traction force microscopy.} Traction forces were computed using Fourier-transform traction microscopy with finite gel thickness from a gel displacements field\cite{Trepat2009}. Gel displacements were obtained using a custom-made particle image velocimetry (PIV). In brief, the fluorescent beads in any experimental timepoint were compared to a reference image obtained after cell trypsinization at the end of the experiment.

\textbf{Monolayer stress microscopy.} Monolayer tension was obtained using Monolayer Stress Microscopy as described previously\cite{Tambe2011,Tambe2013}. Force balance with tractions yields the tension field in the monolayer, as a second rank symmetric tensor. We computed the average normal stress as the mean of the xx and yy components. In this two dimensional approximation, tension has units of surface tension, namely N/m.

\textbf{Western blot}. $\sim 500000$ cells were seeded on $12$ kPa (Young Modulus) PAA gels (for MLC and ppMLC) or plastic (for E-cadherin). After 3 hours, E-cadherin expression was induced and cells were sequentially lysed with Laemli 1x at the desired times post induction. Samples were then mechanically disaggregated using a syringe and centrifuged at $20000$ g for 15 minutes. Samples were heated at $95^\circ$C for 5 minutes and loaded on polyacrylamide gels (Any kd, Bio-rad) for electrophoresis. After that, proteins were transferred to a nitrocellulose membrane (Whatman, GE Healthcare Life Sciences) overnight. Membranes were blocked with $5\%$ dry milk-Tris buffer saline-$0.2\%$ Tween, incubated with primary antibodies overnight at $4^\circ$C and, later, incubated with horseradish-peroxidase-coupled secondary antibodies for 1 hour at room temperature. Bands were revealed using LimiLight kit (Roche), imaged with ImageQuant LAS 4000 and quantified using ImageJ software. Tubulin was used as an endogenous control for normalization.

\textbf{Immunostaining.} MDA-MB-231 cells were washed with PBS, fixed with $4\%$ paraformaldehyde (PFA) for 10 minutes and permeabilized in $0,1\%$ Triton X-100 for 5 minutes. After washing, cells were blocked in $10\%$ FBS for 1 hour and incubated with primary antibodies for 3 hours. Cells were then washed and incubated with the appropriate secondary antibody for 1 hour. After washing, cells were mounted in Mowiol reagent. Images were acquired using a Nikon microscope with a spinning disk confocal unit (CSU-W1, Yokogawa) using a 60X objective (NA 1.40, oil).

\textbf{Antibodies.} The primary antibodies used were: anti-E-cadherin monoclonal antibody (clone 36, BD Transduction Laboratories, no. 610181), anti-$\alpha$-tubulin (clone B-5-1-2, Sigma-Aldrich, no. T5168), anti-$\beta$-catenin (clone 14, BD Transduction Laboratories, no. 610154), anti-paxillin (clone 349, BD Transduction Laboratories, no. 610051), anti-rat collagen type I (EMD Millipore, AB755P), anti-ppMLC (Cell Signaling Technology, $\#3674$), and anti-MLC (Cell Signaling Technology, $\#3672$). The secondary antibodies were: peroxidase-conjugated anti-mouse IgG (Jackson Immuno Research, no. 715-035-151) and peroxidase-conjugated anti-rabbit IgG (Jackson Immuno Research, no. 211-032-171) for western blot and Alexa Fluor 488 anti-rabbit (Invitrogen, Molecular Probes, no. A-21206), Alexa Fluor 488 anti-mouse (Invitrogen, Molecular Probes, no. A-11029), Alexa Fluor 555 anti-mouse (Invitrogen, Molecular Probes, no. A-28180), Alexa Fluor 640 anti-rabbit (Invitrogen, Molecular Probes, no. A-21245), Alexa Fluor 405 anti-mouse (Invitrogen, Molecular Probes, no. A-31553) for immunostaining. For western blot, anti-E-cadherin was diluted 1:2000 and anti-$\alpha$-Tubulin was diluted 1:5000; anti-ppMLC was diluted 1:500; anti-MLC was diluted 1:200; secondary antibody was diluted 1:5000. For immunofluorescence, primary antibodies were diluted 1:200 and secondary antibodies were diluted 1:400. F-actin was labelled with Phalloidin-TRITC (Sigma-Aldrich, no P1951) diluted 1:2000.

\textbf{Cell island segmentation.} At every timepoint, cell islands were semi-automatically segmented using custom-made Matlab software. First, a preliminary mask of the island contour was performed automatically based on changes in contrast of phase contrast images. The errors in the automatic segmentation were manually corrected.

\textbf{Immunostaining fluorescence intensity quantification.} Both for ppMLC and collagen intensity quantifications, the region of interest (ROI) was segmented as explained above. The mean or median intensy in the ROI was calculated and the background intensity was substracted to every individual measurement.

\textbf{Kymography.} We obtained the radial coordinates of each pixel of the cell island masks by calculating its shortest distance to the edge. The radial direction of the edge was calculated and expanded to the inner pixels of the mask to decompose traction vectors in radial and tangential components. Finally, tractions or tensions were averaged according to their distance to the edge at every timepoint to build spatiotemporal kymographs.

\textbf{Wetting transition definition.} We defined an objective criterion to detect the wetting transition in different experimental conditions. First, cell islands are automatically segmented based on changes in contrast of the phase contrast images, followed by a manual correction of the errors in segmentation. Every cell island mask is divided in a specific number of circular sectors based on its initial radius (24 for $200$ $\mu$­m radius, 18 for $150$ $\mu$­m, 12 for $100$ $\mu$­m, and 6 for $50$ $\mu$­m). Using this each strategy sector has an approximately equal arc length at time 0 ($\sim 52$ $\mu$­m). The average radius of every sector is computed over time, obtaining a characteristic curve with a roughly constant value at the first time points that suddenly drops upon the onset of dewetting (\cref{FigS13c}). This curve is fitted with a negative sigmoidal function
\begin{equation}
R(t)= a +\frac{b}{1+\exp[ (d + \ln(2-\sqrt{3})) t/c + d]}
\end{equation}
using the non-linear least squares method. The transition time for every segment is defined as the time point at which the fitted function reaches the $95\%$ of its initial value (open circles in \cref{FigS13b,FigS13c}). For the whole island, we define the onset of dewetting as the moment at which one sixth of the circular sectors are dewetting according to the criterion above.

\textbf{Collagen amount quantification.} Rat tail type I collagen immunostainings were performed on patterns made on polyacrylamide gels coated with three different collagen concentrations ($100$ $\mu$­g/mL, $10$ $\mu$­g/mL, and $1$ $\mu$­g/mL). The patterns were automatically segmented. Their mean intensity was calculated and corrected by subtracting the mean background intensity.

\textbf{Model parameters fit.} We fit the predicted radial traction force profile
\begin{equation}
T_r(r) = -T_0 p_r(r) = - T_0 \frac{I_1(r/L_c)}{I_1(R/L_c)},
\end{equation}
where $I_1$ is the modified Bessel function of the first kind and first order (Supplementary Note), to the experimentally measured profiles at different times, as represented in kymographs as in \cref{FigS3a}. At each time point, the fitting algorithm searches for the radial position of the maximum of the experimental traction force profile, which sets the monolayer radius $R(t)$. Then, the theoretical prediction is fit up to this point, discarding the outer region where the traction force progressively vanishes (\cref{Fig3d}). Traction forces measured in this outer region may arise from poorly attached protrusions or be an artefact due to the long-range propagation of deformations in the elastic substrate used for traction force microscopy. These effects are not described by the model. From the fits, we obtain the time evolution of the maximal traction stress $T_0 (t)$ and the nematic length $L_c (t)$. Finally, the contractility $-\zeta(t)$ during the wetting phase is given by (Supplementary Note)
\begin{equation} \label{eq contractility-methods}
-\zeta=2T_0 \frac{L_c}{h}\frac{I_2\left(R/L_c\right)}{I_1\left(R/L_c\right)-I_0\left(R/L_c\right)\left[\frac{I_0\left(R/L_c\right)}{I_1\left(R/L_c\right)}-\frac{2L_c}{R}\right]}.
\end{equation}
To check the values of the contractility given by \cref{eq contractility-methods}, we extracted the contractility via two other methods. First, this parameter can be obtained from fits of the radial tension profile in the monolayer (Supplementary Note):
\begin{multline}
\sigma_{rr}\left(r\right)=T_0 L_c\left[\frac{I_0^2\left(R/L_c\right)}{I_1^2\left(R/L_c\right)} - \frac{L_c}{R}\right] - T_0 L_c\frac{I_0\left(r/L_c\right)}{I_1\left(R/L_c\right)} \\
+ \frac{\zeta h}{2}\left[1+\frac{L_c}{R}\frac{I_0\left(R/L_c\right)}{I_1\left(R/L_c\right)} - \frac{I_0^2\left(R/L_c\right)}{I_1^2\left(R/L_c\right)}\right] +\frac{\zeta h}{2}\frac{1}{I_1^2\left(R/L_c\right)}\\
\left[\frac{1}{2}I_0\left(r/L_c\right)\left[I_0\left(r/L_c\right) + I_2\left(r/L_c\right)\right] - I_1^2\left(r/L_c\right)\right].
\end{multline}
In the fits of the tension kymographs, the monolayer radius $R(t)$ is determined from the radial coordinate at which the stress vanishes, $\sigma_rr (R)=0$. Second, the contractility can also be obtained from the average radial tension
\begin{multline}
\sigma=\frac{1}{\pi R^2}\int_0^{2\pi} d\theta \int_0^R \sigma_{rr}\, r\, dr =\\
 T_0 L_c \left[\frac{I_0\left(R/L_c\right)}{I_1\left(R/L_c\right)} - 3\frac{L_c}{R} + 2\frac{L_c^2}{R^2}\frac{I_0\left(R/L_c\right) I_1\left(R/L_c\right) - 1}{I_1^2\left(R/L_c\right)}\right] \\
-\frac{\zeta h}{2}\left[1 - \frac{I_0^2\left(R/L_c\right)}{I_1^2\left(R/L_c\right)} \right.\\
\left.+ \frac{L_c}{R}\frac{I_0\left(R/L_c\right)}{I_1\left(R/L_c\right)} + \frac{L_c^2}{R^2}\frac{I_0^2\left(R/L_c\right)-1}{I_1^2\left(R/L_c\right)}\right]
\end{multline}
All three methods yield fully compatible results. Note that, at the lowest order in the small dimensionless parameter $L_c/R$, the average tension is completely given by traction forces: $\sigma=T_0 L_c+\mathcal{O}(L_c/R)$. Therefore, the contractility only contributes to the average stress at the first-order level in $L_c/R$, which explains the large values of this parameter compared to the stress in the monolayer.

\textbf{Monolayer boundary Fourier transform.} The local monolayer radius as function of the polar angle, $R(\theta)$, was computed via the same method than for wetting-dewetting transition definition. However, in this case, the number of segments was systematically multiplied by 8 to increase the spatial resolution. We obtained the radius perturbations as $\delta R(\theta)=R(\theta)-R_0$, where $R_0$ is the average initial radius. This function was Fourier-transformed to obtain the amplitude of every Fourier mode. Two Fourier modes were calculated in a different way. To obtain the evolution of mode $n=0$, we systematically subtracted the mean radius of the island during the last 7 time points before wetting-dewetting transition from the current average radius. Respectively, mode $n=1$ is the direct measure of the centroid motion. To average different replicates, we referred all times to the transition time of each island, namely that we used shifted times $t-t^*$. Theoretical predictions for the growth rates are only valid in a linear regime of the instability, which is characterized by small amplitude perturbations with respect to the wavelength of the specific mode. We consider that a mode is in its linear regime when its amplitude does not exceed $10\%$ of its wavelength. Once this threshold is reached, the mode is excluded from the analysis. Furthermore, islands with high mode amplitudes before dewetting (a specific mode whose amplitude exceeds 6 times the mean amplitude of all the other modes) were also excluded to avoid biases coming from irregularities in the patterning.

\textbf{Retraction rate calculation.} The growth rate of the perturbation mode $n=0$ is obtained by fitting the exponential function $\delta\tilde{R}ƒ_0 (t)=\delta\tilde{R}ƒ_0 (t^*)e^{\omega‰_0 (t-t^*)}$ to the evolution of its amplitude, from the last timepoint before the transition to 7 hours after the onset of dewetting. By choosing this time span, we ensured to have enough time points to perform reliable fits (\cref{FigS15}) while still having most of the perturbation modes in almost all monolayers within the linear regime of the instability. The error of $\omega_0$ is defined as the $95\%$ confidence interval.

\textbf{Data availability.} All the data used for this study is available upon request.

\textbf{Code availability.} All computer codes used for this study are available upon request to the corresponding authors.

\onecolumngrid 

\clearpage

\setcounter{equation}{0}
\setcounter{figure}{0}
\renewcommand{\theequation}{S\arabic{equation}}
\renewcommand{\thefigure}{S\arabic{figure}}

\onecolumngrid
\begin{center}
\textbf{\large Supplementary Note}
\end{center}

\subsection{Active polar fluid model of epithelial spreading} \label{active-polar-fluid-model}

Instead of formulating a model based on adhesion energies, similar to those previously proposed to describe tissue wetting \cite{Douezan2011}, our aim is to see how the wetting transition arises from mechanical models of collective cell migration. To this end, we extend a previously introduced continuum model of epithelial spreading \cite{Blanch-Mercader2017} to the present problem. This continuum model takes a coarse-grained approach that describes the long-time and large-scale dynamics of the tissue as those of an active polar liquid, namely in terms of a polarity field $\bm{p}\left(\bm{r},t\right)$ and a flow field $\bm{v}\left(\bm{r},t\right)$. Below, we briefly justify this description, which has already been applied to the spreading of tissue monolayers\cite{Blanch-Mercader2017,Lee2011a,Lee2011,Vig2017}. A very similar model was also proposed for traction force and velocity profiles of single crawling cells \cite{Roux2016}.


\subsubsection{Polarity dynamics} \label{polarity-dynamics}
In our monolayer, cells at the center exert weak and random traction forces. In contrast, cells at the edge extend large lamellipodia towards the outside and exert strong inward-pointing traction forces on the substrate, indicating that they are polarized (\cref{Fig1h,FigS3}). The outwards polarization of cells at the border is likely due to contact inhibition of locomotion, a cell-cell interaction whereby cells repolarize in opposite directions upon contact \cite{Mayor2010,Stramer2017}. In fact, this interaction is mediated by cell-cell adhesion, with front-rear differences in cadherin-based junctions acting as a cue for the repolarization \cite{Desai2009,Khalil2010,Weber2012,Theveneau2013,Vedula2013a,Ladoux2016}. Although originally proposed for mesenchymal cells, contact inhibition of locomotion is being increasingly recognized to play a key role in orchestrating the collective migration of epithelial monolayers \cite{Mayor2010,Theveneau2013,Vedula2013a,Ladoux2016,Mayor2016,Hakim2017,Zimmermann2016,Coburn2016,Smeets2016}. In a cohesive monolayer, this interaction naturally leads to polarization of cells at the edge towards free space, leaving the inner region of the monolayer unpolarized. Such a polarity profile, in turn, explains the localization of traction forces at the edge and the build-up of tension at the center of epithelial monolayers \cite{Zimmermann2016,Coburn2016}. Therefore, upon the expression of E-cadherin, we expect the polarity field $\bm{p}\left(\bm{r},t\right)$ to be set by an autonomous cellular mechanism such as contact inhibition of locomotion, which polarizes cells within a time scale $\tau_{\text{CIL}}\sim 10$ min \cite{Smeets2016,Weber2012}. Hence, $\bm{p}\left(\bm{r},t\right)$ should remain essentially independent of flows in the monolayer, which occur over a longer time scale given by the strain rate, at least of order $\tau_s\sim 100$ min \cite{Blanch-Mercader2017,Vincent2015}. Consequently, within a phenomenological approach, we propose the polarity field to follow a purely relaxational dynamics given by
\begin{equation} \label{eq polarity-dynamics}
\frac{\partial p_\alpha}{\partial t}=-\frac{1}{\gamma_1}\frac{\delta F}{\delta p_\alpha},
\end{equation}
where $F\left[\bm{p}\right]$ is the coarse-grained free energy functional of the orientational degrees of freedom, and $\gamma_1$ is a kinetic coefficient (the rotational viscosity for the angular degrees of freedom). With respect to the most general dynamics of the polarity field in an active polar fluid, \cref{eq polarity-dynamics} neglects polarity advection and corotation, as well as flow alignment and active spontaneous polarization effects.

Then, since the bulk of the monolayer remains mechanically unpolarized, the coarse-grained free energy $F$ includes a Landau expansion around the isotropic state $\bm{p}=0$, and gradient terms resulting from nematic elasticity \cite{DeGennes-Prost}:
\begin{equation} \label{eq nematic-free-energy}
F=\int_V \left[\frac{a}{2} p_\alpha p_\alpha + \frac{K}{2} \left(\partial_\alpha p_\beta\right)\left(\partial_\alpha p_\beta\right)\right] d^3\bm{r},
\end{equation}
where $a>0$ is a restoring coefficient of the polarity, and $K$ is the Frank elastic constant in the usual one-constant approximation. The dynamics of the polarity is thus given by
\begin{equation}
\partial_t p_\alpha=\frac{1}{\gamma_1}\left(-a p_\alpha + K\nabla^2 p_\alpha\right).
\end{equation}
In the limit of fast polarity dynamics compared to the spreading dynamics, the polarity field is always at equilibrium, $\partial_t p_\alpha=0$, adiabatically adapting to the shape of the monolayer. Under this approximation, the polarity field is given by
\begin{equation} \label{eq polarity-field-monolayer}
L_c^2 \nabla^2 p_\alpha= p_\alpha,
\end{equation}
where we have defined the characteristic length $L_c\equiv \sqrt{K/a}$ of the polar order in the monolayer.


\subsubsection{Force balance} \label{force-balance-monolayer}
Flows in cell monolayers occur at very low Reynolds numbers. Therefore, inertial forces are negligible, and hence momentum conservation reduces to the force balance condition
\begin{equation}
0 = \partial_\beta \left(\sigma_{\alpha\beta}^s+\sigma_{\alpha\beta}^a+\sigma_{\alpha\beta}^{E,s}\right) + f_\alpha,
\end{equation}
where $\sigma_{\alpha\beta}^s$ and $\sigma_{\alpha\beta}^a$ are the symmetric and antisymmetric parts of the deviatoric stress tensor, and $f_\alpha$ is the external force density. Respectively, $\sigma_{\alpha\beta}^{E,s}$ is the symmetric part of the Ericksen tensor. This tensor generalizes the pressure $P$ to include anisotropic elastic stresses associated to the orientational degrees of freedom in liquid crystals \cite{DeGennes-Prost}:
\begin{equation} \label{eq Ericksen-tensor}
\sigma_{\alpha\beta}^{E}=-P\delta_{\alpha\beta} - \frac{\partial f}{\partial\left(\partial_\beta p_\gamma\right)}\partial_\alpha p_\gamma,
\end{equation}
where $f$ is the Frank free energy density, namely the integrand of \cref{eq nematic-free-energy}. Thus, the orientational contribution to the Ericksen tensor is of second order in gradients of the polarity field, and hence we neglect it, so that force balance reads
\begin{equation}
0 = - \partial_\alpha P + \partial_\beta \left(\sigma_{\alpha\beta}^s+\sigma_{\alpha\beta}^a\right) + f_\alpha.
\end{equation}

Then, the pressure is related to the cell number surface density $\rho$ by the equation of state of the monolayer. For the sake of an estimate, we assume the simplest form for an equation of state, $P\left(\rho\right)=B \left(\rho-\rho_0\right)/\rho_0$, where $B$ is the bulk modulus of the monolayer, and $\rho_0$ is a reference density defined by $P\left(\rho_0\right)=0$. Taking the pressure origin at the monolayer edge, $\rho_0\sim 3\cdot 10^3$ cells/mm$^2$ (\cref{FigS10}). Respectively, density differences in the monolayer are, at most, $\rho-\rho_0\sim 10^3$ cells/mm$^2$ (\cref{FigS10}). Then, the monolayer is expected to be highly compressible because area changes can in principle be accommodated by changes in height, resisted only by the shear modulus of the tissue. Hence, we estimate the bulk modulus of the monolayer by typical shear moduli of cell aggregates, which are in the range $G\sim 10^2-10^3$ Pa \cite{Forgacs1998,Marmottant2009,Guevorkian2010}. Thus, the pressure in the monolayer should be $P\lesssim 30-300$ Pa. In fact, isotropic compressive stresses (pressures) of $\sim 50$ Pa were shown to induce cell extrusion \cite{Saw2017}. In conlusion, if tissue spreading is not dominated by cell proliferation \cite{Basan2013,Recho2016,Yabunaka2017a}, the magnitude of the pressure in the monolayer is expected to be much smaller than the tensile stress (tension) induced by traction forces, as measured by monolayer stress microscopy, which is of the order of several kPa (\cref{Fig1i}), with a monolayer height of $h\sim 5$ $\mu$m. Hence, we neglect the pressure in the force balance:
\begin{equation}
0 = \partial_\beta \left(\sigma_{\alpha\beta}^s+\sigma_{\alpha\beta}^a\right) + f_\alpha.
\end{equation}

Now, for a nematic medium, the antisymmetric part of the stress tensor is given by $\sigma_{\alpha\beta}^a=1/2\left(p_\alpha h_\beta - h_\alpha p_\beta\right)$, where $h_\alpha=-\delta F/\delta p_\alpha$ is the molecular field. From \cref{eq polarity-dynamics}, the adiabatic approximation for the polarity dynamics, $\partial_t p_\alpha=0$, implies $h_\alpha=0$. Therefore, the antisymmetric part of the stress tensor vanishes under this approximation, $\sigma_{\alpha\beta}^a=0$. Thus, force balance reduces to
\begin{equation} \label{eq force-balance-monolayer}
0 = \partial_\beta \sigma_{\alpha\beta}^s + f_\alpha.
\end{equation}

Finally, multiplying \cref{eq force-balance-monolayer} by the height $h(t)$ of the monolayer, the force balance can be rewritten in terms of the experimentally measured traction stress $T_\alpha\left(\bm{r},t\right)$ and monolayer tension $\sigma_{\alpha\beta}\left(\bm{r},t\right)$ fields:
\begin{equation}
\partial_\beta \sigma_{\alpha\beta}=T_\alpha,
\end{equation}
from where
\begin{equation}
T_\alpha=- f_\alpha h,\qquad \sigma_{\alpha\beta} = \sigma_{\alpha\beta}^s h.
\end{equation}


\subsubsection{Constitutive equations} \label{constitutive-equations-monolayer}
Next, constitutive equations must be given to specify the deviatoric stress tensor $\sigma_{\alpha\beta}^s$ and the external force $f_\alpha$ in terms of the polarity and velocity fields. The generic constitutive equations of an active liquid crystal are provided by active gel theory \cite{Kruse2005,Julicher2011,Marchetti2013,Prost2015}. Here, based on the previous assumptions for the dynamics of the polarity field, we propose a simplified version of the generic constitutive equations of an active polar gel to describe epithelial spreading.

First, the spreading occurs on timescales of the order of $\tau_s\sim 100$ min \cite{Blanch-Mercader2017}, at which the tissue should have a fluid rheology. This time scale is much slower than the turnover time scales of proteins in the cytoskeleton or in cell-cell junctions, which are of the order of tens of minutes at most \cite{Wyatt2016,Khalilgharibi2016}. Intra- or intercellular processes such as cytoskeletal reorganizations or cell-cell slidings dissipate energy over these time scales, so that elastic energy may only be stored in the tissue at shorter times. Therefore, to describe the slow spreading dynamics, we will not consider the elastic response of the tissue at short time scales. Note that incessant cell-cell sliding and neighbour exchanges are observed throughout the experiments (Supplementary Movie 7), which provides further support to the fluid behaviour of the monolayer at the experimentally relevant time scales.

Then, in the viscous limit, the constitutive equations for the internal stress and the interfacial force of an active polar medium are:
\begin{equation} \label{eq constitutive-equation-stress}
\sigma_{\alpha\beta}^s=2\eta \tilde{v}_{\alpha\beta}+\frac{\nu_1}{2}\left(p_\alpha h_\beta + h_\alpha p_\beta -\frac{2}{d} p_\gamma h_\gamma \delta_{\alpha\beta}\right) - \zeta q_{\alpha\beta} + \left(\bar{\eta}\, d\, v_{\gamma\gamma} + \bar{\nu}_1\, d\, p_\gamma h_\gamma - \bar{\zeta} - \zeta' p_\gamma p_\gamma\right) \delta_{\alpha\beta},
\end{equation}
\begin{equation} \label{eq constitutive-equation-interfacial-force}
f_\alpha = - \xi v_\alpha + \nu_i \dot{p}_\alpha + \zeta_i p_\alpha,
\end{equation}
where, $q_{\alpha\beta}=p_\alpha p_\beta - p_\gamma p_\gamma/d\; \delta_{\alpha\beta}$ is the traceless symmetric nematic order parameter tensor, with $d$ the system dimensionality, and $v_\alpha$ is the velocity of the fluid with respect to the substrate. The coefficients $\eta$ and $\bar{\eta}$ are the shear and bulk viscosities of the medium, $\zeta$ is the anisotropic active stress coefficient, and $\bar{\zeta}$ and $\zeta'$ are two isotropic active stress coefficients. Finally, $\xi$, $\nu_i$, and $\zeta_i$ are the corresponding interfacial versions of the viscosity (viscous friction), flow alignment (polar friction), and active stress (active force) coefficients. The constitutive equation for the internal stress, \cref{eq constitutive-equation-stress}, is that of an active polar gel with a variable modulus of the polarity \cite{Julicher2011}. In turn, the constitutive equation for the interfacial force, \cref{eq constitutive-equation-interfacial-force}, is less conventional \cite{Julicher2009}, but it was derived from a mesoscopic model of an active polar gel \cite{Oriola2017}.

Now, the adiabatic approximation for the polarity dynamics implies $\dot{p}_\alpha=h_\alpha=0$, so that flow alignment terms contribute neither to the stress tensor nor to the interfacial force. Next, we assume that polarized cells generate much larger active stresses than unpolarized cells. Hence, we neglect the active stress coefficient $\bar{\zeta}$ in front of $\zeta$ and $\zeta'$. Note that, to capture the wetting transition with a model for a two-dimensional fluid layer, the fluid must be compressible, meaning that bulk coefficients must be retained. Then, for simplicity, we assume $\zeta=\zeta'\,d=2\zeta'$ and $2\eta=\bar{\eta}\, d=2\bar{\eta}$. Under these simplifications, the constitutive equations reduce to
\begin{equation} \label{eq bulk-constitutive-equation-monolayer}
\sigma_{\alpha\beta}^s=\eta \left(\partial_\alpha v_\beta + \partial_\beta v_\alpha\right) - \zeta p_\alpha p_\beta,
\end{equation}
\begin{equation} \label{eq interfacial-constitutive-equation-monolayer}
f_\alpha= - \xi v_\alpha + \zeta_i p_\alpha,
\end{equation}
which close the set of equations defining the active polar fluid model of the spreading of an epithelial monolayer.


\subsection{Traction and flow profiles} \label{traction-flow-profiles}

In this section, the model is solved in a circular geometry. There are two unknown fields: the polarity field $\bm{p}\left(\bm{r},t\right)$ and the flow field $\bm{v}\left(\bm{r},t\right)$. The polarity field is completely specified by \cref{eq polarity-field-monolayer}. Once the polarity profile is known, introducing the constitutive equations \cref{eq bulk-constitutive-equation-monolayer,eq interfacial-constitutive-equation-monolayer} into the force balance condition \cref{eq force-balance-monolayer} sets a closed equation for the flow field. The equations for both the polarity and the flow field are time-independent. Therefore, the time dependence of these fields arises solely from the boundary conditions at the free interface, which moves according to $dR/dt=v_r\left(R\right)$.


\subsubsection{Traction profile} \label{traction-profile}
Since traction forces are mainly along the radial direction (\cref{Fig1h}, \cref{FigS3}), we assume the polarity field to be radial: $\bm{p}=p\left(r\right)\bm{r}$. Hence, in polar coordinates, \cref{eq polarity-field-monolayer} reads
\begin{equation}
r^2 p''\left(r\right) + r p'\left(r\right) - \left[1+\frac{r^2}{L_c^2}\right]p\left(r\right)=0.
\end{equation}
Because of the strong outwards polarization of cells at the edge of the cell island, we impose $p\left(R\right)=1$, namely the maximal polarity value, as a boundary condition. Finitude and symmetry of the profile also require $p\left(0\right)=0$. Hence, the solution for the radial polarity profile is
\begin{equation} \label{eq polarity-profile-monolayer}
p\left(r\right)=\frac{I_1\left(r/L_c\right)}{I_1\left(R/L_c\right)},
\end{equation}
where $I_1$ is the modified Bessel function of the first kind and first order. Therefore, the nematic length $L_c$ characterizes the decay of the tissue polarity from its maximal value at the boundary towards its vanishing value in the bulk (red gradient in \cref{Fig3a}).

Next, we may compare the two sources of dissipation, the viscosity and the friction coefficient, whose ratio defines the hydrodynamic screening length $\lambda=\sqrt{\eta/\xi}$. For monolayers smaller than this length, $R<\lambda$, viscosity dominates over friction, and the monolayer stress profile features a central plateau of maximal stress. In contrast, for monolayers larger than this length, $R>\lambda$, friction dominates over viscosity, and the monolayer stress decays at the center, thus featuring its maximum close to the monolayer edge \cite{Blanch-Mercader2017}. In our case, the stress is always maximal at the center of the monolayer (\cref{Fig1i}), meaning that $R>\lambda$. Hence, we neglect cell-substrate friction hereafter. This corresponds to the so-called wet limit, $\lambda\rightarrow \infty$, in which the flows in the monolayer are fully hydrodynamically coupled, with no screening effects due to the release of stress to the substrate through friction \cite{Marchetti2013}. In this limit, the force balance reduces to
\begin{equation} \label{eq force-balance-simplified}
\partial_\beta \sigma_{\alpha\beta}=-T_0 p_\alpha,
\end{equation}
where we have defined the active traction stress coefficient $T_0=\zeta_i h$, which gives the maximal traction stress exerted at the edge of the monolayer.

Then, we fit the predicted radial traction force profile $T_r\left(r\right)=-T_0 p\left(r\right)$ to the experimentally measured profiles at different times, as represented in kymographs as in \cref{FigS3} (see \cref{Fig3d} and Methods). From the fits, we obtain the time evolution of the maximal traction stress $T_0\left(t\right)$ and the nematic length $L_c\left(t\right)$ (\cref{Fig3e},\cref{Fig3f}). After an initial transient, the nematic length remains essentially constant throughout the experiment (\cref{Fig3f}), taking a value $L_c\sim 25$ $\mu$m. This gives support to the assumption that the polarity field is set by a flow-independent mechanism, and that its dynamics is quasi-static.

Now, by combining the inferred value of the nematic length $L_c$ with estimates for typical traction forces and cell migration velocities, we can estimate all the parameters of the polarity dynamics, namely the rotational viscosity $\gamma_1$, the restoring force coefficient $a$, and the Frank elastic constant $K$. To this end, we start by estimating the cell-substrate friction coefficient as $\xi\sim T/\left(v h\right)$. Taking typical values of traction stresses $T\sim 100$ Pa and speeds $v\sim 10$ $\mu$m/min for cell migration \cite{Basan2013}, and estimating the cell height $h\sim 5$ $\mu$m (\cref{FigS4}), we get $\xi\sim 100$ Pa$\cdot$s/$\mu$m$^2$, consistent with previous estimates \cite{Cochet-Escartin2014}. Then, we assume that the rotational viscosity mainly arises from the friction between the substrate and polarized cytoskeletal structures such as the lamellipodia \cite{Lee2011a}. Thus, considering the polarized structures to be rods of length $\ell\sim 10$ $\mu$m comparable to cell length, the rotational friction may be estimated as $\gamma_1\sim \xi \ell^2\sim 10$ kPa$\cdot$s. Now, together with the restoring force coefficient $a$, the rotational friction $\gamma_1$ determines the time scale of the polarity field: $\tau_p\sim \gamma_1/a$. As argued above, the polarity field should be essentially set by contact inhibition of locomotion interactions, so that the time scale of the polarity field may be estimated by that of contact inhibition events, $\tau_p\sim \tau_{\text{CIL}}\sim 10$ min \cite{Smeets2016,Weber2012}. This gives an estimate for the polarity restoring force coefficient $a\sim 20$ Pa. Finally, we estimate the Frank constant as $K= a L_c^2\sim 10$ nN. The estimates of model parameters are collected in \cref{t monolayer-parameter-estimates}.

\begin{table}[bt]
\begin{center}
\begin{tabular}{clc}
Symbol&Description&Estimate\\\hline
$h$&monolayer height&$5$ $\mu$m\\
$\xi$&friction coefficient&$100$ Pa$\cdot$s/$\mu$m$^2$\\
$T_0$&maximal traction&$0.2-0.8$ kPa\\
$L_c$&nematic length&$25$ $\mu$m\\
$-\zeta$&contractility&$5-50$ kPa\\
$\gamma_1$&rotational viscosity&$10$ kPa$\cdot$s\\
$a$&polarity restoring coefficient&$20$ Pa\\
$K$&Frank constant&$10$ nN\\
$\eta$&monolayer viscosity&$3-30$ MPa$\cdot$s\\
$\lambda$&hydrodynamic screening length&$0.2-0.6$ mm\\
$D$&noise intensity of monolayer shape fluctuations&$0.05-1.5$ $\mu$m$^2$/h
\end{tabular}
\end{center}
\caption{Estimates of model parameters.} \label{t monolayer-parameter-estimates}
\end{table}

Finally, knowing the value of $K$ allows to check that the orientational contribution of the Ericken tensor in \cref{eq Ericksen-tensor} is negligible as argued above. Using \cref{eq nematic-free-energy} and the polarity profile in \cref{eq polarity-profile-monolayer}, this contribution can be estimated as $K\left(p'\right)^2\sim K/L_c^2=a\sim 20$ Pa. Therefore, it is much smaller than the typical tensile stresses measured in the monolayer, of the order of several kPa, and it can be safely neglected.


\subsubsection{Flow profile} \label{flow-profile}
The next step is to solve the force balance equation to obtain the velocity field. As for the traction field, we also consider a radial velocity field, $\bm{v}=v\left(r\right)\bm{r}$. Thus, in polar coordinates, the nonvanishing components of the stress tensor are
\begin{equation} \label{eq stress-components-radial}
\frac{1}{h}\sigma_{rr} = \eta v' - \zeta p^2,\qquad \frac{1}{h} \sigma_{\theta\theta}=\eta\frac{v}{r},
\end{equation}
and the force balance reads
\begin{equation}
\sigma_{rr}' + \frac{\sigma_{rr}-\sigma_{\theta\theta}}{r}=-T_0 p.
\end{equation}
Hence, the equation for the velocity profile is
\begin{equation}
\eta\left[v''+\frac{1}{r}v'-\frac{1}{r^2}v\right]=-\frac{T_0}{h} p + \zeta\left[\frac{1}{r}p^2 + 2 p p'\right].
\end{equation}
Finitude and symmetry of the velocity profile impose $v\left(0\right)=0$. In addition, in agreement with the experimental measurements, we impose normal stress-free boundary conditions at the tissue boundary: $\left. n_\alpha \sigma_{\alpha\beta} n_\beta\right|_{r=R}=0$. This translates into $\sigma_{rr}\left(R\right)=0$, which is the same condition employed to compute the monolayer tension via monolayer stress microscopy. Under these conditions, the velocity profile reads
\begin{multline} \label{eq velocity-profile-monolayer}
v\left(r\right)=\frac{1}{2\eta}\left[\left[\zeta-2 T_0 \frac{L_c^2}{hR}+\left[\zeta\frac{L_c}{R}+2 T_0 \frac{L_c}{h}\right]\frac{I_0\left(R/L_c\right)}{I_1\left(R/L_c\right)}-\zeta\frac{I_0^2\left(R/L_c\right)}{I_1^2\left(R/L_c\right)}\right]r \right.\\
\left.+ \left[\zeta\frac{I_0\left(r/L_c\right)}{I_1\left(R/L_c\right)}-2 T_0 \frac{L_c}{h}\right]L_c \frac{I_1\left(r/L_c\right)}{I_1\left(R/L_c\right)}\right],
\end{multline}
which is plotted in \cref{FigS9a} (red curve).

Now, the previous solution is general for a freely spreading cell monolayer. However, the monolayers in our experiments are confined within circular adherent regions. In the wetting phase, this confinement imposes $v\left(R\right)=0$. With no integration constants left, this extra boundary condition sets a relationship between model parameters. Since the values of $T_0\left(t\right)$ and $L_c\left(t\right)$ are set by traction profiles, this condition directly determines the active stress coefficient $\zeta\left(t\right)$ in terms of the other parameters:
\begin{equation} \label{eq contractility}
\zeta=-2T_0 \frac{L_c}{h}\frac{I_2\left(R/L_c\right)}{I_1\left(R/L_c\right)-I_0\left(R/L_c\right)\left[\frac{I_0\left(R/L_c\right)}{I_1\left(R/L_c\right)}-\frac{2L_c}{R}\right]}.
\end{equation}
Thus, whereas all model parameters are free in a spreading or retracting monolayer, they are not independent in a confined monolayer. \Cref{eq contractility} shows that, under confinement, the active stress coefficient is negative, hence corresponding to a so-called contractile active stress. Accordingly, the coefficient $-\zeta$ is called \emph{contractility} hereinafter. During the dewetting phase, the confinement restriction is released, and hence the contractility becomes an independent parameter, not governed by \cref{eq contractility} anymore.

Both for confined and free monolayers, a general feature of the predicted velocity profiles is their nonmonotonicity (\cref{FigS9a}). The model predicts an outwards flow at a velocity that, close to the center, has a linearly increasing profile, with a slope controlled by traction forces: $v\left(r\right)\approx T_0L_c/(\eta h)\;r$; $r\ll R$, as obtained from \cref{eq velocity-profile-monolayer} in the limit $L_c\ll R$. In contrast, through \cref{eq stress-components-radial}, the stress-free boundary condition $\sigma_{rr}\left(R\right)=0$ imposes the slope of the velocity at the boundary to be $v'\left(R\right)=\zeta/\eta<0$. Hence, the contractility causes the velocity to drop at the peripheral polarized region of width $L_c$. Therefore, the velocity a bit behind the boundary is always higher than at the very boundary (\cref{FigS9a}). As a consequence, cells are expected to accumulate close to the monolayer edge as they flow outwards. Experimentally, a gradient of increasing cell density towards the edge develops in the monolayer (\cref{FigS10}), consistently with the predicted flow profile.

The increase in peripheral cell number density might promote the extrusion of live cells from the monolayer \cite{Eisenhoffer2012,Marinari2012,Eisenhoffer2013}, eventually leading to the formation of 3D structures at the monolayer edge. This is indeed what seems to occur in our monolayers (\cref{FigS4}, Supplementary Movie 4). In fact, 3D structures in the form of cell rims were previously observed both in confined and unconfined monolayers \cite{Deforet2014,Kaliman2014}. We suggest that the formation of these structures might partially stem from the predicted flow-induced accumulation of cells at the tissue edge.

\subsection{Critical size for tissue wetting} \label{critical-size-wetting}

We now focus on deriving the wetting transition, defined by a vanishing spreading parameter, $S=0$. The spreading parameter is directly related to the spreading velocity \cite{Beaune2014} $V=v\left(R\right)$ by $S=\eta V$, so that using \cref{eq velocity-profile-monolayer} it reads
\begin{equation} \label{eq spreading-parameter}
S=\frac{\zeta R}{2}-2\frac{T_0 L_c^2}{h}+\left[\zeta L_c + \frac{T_0 L_c R}{h}\right]\frac{I_0\left(R/L_c\right)}{I_1\left(R/L_c\right)}-\frac{\zeta R}{2}\frac{I_0^2\left(R/L_c\right)}{I_1^2\left(R/L_c\right)}.
\end{equation}
In the experimentally relevant limit $L_c\ll R$, it reduces to
\begin{equation} \label{eq spreading-parameter-limit}
S\approx \frac{T_0 L_c}{h} R + \left(\zeta - \frac{3 T_0 L_c}{h}\right)\frac{L_c}{2}.
\end{equation}
This result gives the spreading parameter in terms of the active forces responsible for collective cell migration, showing that the wetting transition results from a competition between traction forces and tissue contractility. Thus, note that, in contrast to other studies of wetting phenomena in active liquid crystal films \cite{Joanny2012,Joanny2013}, our treatment of the wetting transition in tissues based on an active polar fluid model crucially accounts for active traction forces, which give rise to the distinct physics of the active wetting transition.

From \cref{eq spreading-parameter}, monolayer dewetting will occur whenever the contractility exceed the critical value
\begin{equation} \label{eq critical-contractility}
-\zeta^*=2T_0 \frac{L_c}{h}\frac{I_2\left(R/L_c\right)}{I_1\left(R/L_c\right)-I_0\left(R/L_c\right)\left[\frac{I_0\left(R/L_c\right)}{I_1\left(R/L_c\right)}-\frac{2L_c}{R}\right]}\approx\frac{T_0}{h}\left(2R - 3L_c\right),
\end{equation}
which increases with the radius of the monolayer as shown in \cref{Fig3c}. Therefore, larger monolayers require a higher contractility to induce the dewetting. This can be understood by looking at the velocity profiles of monolayers of different radii. As explained above, traction forces at the edge impose a linearly increasing velocity profile at the central region of the monolayer. As a result, larger monolayers reach higher velocities right behind the narrow polarized peripheral region of width $L_c$ (\cref{FigS9b}). In turn, the wetting transition condition imposes a vanishing velocity at the boundary. As also explained above, the contractility is responsible for the velocity drop across the strongly polarized peripheral layer. Thus, larger monolayers require a higher contractility to bring the velocity down to zero at the boundary.

Note that the critical contractility $-\zeta^*$ is precisely the contractility under confinement (\cref{eq contractility}), since confinement also imposes the condition $V=0$ in the wetting phase. Therefore, while fully spread, our confined monolayers are in a resting state ($V=0$) maintained by the parallel increase of traction $T_0\left(t\right)$ and contractility $-\zeta\left(t\right)$, continuously fulfilling \cref{eq critical-contractility}.

\subsection{Morphological instability during monolayer de\-wet\-ting} \label{morphological-instability-dewetting}


\subsubsection{Linear stability analysis} \label{linear-stability-analysis}
In this section, we study the morphological stability of the retracting tissue front during monolayer dewetting. To allow for the loss of the circular tissue shape, we include the ortoradial components of the polarity and velocity fields. Thus, \cref{eq polarity-field-monolayer} reads
\begin{subequations} \label{eq polarity-components-instability-dewetting}
\begin{align}
\left(\partial_r^2+\frac{1}{r}\partial_r - \frac{1}{r^2}\right) p_r + \frac{1}{r^2} \partial_\theta^2 p_r - \frac{2}{r^2}\partial_\theta p_\theta &= \frac{1}{L_c^2} p_r,\\
\left(\partial_r^2+\frac{1}{r}\partial_r - \frac{1}{r^2}\right) p_\theta + \frac{1}{r^2} \partial_\theta^2 p_\theta + \frac{2}{r^2}\partial_\theta p_r &= \frac{1}{L_c^2} p_\theta.
\end{align}
\end{subequations}
Force balance is expressed as
\begin{subequations} \label{eq force-balance-components-instability-dewetting}
\begin{align}
\frac{1}{r}\partial_r\left(r \sigma^s_{rr}\right)+\frac{1}{r}\partial_\theta \sigma^s_{\theta r} - \frac{1}{r}\sigma^s_{\theta\theta} &= -T_0/h \, p_r,\\
\frac{1}{r}\partial_r\left(r \sigma^s_{r\theta}\right)+\frac{1}{r}\partial_\theta \sigma^s_{\theta\theta} + \frac{1}{r}\sigma^s_{\theta r} &= -T_0/h \, p_\theta,
\end{align}
\end{subequations}
with the components of the stress tensor given by
\begin{subequations} \label{eq stress-tensor-components-instability-dewetting}
\begin{align}
\sigma^s_{rr}&=2\eta \, \partial_r v_r - \zeta p_r^2,\\
\sigma^s_{r\theta}&=\sigma^s_{\theta r}=\eta\left[r\partial_r\left(\frac{v_\theta}{r}\right)+\frac{1}{r}\partial_\theta v_r\right]-\zeta p_r p_\theta,\\
\sigma^s_{\theta\theta}&=\frac{2\eta}{r}\left(v_r+\partial_\theta v_\theta\right)-\zeta p_\theta^2.
\end{align}
\end{subequations}
The solution for the unperturbed state, which preserves circular symmetry, is given by \cref{eq polarity-profile-monolayer,eq velocity-profile-monolayer}:
\begin{equation}
p_r^0\left(r\right)=\frac{I_1\left(r/L_c\right)}{I_1\left(R_0/L_c\right)},
\end{equation}
\begin{multline}
v_r^0\left(r\right)=\frac{1}{2\eta}\left[\left[\zeta-2 T_0 \frac{L_c^2}{hR_0}+\left[\zeta\frac{L_c}{R_0}+2 T_0 \frac{L_c}{h}\right]\frac{I_0\left(R_0/L_c\right)}{I_1\left(R_0/L_c\right)}-\zeta\frac{I_0^2\left(R_0/L_c\right)}{I_1^2\left(R_0/L_c\right)}\right]r \right.\\
\left.+ \left[\zeta\frac{I_0\left(r/L_c\right)}{I_1\left(R_0/L_c\right)}-2 T_0 \frac{L_c}{h}\right]L_c \frac{I_1\left(r/L_c\right)}{I_1\left(R_0/L_c\right)}\right],
\end{multline}
where the superindex indicates the zeroth order in the front perturbations, and $R_0$ is the monolayer radius, which changes according to $dR_0/dt = v_r^0\left(R_0\right)$.

Next, we introduce small-amplitude perturbations of the circular monolayer boundary (\cref{Fig5b}):
\begin{equation}
R\left(\theta\right)=R_0 + \delta R\left(\theta\right).
\end{equation}
Then, the polarity and velocity fields are correspondingly perturbed:
\begin{equation}
p_r\left(r,\theta\right) = p_r^0\left(r\right) + \delta p_r\left(r,\theta\right),\qquad p_\theta\left(r,\theta\right)=\delta p_\theta\left(r,\theta\right),
\end{equation}
\begin{equation}
v_r\left(r,\theta\right) = v_r^0\left(r\right) + \delta v_r\left(r,\theta\right),\qquad v_\theta\left(r,\theta\right)=\delta v_\theta\left(r,\theta\right).
\end{equation}
To impose boundary conditions, we define the normal and tangential vectors at the boundary,
\begin{subequations}
\begin{align}
\bm{n}&=\cos\alpha\;\bm{r} + \sin\alpha\;\bm{\theta}\approx \bm{r}-\frac{1}{R_0}\frac{d\delta R}{d\theta} \, \bm{\theta},\\
\bm{t}&=-\sin\alpha\;\bm{r}+\cos\alpha\;\bm{\theta}\approx \frac{1}{R_0}\frac{d\delta R}{d\theta} \, \bm{r}+\bm{\theta},
\end{align}
\end{subequations}
where $\alpha$ is the angle between the normal directions of the perturbed and unperturbed interfaces. In terms of the normal and tangential vectors, the conditions that impose a normal and maximal polarity at the boundary read
\begin{equation}
\left.\bm{p}\cdot\bm{n}\right|_{r=R}=1,\qquad \left.\bm{p}\cdot\bm{t}\right|_{r=R}=0.
\end{equation}
For the radial component, these conditions imply $p_r\left(R\right)\approx 1$, which expands into
\begin{equation}
p_r\left(R\right)=p_r^0\left(R\right)+\delta p_r\left(R\right)\approx p_r^0\left(R_0\right)+\partial_r p_r^0\left(R_0\right)\delta R+\delta p_r\left(R\right)\approx 1,
\end{equation}
so that
\begin{equation}
\delta p_r\left(R\right)=-\partial_r p_r^0\left(R_0\right)\delta R.
\end{equation}
For the ortoradial component,
\begin{equation}
\delta p_\theta\left(R\right)= -\frac{1}{R_0}\frac{d\delta R}{d\theta}.
\end{equation}
In turn, the boundary conditions on the stress impose a vanishing normal and shear stress at the interface:
\begin{equation}
\left.\bm{n}\cdot\bm{\sigma}\cdot \bm{n}\right|_{r=R}=0,\qquad \left.\bm{t}\cdot\bm{\sigma}\cdot \bm{n}\right|_{r=R}=0.
\end{equation}
The condition on the normal stress gives $\sigma_{rr}\left(R\right)=0$ which, after expanding as previously, yields
\begin{equation}
\delta\sigma_{rr}\left(R\right)=- \partial_r\sigma_{rr}^0\left(R_0\right)\delta R.
\end{equation}
Finally, the condition on the shear stress imposes $\sigma_{r\theta}\left(R\right)=0$, which translates into
\begin{equation}
\delta\sigma_{r\theta}\left(R\right)=\frac{1}{R_0}\frac{d\delta R}{d\theta}\sigma_{\theta\theta}^0\left(R_0\right)=\frac{2\eta}{h R_0^2} v_r^0\left(R_0\right)\frac{d\delta R}{d\theta}.
\end{equation}

Next, we decompose the perturbations in their Fourier modes, identified by an index $n$ (\cref{Fig5b}):
\begin{subequations}
\begin{align}
\delta R\left(\theta,t\right)=\sum_{n=0}^\infty \delta \tilde{R}_n\left(t\right) e^{i n \theta},\\
\delta p_\alpha\left(r,\theta,t\right)=\sum_{n=0}^\infty \delta \tilde{p}_{\alpha,n}\left(r,t\right) e^{i n \theta},\\
\delta v_\alpha\left(r,\theta,t\right)=\sum_{n=0}^\infty \delta \tilde{v}_{\alpha,n}\left(r,t\right) e^{i n \theta}.
\end{align}
\end{subequations}
In terms of the Fourier modes, the equations for the polarity components read
\begin{subequations} \label{eq polarity-Fourier-instability-dewetting}
\begin{align}
\left(\partial_r^2+\frac{1}{r}\partial_r - \frac{1+n^2}{r^2}-\frac{1}{L_c^2}\right)\delta\tilde{p}_r&=\frac{2i n}{r^2}\delta\tilde{p}_\theta,\\
\left(\partial_r^2+\frac{1}{r}\partial_r - \frac{1+n^2}{r^2}-\frac{1}{L_c^2}\right)\delta\tilde{p}_\theta&=-\frac{2i n}{r^2}\delta\tilde{p}_r.
\end{align}
\end{subequations}
In turn, the components of the force balance, once the constitutive equation has been introduced, are expressed as
\begin{subequations} \label{eq velocity-Fourier-instability-dewetting}
\begin{gather}
\begin{multlined}
2\eta\left(\partial_r^2+\frac{1}{r}\partial_r - \frac{1+n^2/2}{r^2}\right)\delta \tilde{v}_r+\frac{i n \eta}{r}\left(\partial_r-\frac{3}{r}\right)\delta \tilde{v}_\theta\\
+ \left[T_0/h-2\zeta\left(\left(\partial_r + \frac{1}{r}\right) p_r^0 + p_r^0\partial_r\right)\right]\delta \tilde{p}_r - \frac{i n \zeta}{r} p_r^0 \delta \tilde{p}_\theta=0,
\end{multlined}
\\
\begin{multlined}
\frac{i n \eta}{r}\left(\partial_r + \frac{3}{r}\right)\delta \tilde{v}_r + \eta \left(\partial_r^2 + \frac{1}{r}\partial_r - \frac{1+ 2n^2}{r^2}\right)\delta\tilde{v}_\theta\\
+ \left[T_0/h-\zeta\left(\left(\partial_r + \frac{2}{r}\right)p_r^0 + p_r^0\partial_r\right)\right]\delta \tilde{p}_\theta=0.
\end{multlined}
\end{gather}
\end{subequations}
Finally, in Fourier space, the boundary conditions read
\begin{equation} \label{eq bc-polarity-Fourier}
\delta\tilde{p}_{r,n}\left(R\right)= -\partial_r p_r^0\left(R_0\right)\delta\tilde{R}_n,\qquad \delta\tilde{p}_{\theta,n}\left(R\right)=-\frac{in}{R_0}\delta\tilde{R}_n,
\end{equation}
\begin{equation} \label{eq bc-velocity-Fourier}
\delta\tilde{\sigma}_{rr,n}\left(R\right) = - \partial_r\sigma_{rr}^0\left(R_0\right)\delta\tilde{R}_n,\qquad \delta\tilde{\sigma}_{r\theta,n}\left(R\right)=\frac{2 i n \eta}{h R_0^2} v_r^0\left(R_0\right)\delta\tilde{R}_n.
\end{equation}

At this point, the four coupled ordinary differential equations \cref{eq polarity-Fourier-instability-dewetting,eq velocity-Fourier-instability-dewetting} are solved for $\delta\tilde{p}_{\alpha,n}\left(r\right)$ and $\delta\tilde{v}_{\alpha,n}\left(r\right)$. The solution is completely analytical for mode $n=0$ and almost analytical for the rest of modes, meaning that it has an analytical expression that involves two integrals that need to be numerically evaluated. Then, from the Fourier modes of the velocity field, the perturbed spreading velocity can be obtained as
\begin{equation}
V=\left.\bm{v}\cdot\bm{n}\right|_{r=R}=\left[\bm{v}\,^0 \cdot \bm{n} + \delta\bm{v}\cdot\bm{n}\right]_{r=R} \approx v_r^0\left(R_0\right) + \partial_r v_r^0\left(R_0\right) \delta R + \delta v_r\left(R_0\right),
\end{equation}
which implies
\begin{equation}
\delta V\left(\theta\right)=V\left(\theta\right)-V_0=\partial_r v_r^0\left(R_0\right) \delta R\left(\theta\right) + \delta v_r\left(R_0,\theta\right).
\end{equation}
Thus, the growth rate $\omega_n$ of the tissue shape perturbations follows from
\begin{equation} \label{eq perturbation-dynamics-instability-dewetting}
\delta\tilde{V}_n=\frac{1}{2\pi}\int_0^{2\pi} \delta V\left(\theta\right) e^{-in\theta} d\theta = \frac{d\delta\tilde{R}_n}{dt}=\omega_n \delta\tilde{R}_n.
\end{equation}
Hence,
\begin{equation} \label{eq growth-rate-instability-dewetting}
\omega_n=\partial_r v_r^0\left(R_0\right) + \frac{\delta \tilde{v}_{r,n}\left(R_0\right)}{\delta\tilde{R}_n}.
\end{equation}
The resulting growth rate is a purely real number under any conditions, showing that there is no oscillatory instability. At the onset of dewetting, namely using typical critical parameter values $T_0^*$, $L_c^*$, and $-\zeta^*$ that define the tissue wetting transition (\cref{Fig4d,Fig4f,FigS14}), the cell monolayer exhibits a long-wavelength morphological instability (\cref{Fig5j}). Several modes corresponding to deformations of the tissue shape ($n\geq 2$, \cref{Fig5b}) are unstable. Hence, we propose that this instability is at the root of the observed shape changes during monolayer dewetting (\cref{Fig5a}, Supplementary Movie 11).

\subsubsection{Monolayer viscosity} \label{monolayer-viscosity}
At the wetting transition point, all the model parameters are known except for the monolayer viscosity. This allows us to estimate the monolayer viscosity at the wetting transition $\eta^*$ from the retraction rate of the monolayer (\cref{Fig5e}):
\begin{multline} \label{eq retraction-rate}
\omega_0=\frac{1}{4\eta}\left[2T_0\frac{R_0}{h} + 3\zeta + 2\left[\zeta\left[\frac{L_c}{R_0} - \frac{R_0}{L_c}\right] + 2T_0\frac{L_c}{h}\right] \frac{I_0\left(R_0/L_c\right)}{I_1\left(R_0/L_c\right)} \right.\\
\left.- \left[2T_0 \frac{R_0}{h} + 5\zeta\right] \frac{I_0^2\left(R_0/L_c\right)}{I_1^2\left(R_0/L_c\right)} + 2\zeta\frac{R_0}{L_c} \frac{I_0^3\left(R_0/L_c\right)}{I_1^3\left(R_0/L_c\right)} \right],
\end{multline}
which is approximated by
\begin{equation}
\omega_0\approx \frac{T_0 L_c}{2\eta h}
\end{equation}
in the limit $L_c\ll R_0$.

\subsubsection{Structure factor of monolayer shape} \label{structure-factor}
To compute the structure factor, we add a noise term to the dynamics of the perturbation modes, \cref{eq perturbation-dynamics-instability-dewetting}. Thus, the corresponding Langevin equation reads
\begin{equation} \label{eq Langevin-perturbations-instability-dewetting}
\frac{d\delta\tilde{R}_n}{dt}=\omega_n \delta\tilde{R}_n + \tilde{\xi}_n\left(t\right).
\end{equation}
Assuming that shape fluctuations are fast compared to the dewetting dynamics, we consider that they are temporally uncorrelated and hence we take a Gaussian white noise:
\begin{equation} \label{eq noise-perturbations-instability-dewetting}
\left\langle\tilde{\xi}_n\left(t\right)\right\rangle=0,\qquad \left\langle\tilde{\xi}_n\left(t\right)\tilde{\xi}_m^*\left(t'\right)\right\rangle=2D\delta_{n,m}\delta\left(t-t'\right),
\end{equation}
where $D$ is the noise intensity, which we assume independent of the mode number $n$. Now, under the approximation of a constant growth rate $\omega_n$ in the short time span $t_f-t^*=7$ h, the solution to \cref{eq Langevin-perturbations-instability-dewetting} can be formally expressed as
\begin{equation}
\delta\tilde{R}_n\left(t\right) = \delta\tilde{R}_n\left(t^*\right) e^{\omega_n(t-t^*)} + e^{\omega_n t} \int_{t^*}^t \tilde{\xi}_n\left(t'\right) e^{-\omega_n t'} dt'.
\end{equation}
Considering no shape perturbations at the onset of dewetting, $\delta\tilde{R}_n\left(t^*\right)=0$, the equal-time structure factor reads
\begin{equation} \label{eq structure-factor-instability-dewetting}
S_n\left(t\right)=\left\langle |\delta\tilde{R}_n\left(t\right)|^2\right\rangle = e^{2\omega_n t} \int_{t^*}^t \int_{t^*}^t \left\langle \tilde{\xi}_n \left(t'\right)\tilde{\xi}_n^*\left(t''\right)\right\rangle e^{-\omega_n\left(t'+t''\right)}\,dt'\,dt'' = \frac{D}{\omega_n}\left[ e^{2\omega_n\left(t-t^*\right)}-1\right],
\end{equation}
where we have employed \cref{eq noise-perturbations-instability-dewetting}.

Finally, once the experimental value for $\omega_0$ is known, the experimental growth rate is determined from the structure factor by numerically inverting the relation
\begin{equation}
\frac{S_n\left(t_f\right)}{S_0\left(t_f\right)}=\frac{\omega_0}{\omega_n}\frac{e^{2\omega_n \left(t_f-t^*\right)}-1}{e^{2\omega_0 \left(t_f-t^*\right)}-1},
\end{equation}
which is independent of the noise intensity $D$.

\clearpage
\appendix

\onecolumngrid
\begin{center}
\textbf{\large Supplementary Figures}
\end{center}

\begin{figure}[h]
\begin{center}
\includegraphics[width=0.4\textwidth]{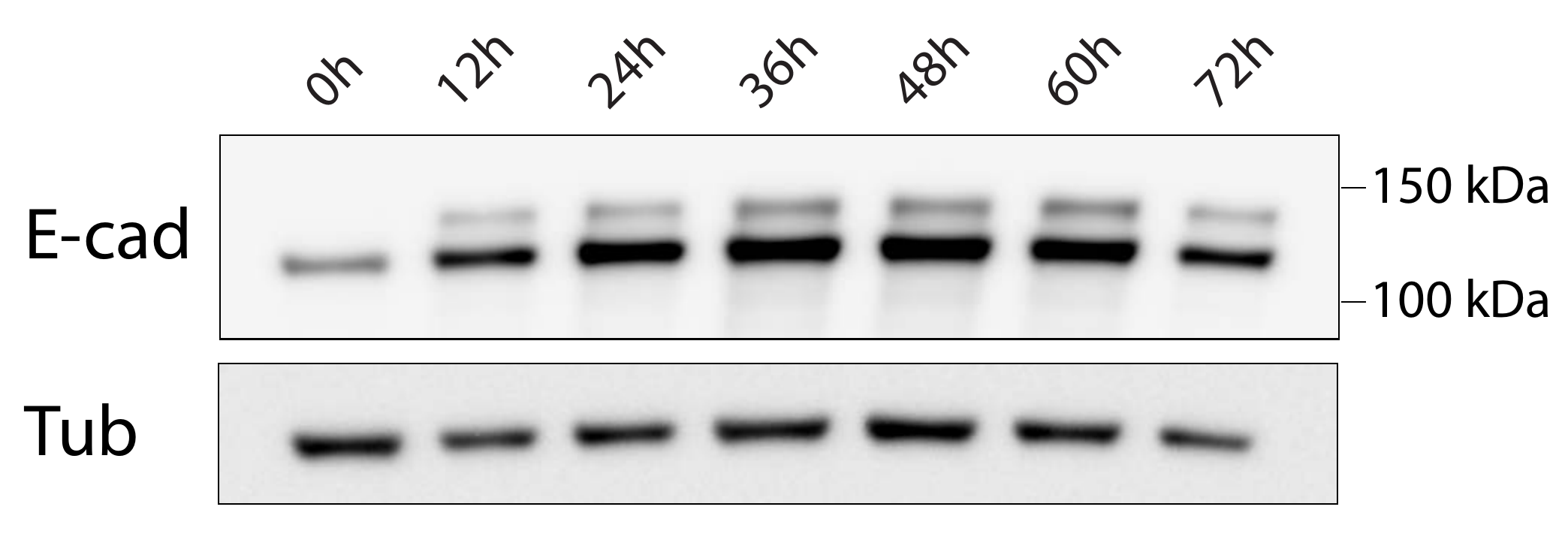}
\end{center}
\bfcaption{E-cadherin western blot during 3 days of induction}{ E-cadherin increases in time during the first 24h and plateaus for at least 2 more days.} \label{FigS1}
\end{figure}

\begin{figure}[tb!]
\begin{center}
\includegraphics[width=0.9\textwidth]{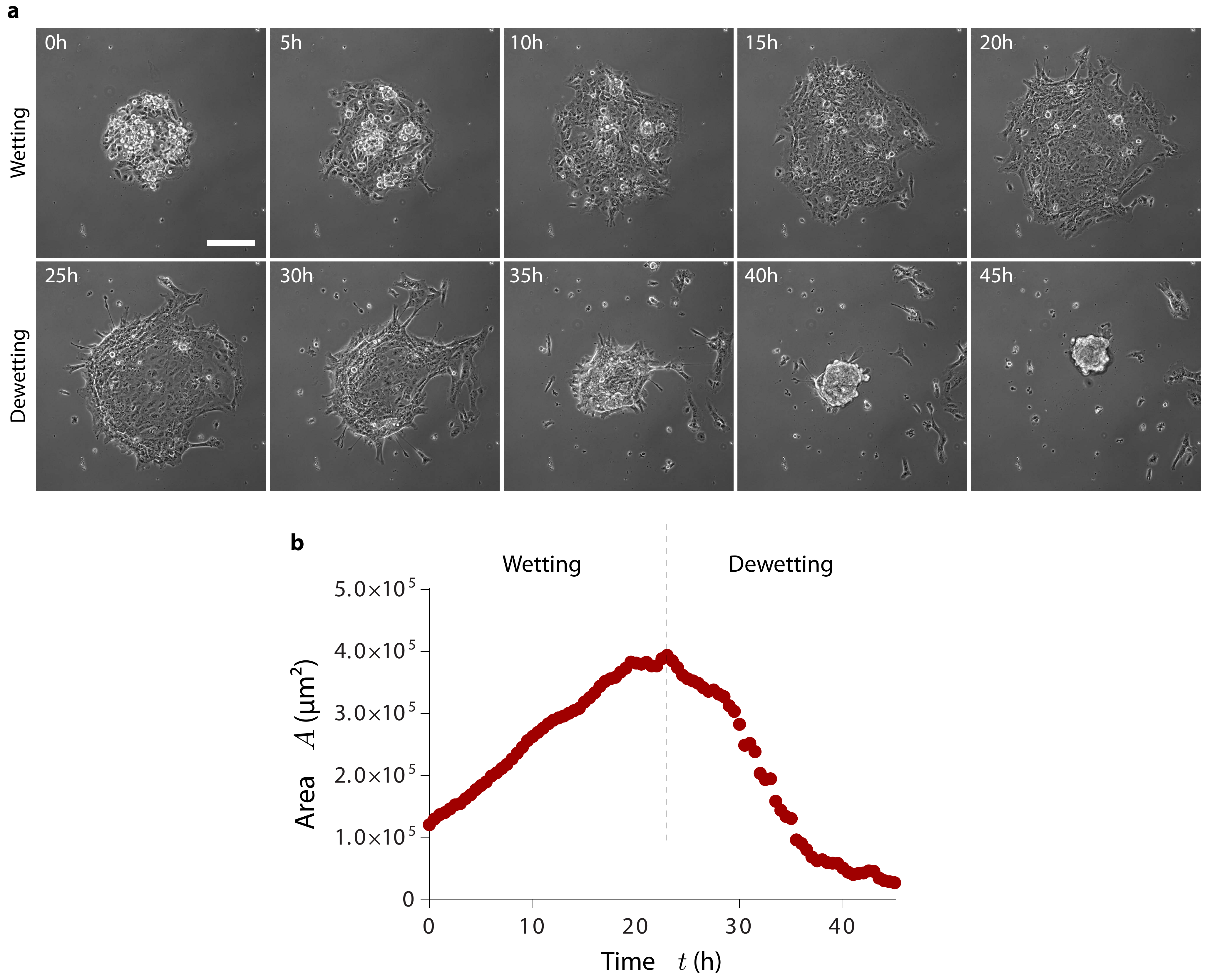}
\end{center}
  {\phantomsubcaption\label{FigS2a}}
  {\phantomsubcaption\label{FigS2b}}
\bfcaption{Spreading cell monolayers exhibit a transition from wetting to dewetting}{ \subref*{FigS2a},\subref*{FigS2b}, Phase contrast images (\subref*{FigS2a}) and time evolution of monolayer area (\subref*{FigS2b}) of a representative unconfined monolayer undergoing a transition between wetting ($\dot{A}>0$) and dewetting ($\dot{A}<0$) at time $t=22$ h. Scale bar $= 150$ $\mu$m.} \label{FigS2}
\end{figure}

\begin{figure}[tb!]
\begin{center}
\includegraphics[width=0.9\textwidth]{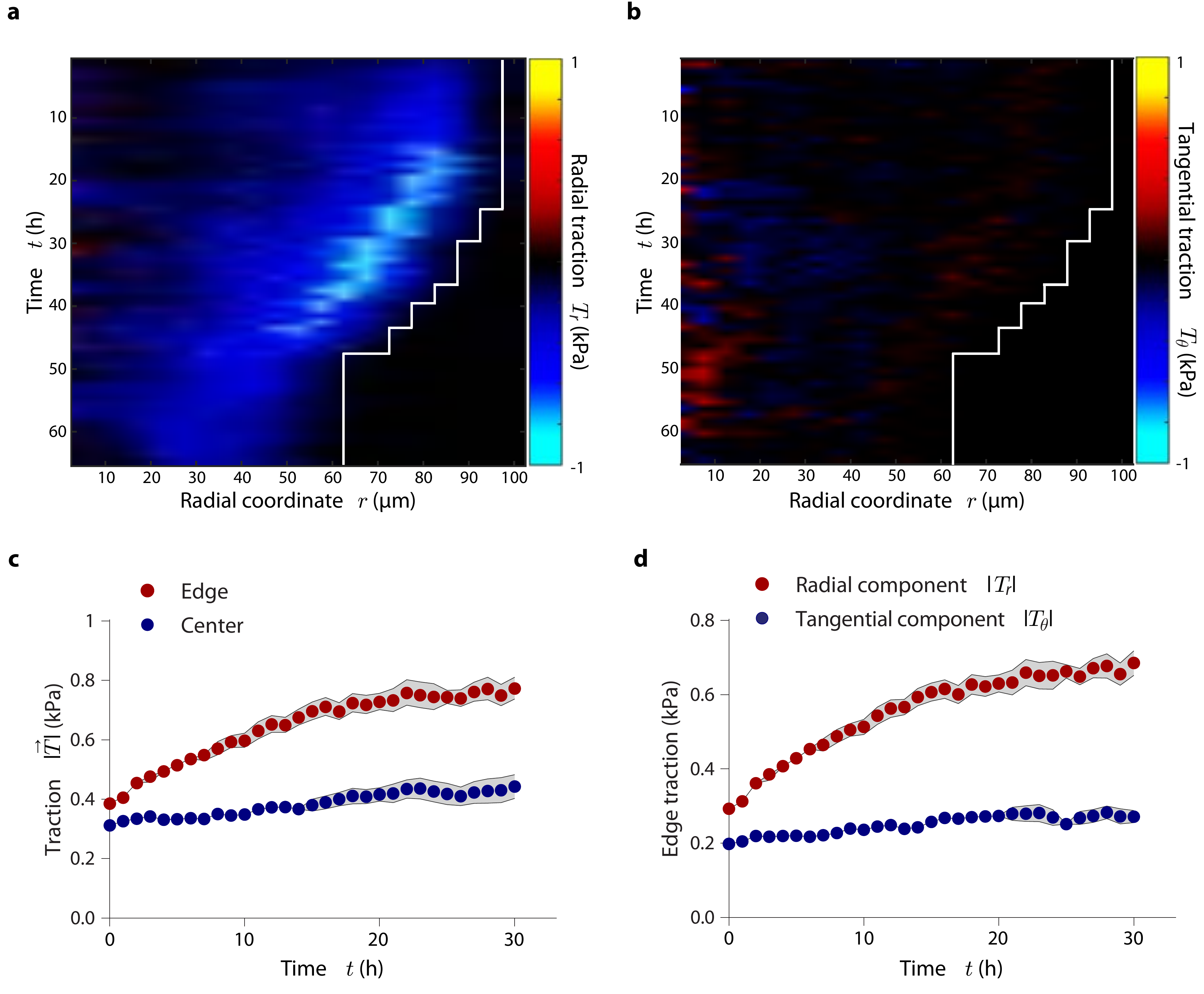}
\end{center}
  {\phantomsubcaption\label{FigS3a}}
  {\phantomsubcaption\label{FigS3b}}
  {\phantomsubcaption\label{FigS3c}}
  {\phantomsubcaption\label{FigS3d}}
\bfcaption{Traction forces are primarily radial and accumulate at the edge of the monolayer upon E-cadherin expression}{ \subref*{FigS3a},\subref*{FigS3b}, Kymographs of the radial (\subref*{FigS3a}) and tangential (\subref*{FigS3b}) components of the traction profile. \subref*{FigS3c}, Evolution of the traction magnitude at the monolayer edge and at the center. \subref*{FigS3d}, Evolution of the radial and tangential components of the traction at the edge of the monolayer. Data are presented as mean $\pm$ s.e.m. $n=18$ cell islands.} \label{FigS3}
\end{figure}

\begin{figure}[tb!]
\begin{center}
\includegraphics[width=\textwidth]{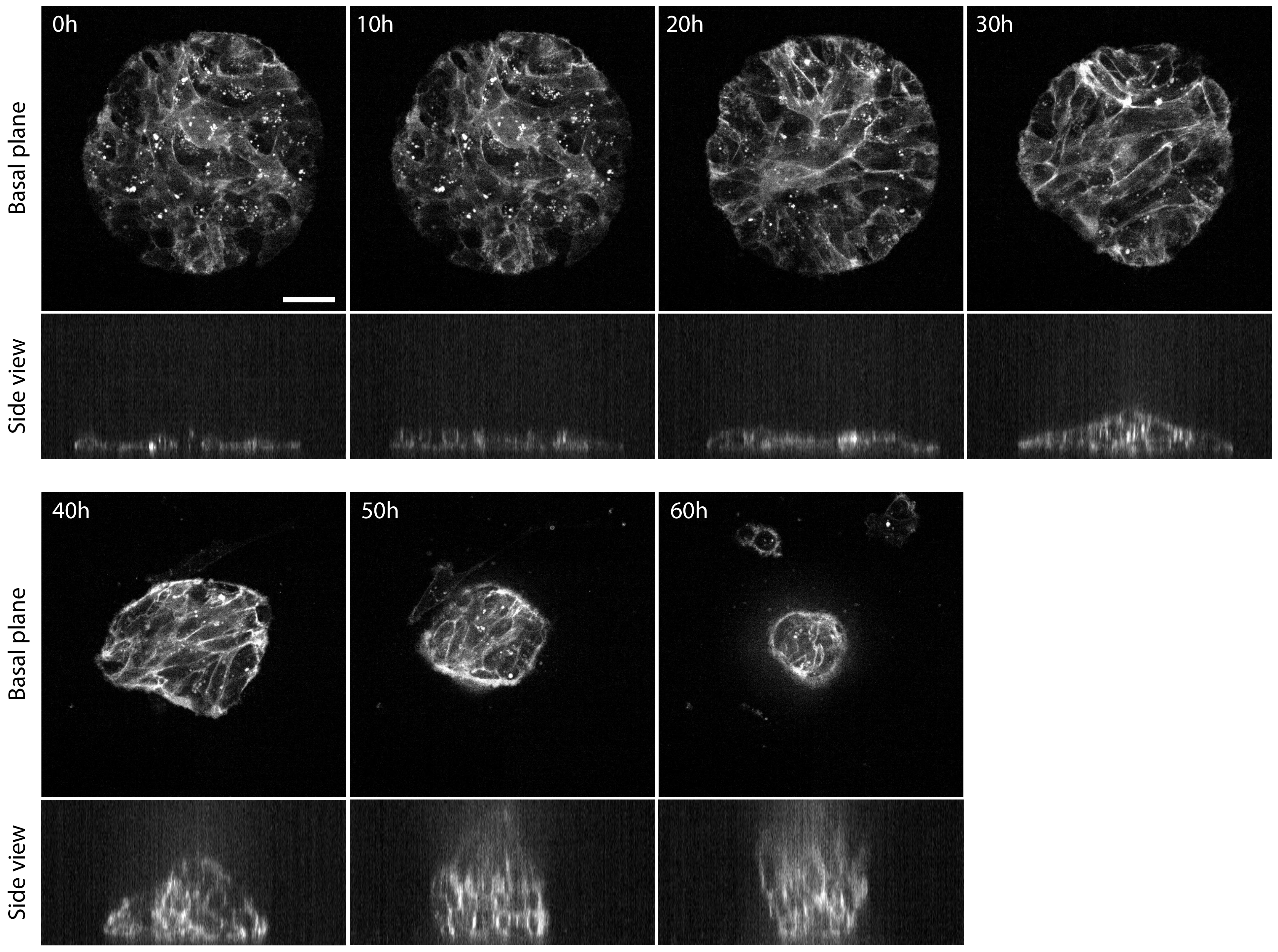}
\end{center}
\bfcaption{Orthogonal views of monolayer dewetting.}{ Life imaging of plasma membrane labelled cells (CAAX-iRFP) allow the visualization of tissue morphology in 3D. The tissue-substrate contact area decreases pronouncedly during dewetting, while the tissue evolves from a monolayer to a spheroidal cell aggregate, resembling a droplet. Scale bar $= 40$ $\mu$m.} \label{FigS4}
\end{figure}

\begin{figure}[tb!]
\begin{center}
\includegraphics[width=0.9\textwidth]{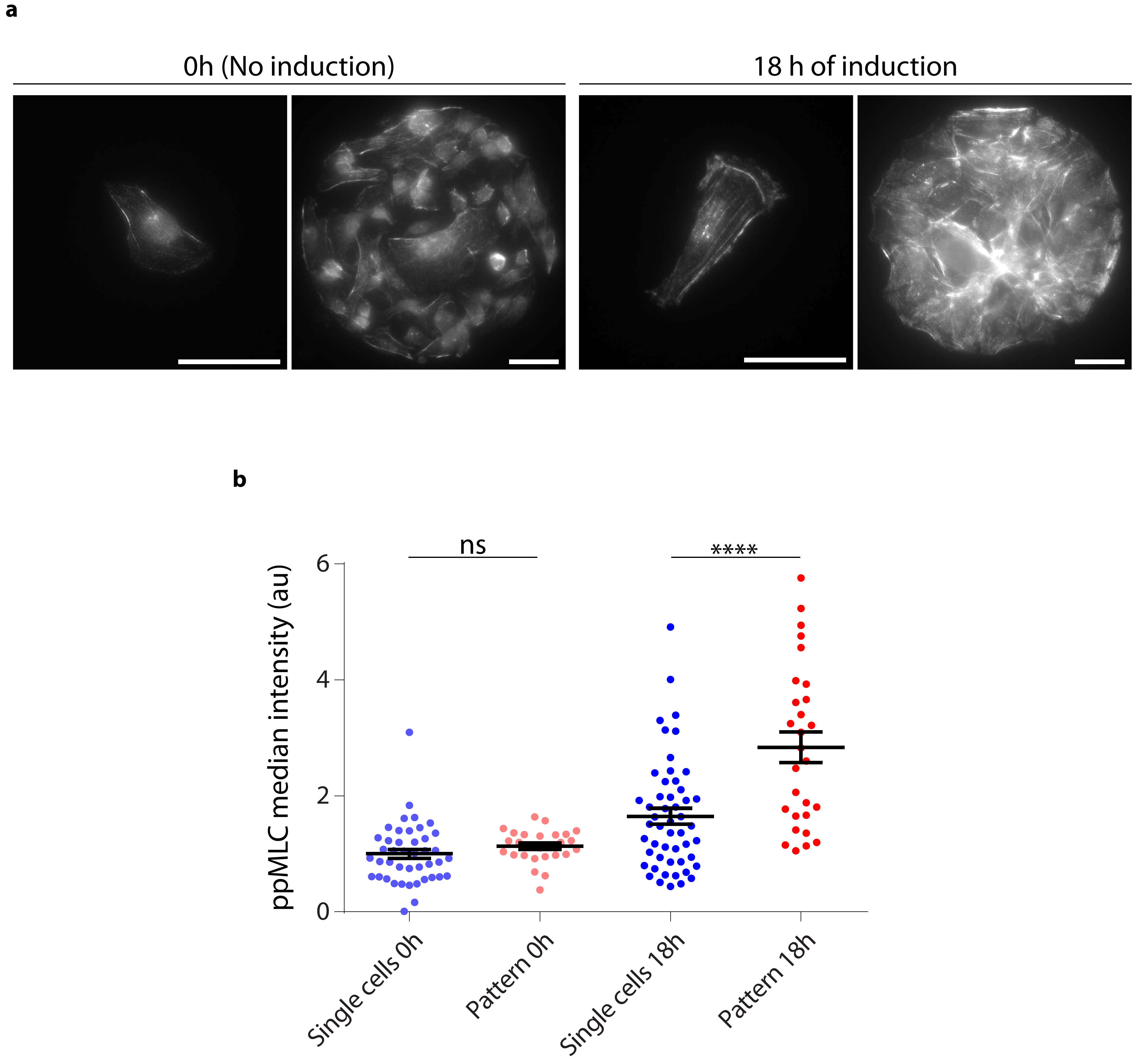}
\end{center}
  {\phantomsubcaption\label{FigS5a}}
  {\phantomsubcaption\label{FigS5b}}
\bfcaption{Monolayers display higher levels of ppMLC than single cells upon E-cadherin induction}{ \subref*{FigS5a}, ppMLC immunostaining of single cells and cell islands before and after 18h of E-cadherin induction. Scale bar $= 40$ $\mu$m. \subref*{FigS5b}, Quantification of median fluorescence intensity shows that single cells and cells in monolayers start with similar levels of ppMLC. After 18h of E-cadherin induction, cell monolayers show significantly higher levels of ppMLC than single cells (Kruskal Wallis test, p-value $< 0.0001$). Data are presented as mean $\pm$ s.e.m. $n=43$ (Single cells 0 h), $n=27$ (Pattern 0 h), $n=51$ (Single cells 18 h) and $n=28$ (Pattern 18 h).} \label{FigS5}
\end{figure}

\begin{figure}[tb!]
\begin{center}
\includegraphics[width=0.85\textwidth]{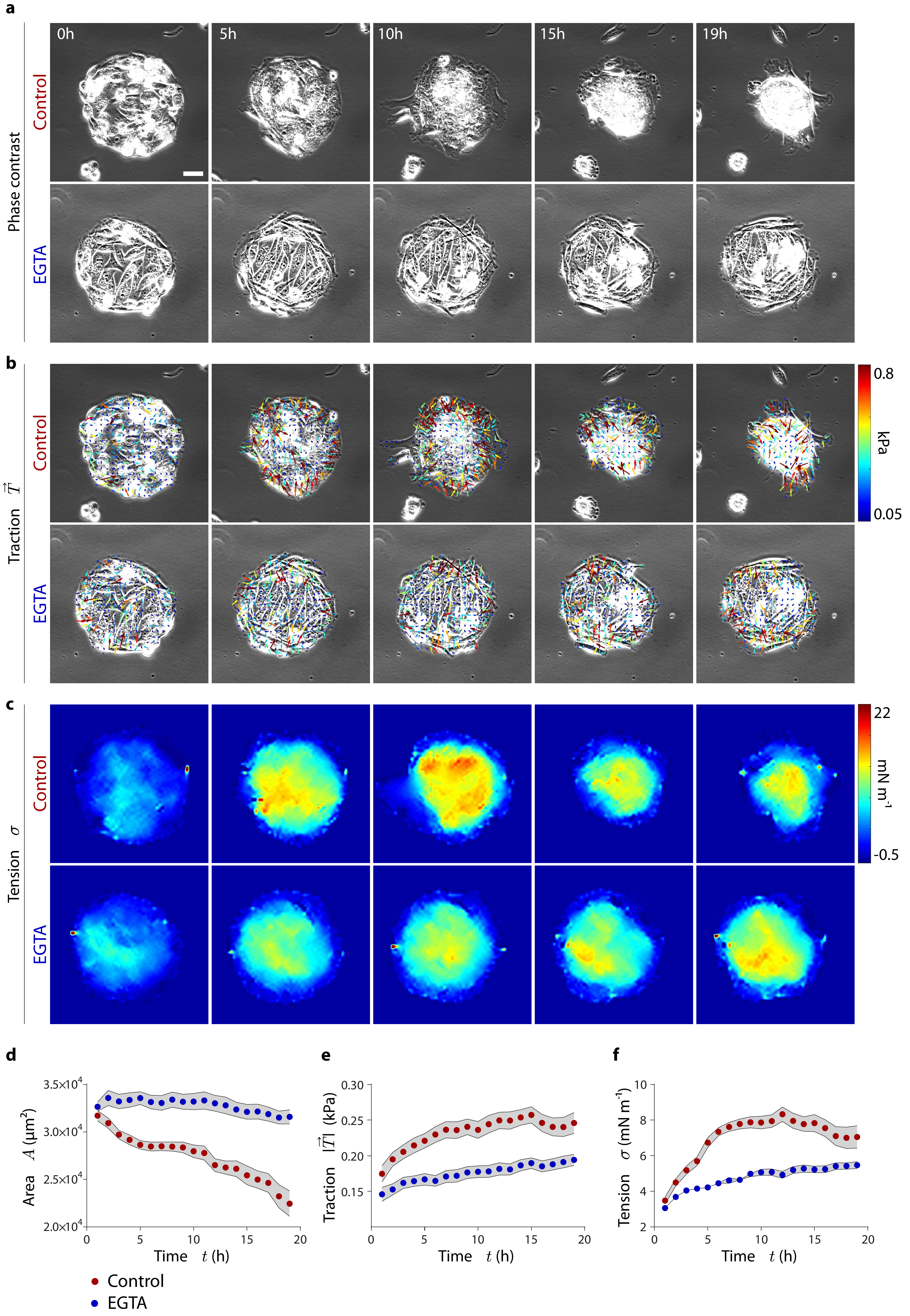}
\end{center}
  {\phantomsubcaption\label{FigS6a}}
  {\phantomsubcaption\label{FigS6b}}
  {\phantomsubcaption\label{FigS6c}}
  {\phantomsubcaption\label{FigS6d}}
  {\phantomsubcaption\label{FigS6e}}
  {\phantomsubcaption\label{FigS6f}}
\bfcaption{Calcium chelation prevents the buildup of tissue forces and the wetting transition}{ \subref*{FigS6a}-\subref*{FigS6c}, Phase contrast images (\subref*{FigS6a}), and maps of traction forces (\subref*{FigS6b}) and average normal monolayer tension (\subref*{FigS6c}) in control and EGTA treated monolayers. EGTA efficiently inhibits cell-cell junction formation, as seen from the lack of cohesiveness in the treated monolayer. \subref*{FigS6d}-\subref*{FigS6f}, EGTA hinders the abrupt increase in tractions (\subref*{FigS6e}) and tension (\subref*{FigS6f}) after E-cadherin induction, thereby preventing the decrease in area caused by tissue dewetting (\subref*{FigS6d}).} \label{FigS6}
\end{figure}

\begin{figure}[tb!]
\begin{center}
\includegraphics[width=0.9\textwidth]{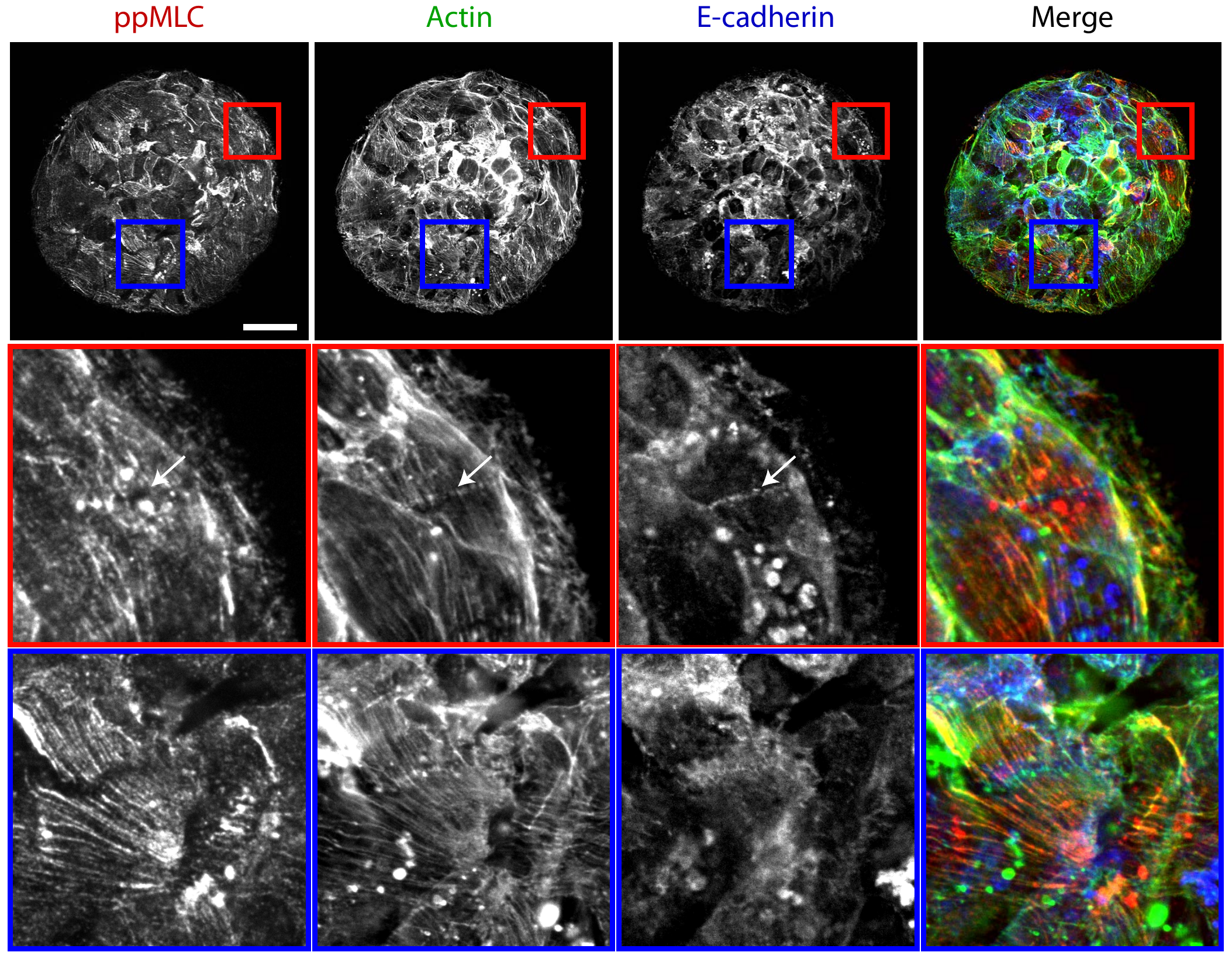}
\end{center}
  {\phantomsubcaption\label{FigS7a}}
  {\phantomsubcaption\label{FigS7b}}
  {\phantomsubcaption\label{FigS7c}}
\bfcaption{ppMLC accumulates at actin stress fibers but not at E-cadherin junctions}{ \subref*{FigS7a}, Immunostaining of ppMLC, Actin, and E-cadherin in cell islands after $18$ h of E-cadherin induction. Scale bar $= 40$ $\mu$m. \subref*{FigS7b},\subref*{FigS7c}, Insets of the labelled regions in (\subref*{FigS7a}). E-cadherin junctions co-localize with actin but not with ppMLC (arrows). Instead, ppMLC accumulates at actin stress fibers.} \label{FigS7}
\end{figure}

\begin{figure}[tb!]
\begin{center}
\includegraphics[width=0.9\textwidth]{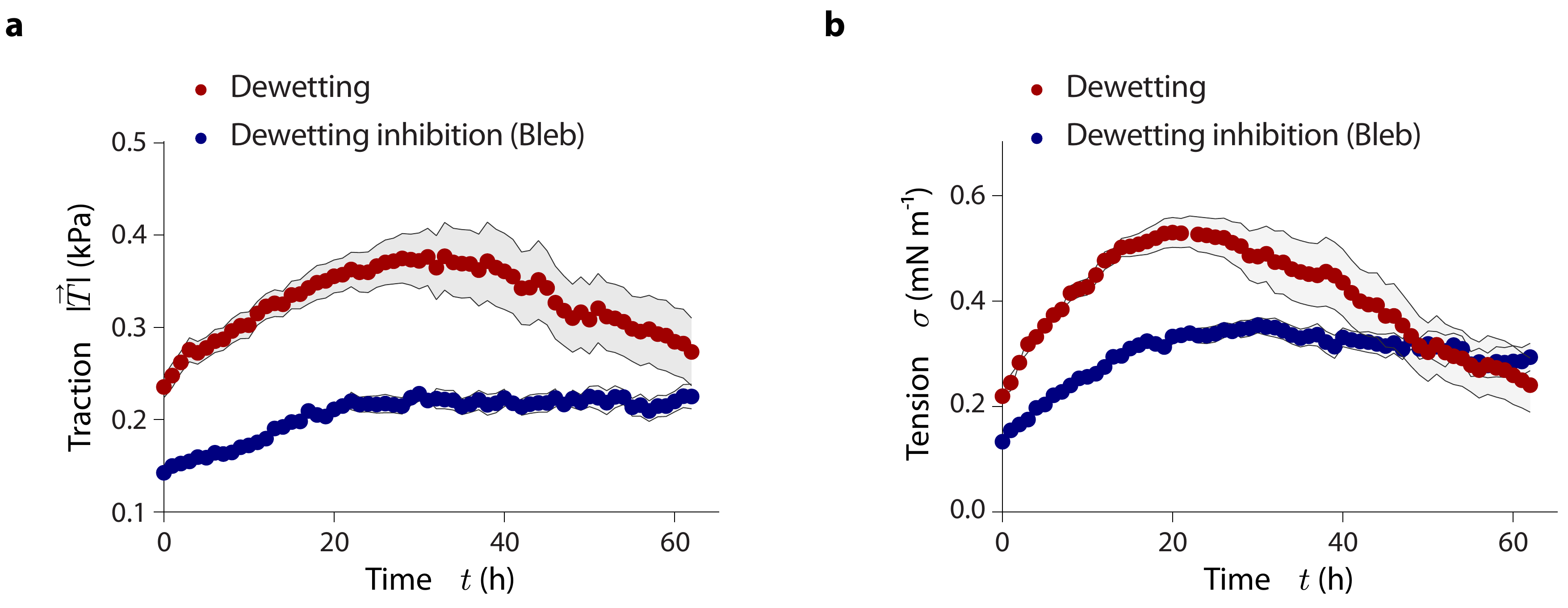}
\end{center}
  {\phantomsubcaption\label{FigS8a}}
  {\phantomsubcaption\label{FigS8b}}
\bfcaption{Blebbistatin decreases traction forces and monolayer tension}{ \subref*{FigS8a},\subref*{FigS8b}, Evolution of the average traction magnitude (\subref*{FigS8a}) and average monolayer tension (\subref*{FigS8a}) for untreated islands, and for islands treated with blebbistatin.} \label{FigS8}
\end{figure}

\begin{figure}[tb!]
\begin{center}
\includegraphics[width=0.9\textwidth]{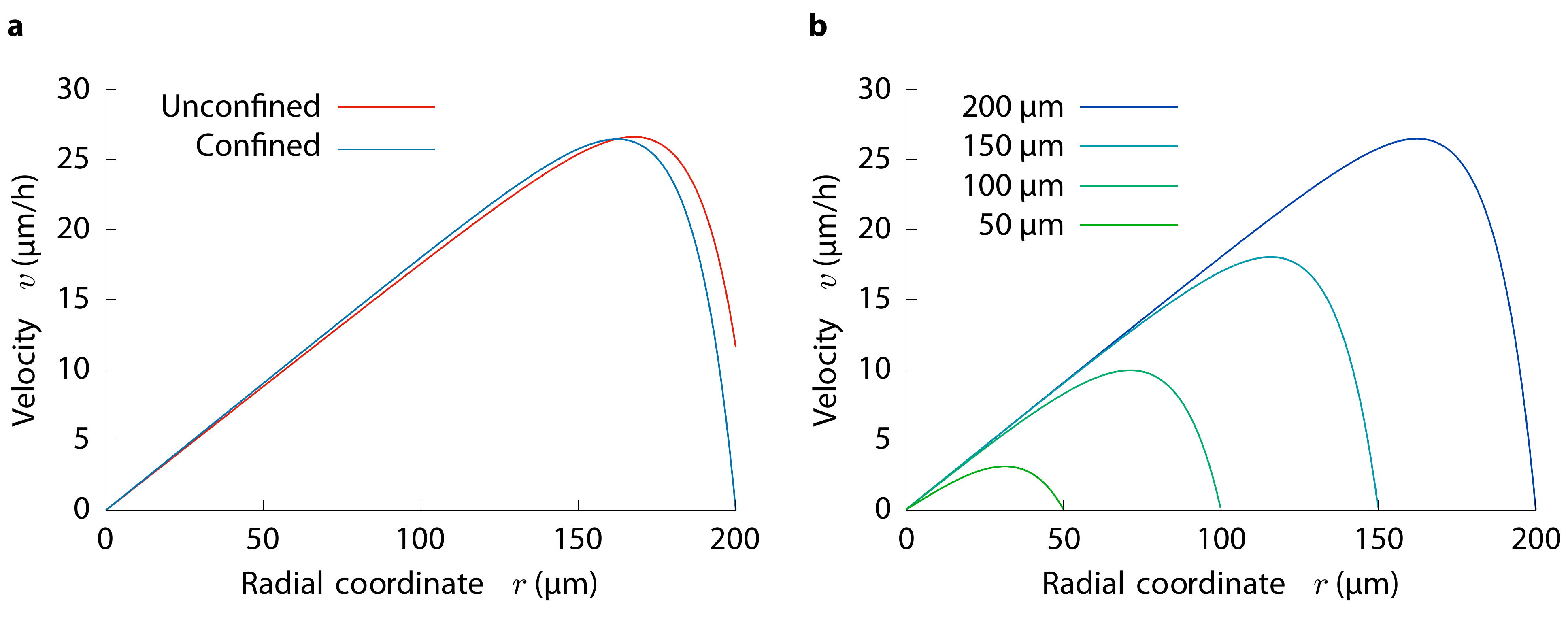}
\end{center}
  {\phantomsubcaption\label{FigS9a}}
  {\phantomsubcaption\label{FigS9b}}
\bfcaption{Predicted velocity profiles in the monolayers}{ \subref*{FigS9a}, Velocity profile in confined and unconfined monolayers, \cref{eq velocity-profile-monolayer}. Parameter values are $T_0 = 0.5$ kPa, $L_c = 25$ $\mu$m, $R = 200$ $\mu$m, $h = 5$ $\mu$m, and $\eta = 50$ MPa·s. For the unconfined case, the contractility is $-\zeta = 20$ kPa, under which the monolayer expands [$v(R) > 0$]. For the confined case, the condition $v(R) = 0$ sets the contractility to be given by \cref{eq contractility}. \subref*{FigS9b}, Velocity profile for monolayers of different radius. Parameter values are $T_0 = 0.5$ kPa, $L_c = 25$ $\mu$m, $h = 5$ $\mu$m, and $\eta = 50$ MPa·s, with the critical contractility $-\zeta^*$ given by \cref{eq critical-contractility}.} \label{FigS9}
\end{figure}

\begin{figure}[tb!]
\begin{center}
\includegraphics[width=0.9\textwidth]{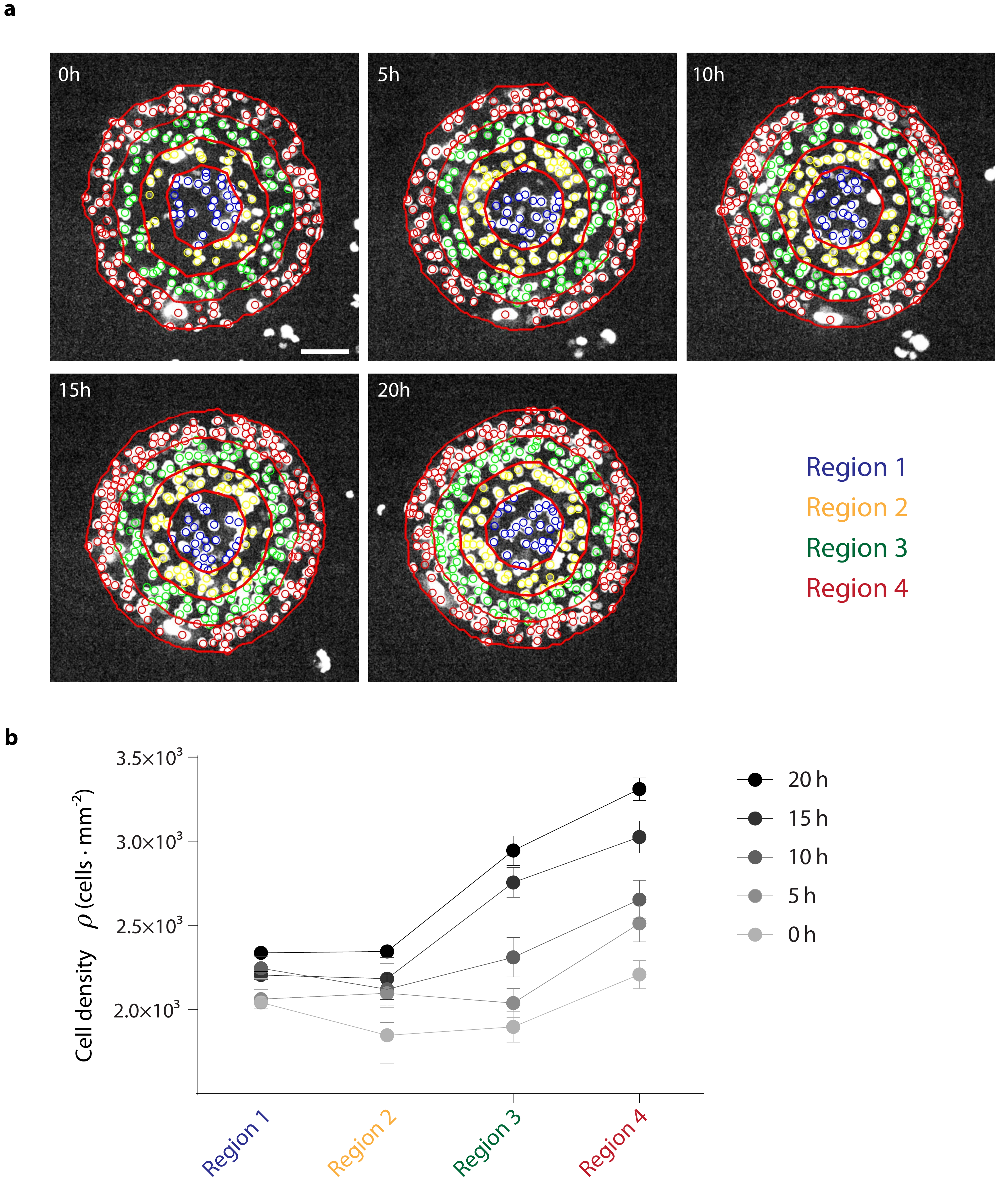}
\end{center}
  {\phantomsubcaption\label{FigS10a}}
  {\phantomsubcaption\label{FigS10b}}
\bfcaption{Cell density increases near the monolayer edge upon induction of E-cadherin expression}{ \subref*{FigS10a}, Expression of H2B-mNeonGreen allows counting cell nuclei in concentric circular coronae. \subref*{FigS10b}, Upon induction of E-cadherin expression, the cell density profile develops a gradient towards the monolayer edge. Data are presented as mean $\pm$ s.e.m. $n=5$ cell islands.} \label{FigS10}
\end{figure}

\begin{figure}[tb!]
\begin{center}
\includegraphics[width=0.8\textwidth]{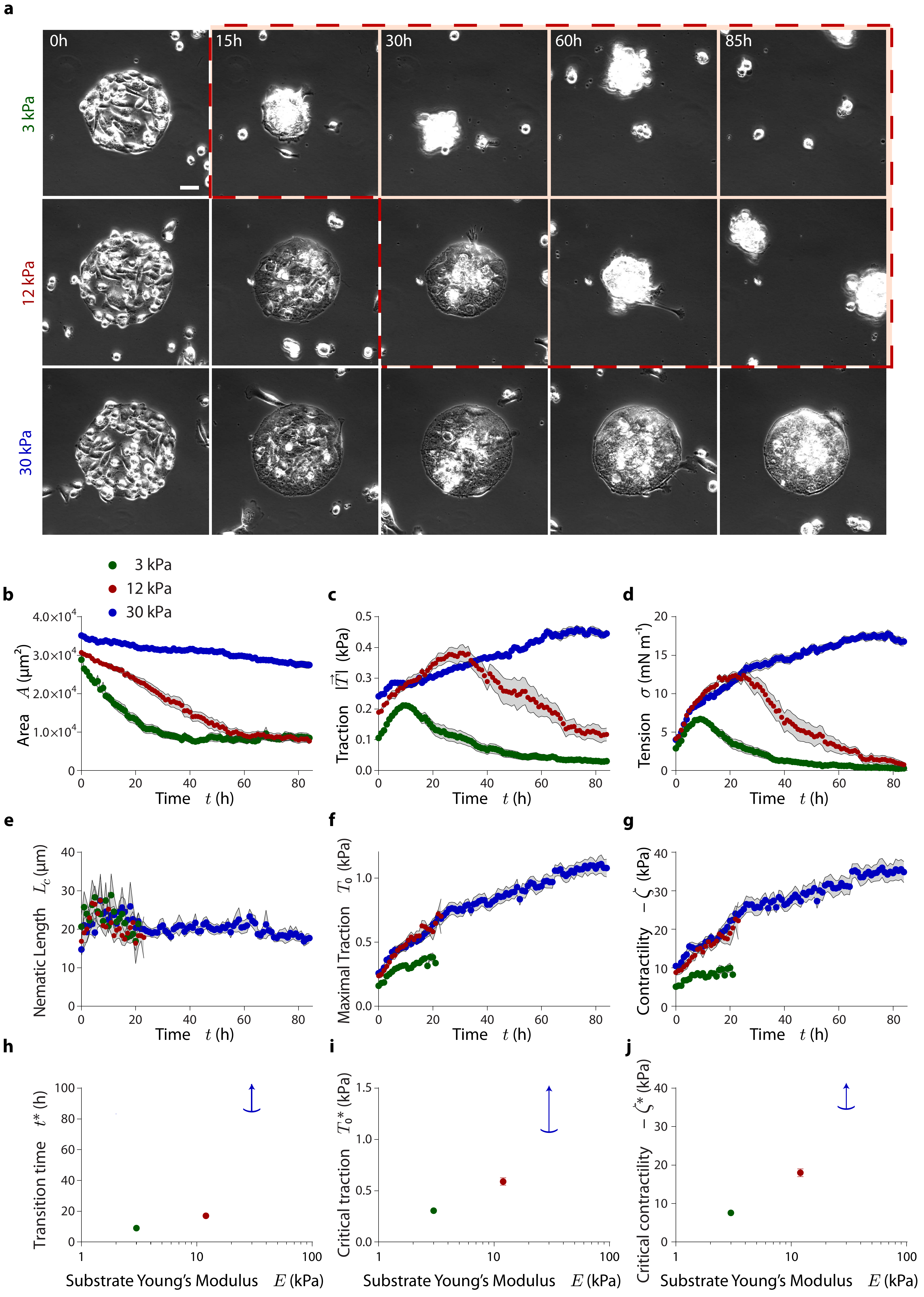}
\end{center}
  {\phantomsubcaption\label{FigS11a}}
  {\phantomsubcaption\label{FigS11b}}
  {\phantomsubcaption\label{FigS11c}}
  {\phantomsubcaption\label{FigS11d}}
  {\phantomsubcaption\label{FigS11e}}
  {\phantomsubcaption\label{FigS11f}}
  {\phantomsubcaption\label{FigS11g}}
  {\phantomsubcaption\label{FigS11h}}
  {\phantomsubcaption\label{FigS11i}}
  {\phantomsubcaption\label{FigS11j}}
\bfcaption{The wetting transition depends on substrate stiffness}{ \subref*{FigS11a}, Time evolution of epithelial monolayers on substrates of different stiffness (Young's modulus of $3$, $12$ and $30$ kPa). Monolayers on stiffer substrates dewet later (red dashed line and shade indicate dewetting). Scale bar $= 40$ $\mu$m. \subref*{FigS11b}-\subref*{FigS11d}, Evolution of monolayer area (\subref*{FigS11b}), mean traction magnitude (\subref*{FigS11c}) and average normal monolayer tension (\subref*{FigS11d}). \subref*{FigS11e}-\subref*{FigS11g}, Evolution of the maximal traction (\subref*{FigS11e}), nematic length (\subref*{FigS11f}), and contractility (\subref*{FigS11g}) for monolayers on substrates of different stiffness. These model parameters were obtained by fitting the model predictions to the experimental data (see Methods). \subref*{FigS11h}-\subref*{FigS11j}, Transition time (\subref*{FigS11h}), critical traction (\subref*{FigS11i}) and critical contractility (\subref*{FigS11j}) for the different substrate stiffnesses. For monolayers on $30$ kPa gels, a wetting transition is not observed within the time of the experiment. Thus, critical parameter values are higher than the maximal value measured at the end of the experiment ($t=84$ h), and hence they are indicated as open intervals (parentheses). Data are presented as mean $\pm$ s.e.m. $n=23$ for $3$ kPa gels, $n=10$ for $12$ kPa gels, and $n=17$ for $30$ kPa gels.} \label{FigS11}
\end{figure}

\begin{figure}[tb!]
\begin{center}
\includegraphics[width=0.95\textwidth]{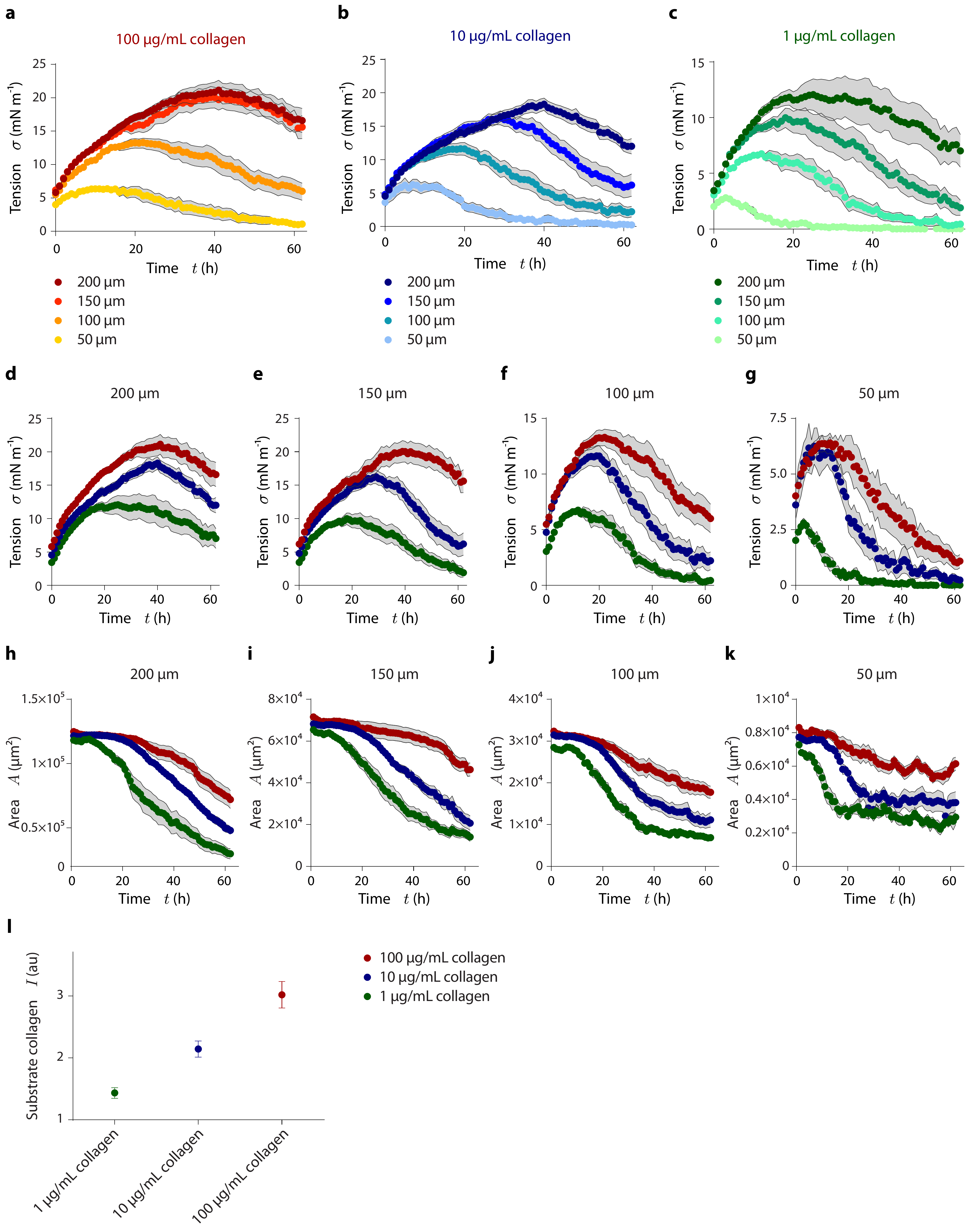}
\end{center}
  {\phantomsubcaption\label{FigS12a}}
  {\phantomsubcaption\label{FigS12b}}
  {\phantomsubcaption\label{FigS12c}}
  {\phantomsubcaption\label{FigS12d}}
  {\phantomsubcaption\label{FigS12e}}
  {\phantomsubcaption\label{FigS12f}}
  {\phantomsubcaption\label{FigS12g}}
  {\phantomsubcaption\label{FigS12h}}
  {\phantomsubcaption\label{FigS12i}}
  {\phantomsubcaption\label{FigS12j}}
  {\phantomsubcaption\label{FigS12k}}
  {\phantomsubcaption\label{FigS12l}}
\bfcaption{Area and monolayer tension evolution in islands of different sizes and substrate ligand densities}{ \subref*{FigS12a}-\subref*{FigS12c}, Evolution of the average normal monolayer tension for islands of different radii on substrates of $100$ $\mu$g/mL of collagen (\subref*{FigS12a}), $10$ $\mu$g/mL of collagen (\subref*{FigS12b}) and $1$ $\mu$g/mL of collagen (\subref*{FigS12c}). \subref*{FigS12d}-\subref*{FigS12g}, Evolution of the average normal monolayer tension for islands seeded on substrates with different collagen densities with radii of $200$ $\mu$m (\subref*{FigS12d}), $150$ $\mu$m (\subref*{FigS12e}), $100$ $\mu$m (\subref*{FigS12f}) and $50$ $\mu$m (\subref*{FigS12g}). \subref*{FigS12h}-\subref*{FigS12k}, Evolution of the mean area for islands seeded on substrates with different collagen density with radii of $200$ $\mu$m (\subref*{FigS12h}), $150$ $\mu$m (\subref*{FigS12i}), $100$ $\mu$m (\subref*{FigS12j}) and $50$ $\mu$m (\subref*{FigS12k}). \subref*{FigS12l}, Relative fluorescence intensity of collagen coating the substrate, quantified by immunostaining for different concentrations of collagen in solution. Data are presented as mean $\pm$ s.e.m. $n=57$ for $100$ $\mu$g/mL of collagen, $n=40$ for $10$ $\mu$g/mL of collagen, and $n=40$ for $1$ $\mu$g/mL of collagen.} \label{FigS12}
\end{figure}

\begin{figure}[tb!]
\begin{center}
\includegraphics[width=0.9\textwidth]{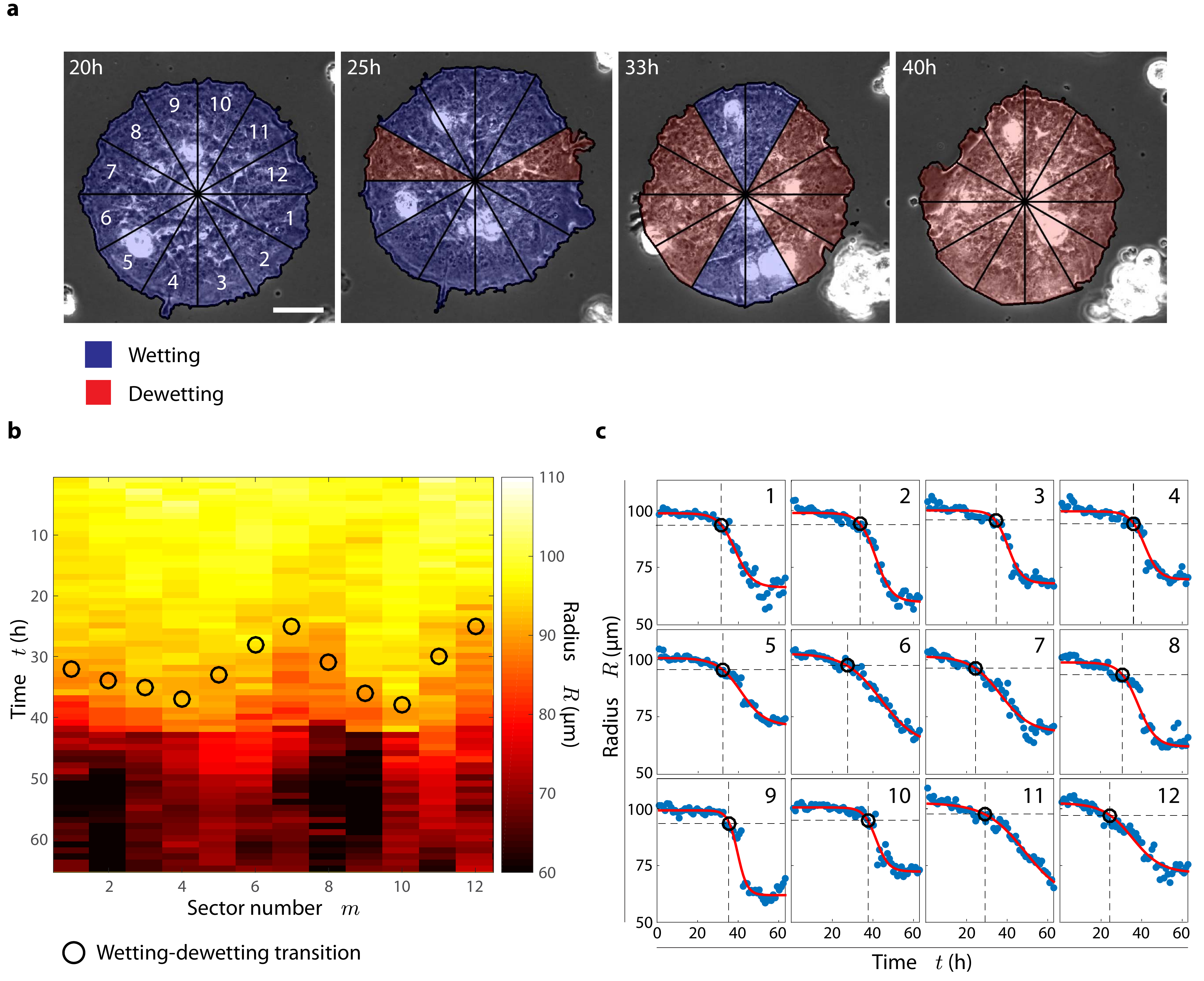}
\end{center}
  {\phantomsubcaption\label{FigS13a}}
  {\phantomsubcaption\label{FigS13b}}
  {\phantomsubcaption\label{FigS13c}}
\bfcaption{User-blind criterion to determine the wetting transition time}{ \subref*{FigS13a}, Cell islands are divided in circular sectors to measure the average radius as a function of the polar angle. Blue = wetting, red = dewetting. \subref*{FigS13b}, Kymograph of the evolution of the average radius of every sector. \subref*{FigS13c}, Fit of a sigmoidal function (red line) to the evolution of the average radius of every sector (blue dots). Open circles indicate the wetting transition for each sector.} \label{FigS13}
\end{figure}

\begin{figure}[tb!]
\begin{center}
\includegraphics[width=0.95\textwidth]{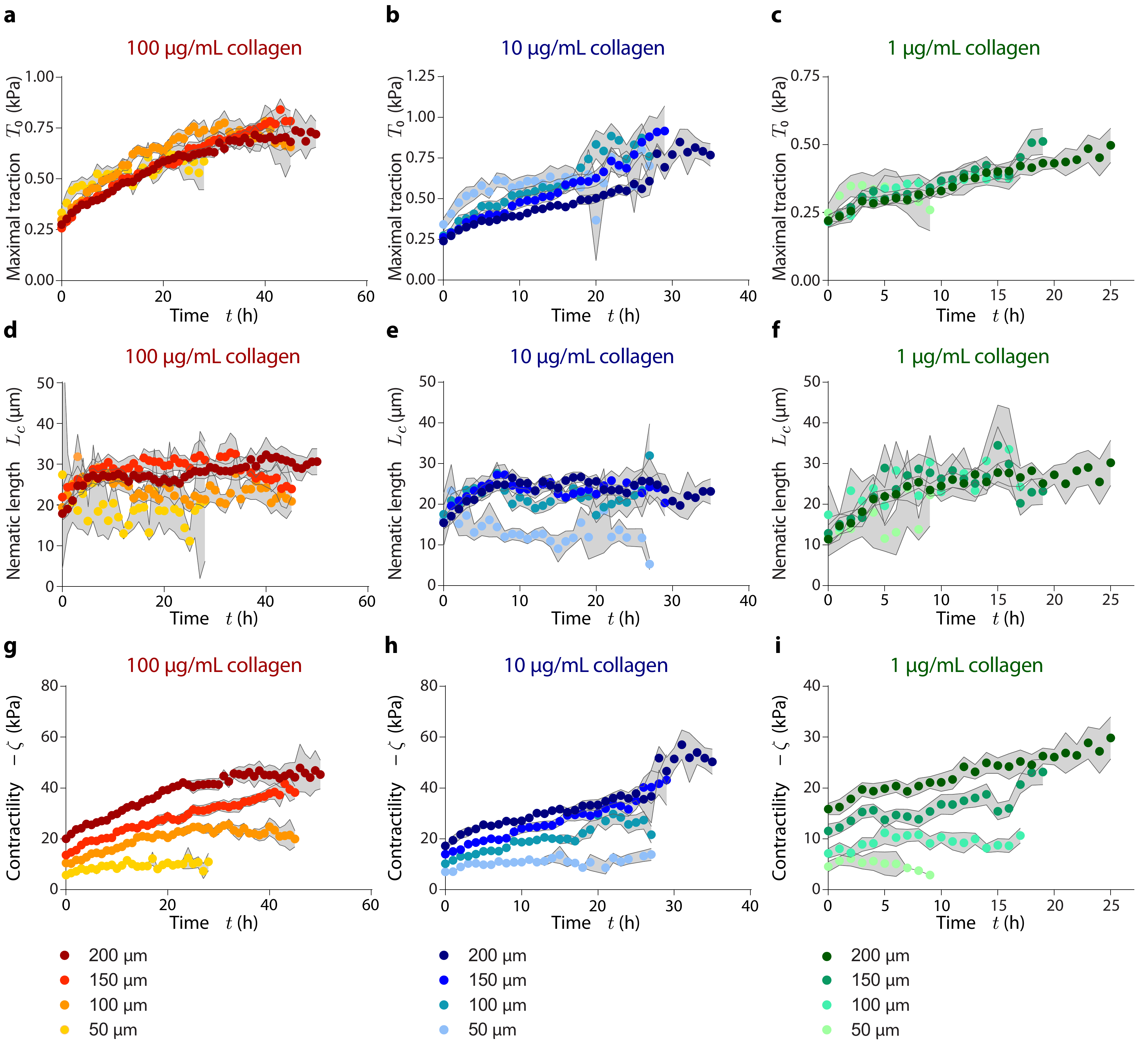}
\end{center}
  {\phantomsubcaption\label{FigS14a}}
  {\phantomsubcaption\label{FigS14b}}
  {\phantomsubcaption\label{FigS14c}}
  {\phantomsubcaption\label{FigS14d}}
  {\phantomsubcaption\label{FigS14e}}
  {\phantomsubcaption\label{FigS14f}}
  {\phantomsubcaption\label{FigS14g}}
  {\phantomsubcaption\label{FigS14h}}
  {\phantomsubcaption\label{FigS14i}}
\bfcaption{Evolution of model parameters in islands of different sizes and substrate ligand densities}{ Evolution of the maximal traction (\subref*{FigS14a}-\subref*{FigS14c}), nematic length (\subref*{FigS14d}-\subref*{FigS14f}) and contractility (\subref*{FigS14g}-\subref*{FigS14i}) in islands of different radii ($50$, $100$, $150$, and $200$ $\mu$m) and substrate ligand densities ($100$, $10$, and $1$ $\mu$g/mL).} \label{FigS14}
\end{figure}

\begin{figure}[tb!]
\begin{center}
\includegraphics[width=0.8\textwidth]{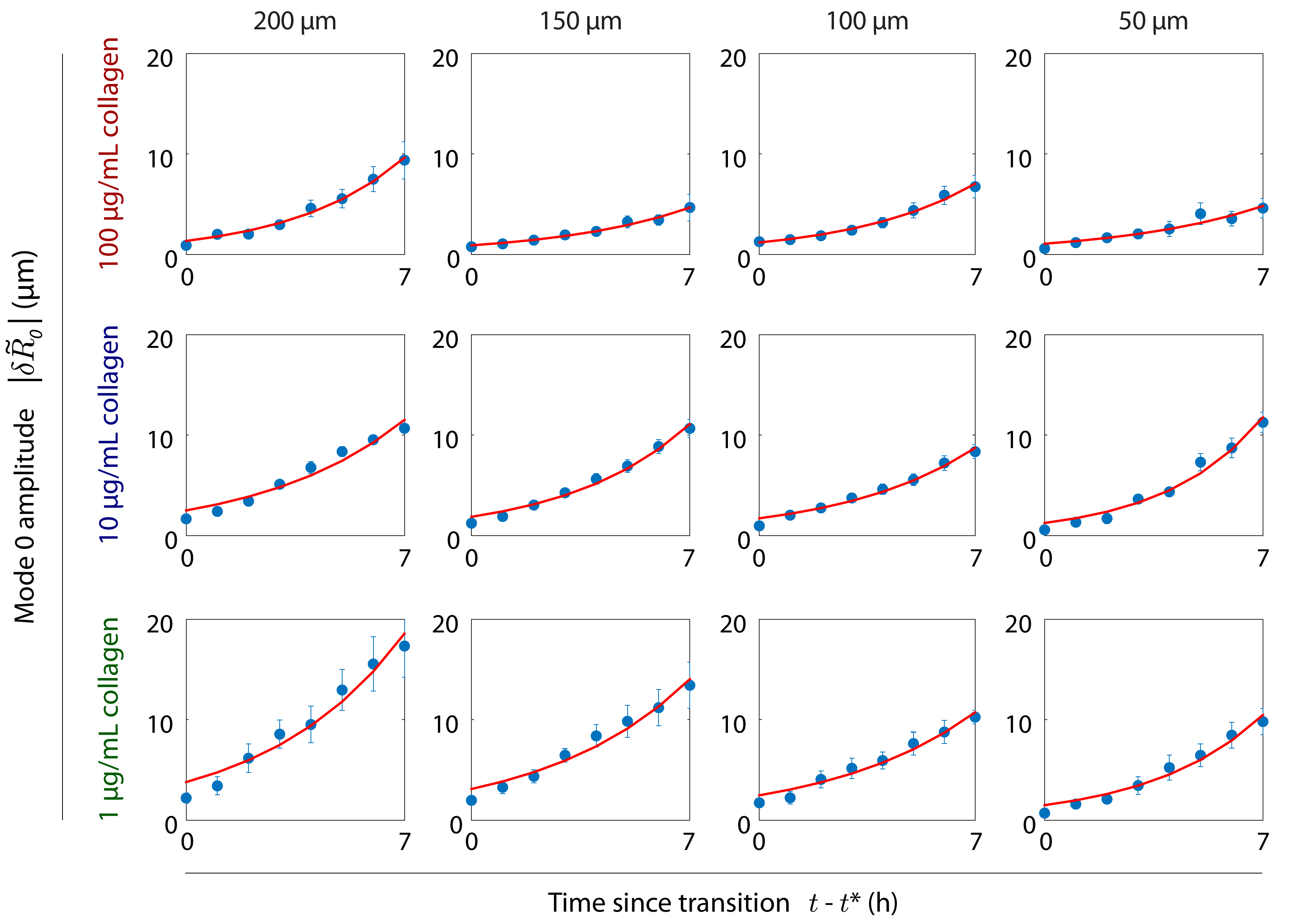}
\end{center}
\bfcaption{Fits of the exponential growth of the perturbation mode $n=0$}{ Evolution of the average
amplitude of the perturbation mode $n=0$ for all different monolayer radii and substrate ligand densities.} \label{FigS15}
\end{figure}

\begin{figure}[tb!]
\begin{center}
\includegraphics[width=0.85\textwidth]{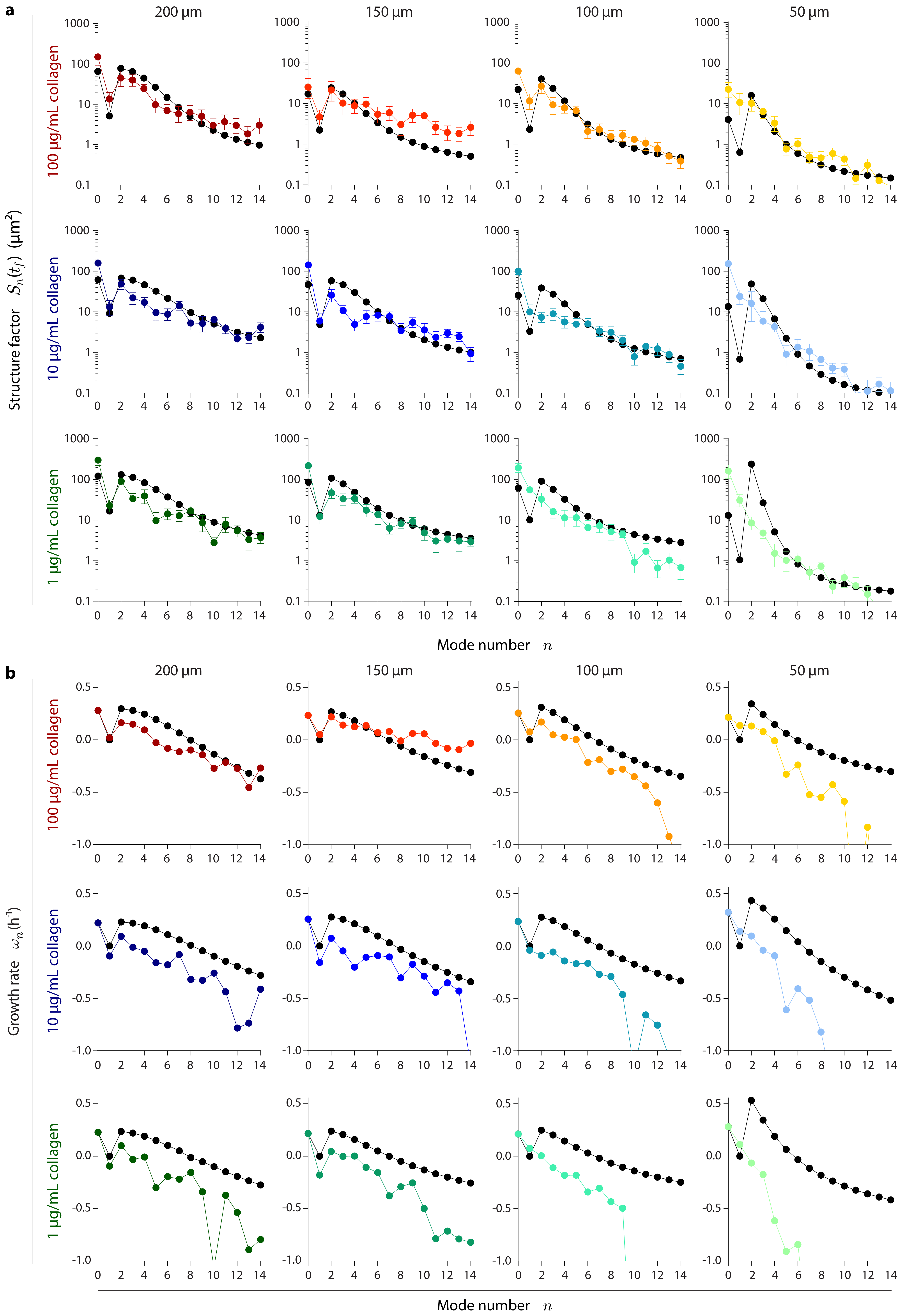}
\end{center}
  {\phantomsubcaption\label{FigS16a}}
  {\phantomsubcaption\label{FigS16b}}
\bfcaption{Structure factors and perturbation growth rates for dewetting monolayers in islands of different sizes and substrate ligand densities}{ \subref*{FigS16a}-\subref*{FigS16b}, Structure factors at time $t_f-t^*=7$ h upon the onset of dewetting (\subref*{FigS16a}) and perturbation growth rates (\subref*{FigS16b}) in islands of different radii ($50$, $100$, $150$, and $200$ $\mu$m) and substrate ligand densities ($100$, $10$, and $1$ $\mu$g/mL). The predicted structure factors (\cref{eq structure-factor}, black symbols) are fitted to the experimental data (color symbols) to obtain the noise intensity $D$ of perturbation mode amplitudes (see \cref{Fig5h}). The predicted growth rates (black symbols) are compared to the experimental data (color symbols).} \label{FigS16}
\end{figure}

\clearpage
\appendix

\onecolumngrid
\begin{center}
\textbf{\large Supplementary Movies}
\end{center}

\textbf{Movie 1. Unconfined monolayer exhibiting a transition from wetting to dewetting.} Representative example of a spreading monolayer (shown in \cref{Fig1f}) undergoing a wetting transition. The release of confinement at $t=0$ h allows the monolayer to freely spread. At $\sim 25$ h, the monolayer spontaneously starts retracting until it collapses into a spheroidal aggregate.

\vskip0.5cm

\textbf{Movie 2. Another example of a wetting transition in a spreading monolayer.} Another spreading monolayer (shown in \cref{FigS2}) undergoing a wetting transition.

\vskip0.5cm

\textbf{Movie 3. Evolution of traction and tension fields during wetting and dewetting.} Videos of phase contrast images (left), maps of traction (center) and monolayer tension (right) in a monolayer with increasing concentration of E-cadherin. A wetting transition is observed at time $t=22$ h.

\vskip0.5cm

\textbf{Movie 4. Orthogonal views of monolayer dewetting.} Time lapse of MDA-MB-231 cells stably expressing a cell membrane marker (CAAX-iRFP). The tissue-substrate contact area decreases pronouncedly during dewetting, while the tissue evolves from a monolayer to a spheroidal cell aggregate, resembling a droplet.

\vskip0.5cm

\textbf{Movie 5. Calcium chelation hinders the increase of tissue forces and prevents dewetting.} Phase contrast, and maps of traction forces and monolayer tension of control (left) and EGTA-treated (right) cell islands. Cells treated with EGTA move individually rather than forming a cohesive monolayer, suggesting that cell-cell junctions are efficiently abrogated. In the presence of EGTA, both tractions and monolayer tension increase much more slowly than in control islands, and the wetting transition does not occur.

\vskip0.5cm

\textbf{Movie 6. Dewetting is inhibited and reversed when tissue contractility is externally decreased.} Dewetting (left), dewetting inhibition (center) and reversibility (right) assays. Partial inhibition of contractility with blebbistatin clearly delays the wetting transition. A sudden inhibition of contractility with Y27632 ($t=46$ h) is enough to revert dewetting, inducing a rewetting of the substrate. The name of the drug indicates its presence in the cell medium.

\vskip0.5cm

\textbf{Movie 7. Cell rearrangements in the monolayer.} Phase contrast (left) and cell nuclei (right) in a $200$ $\mu$m radius island during the wetting phase of the experiment. Cells incessantly exchange neighbors, a fact that provides support to the fluid behavior of the monolayer. Moreover, cells progressively accumulate at the edge of the monolayer, which develops a gentle cell density gradient.

\vskip0.5cm

\textbf{Movie 8. Evolution of traction and monolayer tension fields in islands of different radii.} For all sizes, the magnitude of tractions and monolayer tension increase in time as E-cadherin is progressively expressed. Tractions accumulate at the edges of the monolayers, while monolayer tension has a maximum at the center. Red frames indicate monolayer dewetting.

\vskip0.5cm

\textbf{Movie 9. Evolution of traction and monolayer tension fields in islands on substrates of different stiffnesses.} For monolayer on substrates of Young's modulus $3$ and $12$ kPa, tissue forces increase in time, eventually triggering monolayer dewetting. This transition occurs earlier for the softest substrate. For the stiffest substrate ($30$ kPa), tissue forces keep increasing until the end of the experiment, suggesting that the critical contractility to induce dewetting is not reached. Red frames indicate monolayer dewetting.

\end{document}